\documentclass[12pt]{article}
\usepackage{amsmath}
\usepackage{amssymb}
\usepackage{calc}
\usepackage{amsfonts}
\usepackage{graphicx}
\usepackage[ansinew]{inputenc}
\usepackage{color}

\catcode`\@=11
\def\marginnote#1{}
\def\draftlabel#1{{\@bsphack\if@filesw {\let\thepage\relax
   \xdef\@gtempa{\write\@auxout{\string
      \newlabel{#1}{{\@currentlabel}{\thepage}}}}}\@gtempa
   \if@nobreak \ifvmode\nobreak\fi\fi\fi\@esphack}
        \gdef\@eqnlabel{#1}}
\def\@eqnlabel{}
\def\@vacuum{}
\def\draftmarginnote#1{\marginpar{\raggedright\scriptsize\tt#1}}
\def\draft{\oddsidemargin -.5truein
        \def\@oddfoot{\sl preliminary draft \hfil
        \rm\thepage\hfil\sl\today}
        \let\@evenfoot\@oddfoot \overfullrule 3pt
        \let\label=\draftlabel
        \let\marginnote=\draftmarginnote
   \def\@eqnnum{(\theequation)\rlap{\kern\marginparsep\tt\@eqnlabel}
\global\let\@eqnlabel\@vacuum}  }

\def\preprint{\twocolumn\sloppy\flushbottom\parindent 1em
        \leftmargini 2em\leftmarginv .5em\leftmarginvi .5em
        \oddsidemargin -.5in    \evensidemargin -.5in
        \columnsep 15mm \footheight 0pt
        \textwidth 250mmin      \topmargin  -.4in
        \headheight 12pt \topskip .4in
        \textheight 175mm
        \footskip 0pt
        \def\@oddhead{\thepage\hfil\addtocounter{page}{1}\thepage}
        \let\@evenhead\@oddhead \def\@oddfoot{} \def\@evenfoot{} }
\def\signed{
        \def\@oddfoot{\sl \small MN \&AB  : Effective gravitational action for 2d massive fermions\hfil
        \rm\thepage\hfil\sl \small for arbitrary genus \hskip2.mm \today }
        \let\@evenfoot\@oddfoot 
  \global  }

\def\titlepage{\@restonecolfalse\if@twocolumn\@restonecoltrue\onecolumn
     \else \newpage \fi \thispagestyle{empty}\c@page\z@
        \def\thefootnote{\fnsymbol{footnote}} }

\def\endtitlepage{\if@restonecol\twocolumn \else  \fi
        \def\thefootnote{\arabic{footnote}} \setcounter{footnote}{0}}

\catcode`@=12 \relax
\def\dsp{\displaystyle}
\def\bea{\begin{array}}
\def\eea{\end{array}}
\def\beb{\be\begin{array}{|c|}\hline\\ \dsp}
\def\bebns{\be\begin{array}{|c|}\hline \dsp}
\def\eeb{\\ \\ \hline\end{array}\ee}
\def\eebns{\\ \hline\end{array}\ee}
\def\bab{\be\begin{array}{|ccc|}\hline&&\\ \dsp}
\def\eab{\\ &&\\ \hline\end{array}\ee}
\def\eabns{\\ \hline\end{array}\ee}

\def\bra#1{\left\langle #1\right|}

\def\ket#1{\left| #1\right\rangle}

\def\Im{\mathop{\rm Im}}
\def\ov{\overline}

\relax
\def\be{\begin{equation}}
\def\ee{\end{equation}}
\def\ba{\begin{eqnarray}}
\def\ea{\end{eqnarray}}
\def\del{\partial}
\def\d{{\rm d}}
\def\tr{\,{\rm tr}\,}

\def\Tr{\,{\rm Tr}\,}
\def\Det{\,{\rm Det}\,}

\def\r{\rho}
\def\a{\alpha}
\def\b{\beta}
\def\g{\gamma}

\def\G{\Gamma}
\def\dd{\delta}
\def\D{\Delta}
\def\e{\epsilon}
\def\f{\phi}
\def\F{\Phi}
\def\p{\psi}

\def\zb{\bar z}
\def\P{\Psi}

\def\m{\mu}
\def\n{\nu}
\def\o{\omega}

\def\l{\lambda}
\def\L{\Lambda}
\def\s{\sigma}

\def\vf{\varphi}

\def\cN{{\cal N}}
\def\cM{{\cal M}}

\def\cD{{\cal D}}
\def\cR{{\cal R}}

\def\St{{\overline S}}
\def\Pt{{\overline P}}

\def\t{\theta}

\def\wt{\widetilde}

\def\wh{\widehat}

\def\dsl{{\del\hskip-2.2mm /}}

\def\Nsl{{\nabla\hskip-3.0mm /}}

\numberwithin{equation}{section}

\relax

\renewcommand{\theequation}{\thesection.\arabic{equation}}
\begin{document}
\topmargin-2.4cm
%
%
%
%
\begin{titlepage}
\begin{flushright}
July 2023\\
\end{flushright}
\vskip 0.5cm

\begin{center}
{\Large\bf Effective gravitational action for 2D massive Majorana }\\
\vskip2.mm
{\Large\bf fermions on arbitrary genus Riemann surfaces }\\

\vskip12.mm

{\large\bf Manojna Namuduri  and  Adel Bilal }
\\

\medskip
\it {Laboratoire de Physique de l'\'Ecole Normale Sup\'erieure\\
PSL University, CNRS, Sorbonne Universit\'e, Universit\'e de Paris\\
24 rue Lhomond, F-75231 Paris Cedex 05, France}

\end{center}
\vskip .3cm

\begin{center}
{\bf \large Abstract}
\end{center}
\begin{quote}
We  explore the effective gravitational action for two-dimensional massive Euclidean\break Majorana fermions in a small mass expansion, continuing and completing the study initiated in a previous paper \cite{BDE}.  We perform a detailed analysis of local zeta functions, heat kernels, and Green's functions of the Dirac operator on arbitrary Riemann surfaces. We obtain the full expansion of the effective gravitational action to all orders in $m^2$. For genus one and larger, this requires the understanding of the role of the zero-modes of the (massless) Dirac operator which is worked out.

Besides the Liouville action, at order $m^0$, which only involves the background metric and the conformal factor $\sigma$,  the various contributions to the effective gravitational action at higher orders in $m^2$ can be expressed in terms of integrals of the renormalized Green's function at coinciding points  of the squared (massless) Dirac operator, as well as of higher Green's functions.   In particular, at order $m^2$, these contributions can be re-written as a term $\int e^{2\sigma}\, \sigma$ characteristic of the Mabuchi action, much as for 2D massive scalars, as well as several other terms that are multi-local in the conformal factor $\s$ and involve the  Green's functions of the massless Dirac operator and the renormalized Green's  function, but for the background metric only.
\end{quote}

\end{titlepage}
%
%
%
{\parskip -0.3mm
\small{\tableofcontents}}

\setcounter{footnote}{0}

\newpage
\setcounter{page}{1}

%

\setcounter{section}{0}

\section{Introduction\label{intro}}
\setlength{\baselineskip}{.56cm}
\subsection{Motivation and outline of the paper\label{motivsec}}


Quantum field theory on a fixed curved Lorentzian (space-time) or Euclidean (space) manifold has been extensively studied for more than half a century. Standard references include \cite{QFTcurved}. One of the most prominent early manifestations was the discovery of Hawking radiation near a black hole horizon \cite{HR}. In a full theory of quantum gravity one should, of course, also do the functional integral over all geometries which means summing over topologies and, for each topology, over the inequivalent metrics. This is a notoriously difficult problem. The usual way to decompose the problem is to first do the functional integral over the ``matter" fields giving the so-called matter partition function, or equivalently the effective gravitational action, and deal with the functional integral over the geometries only in a second step. This is, of course, similar to what one often does for non-abelian gauge theories where one first integrates over the matter fields for a fixed gauge field configuration yielding an effective action $S_{\rm eff}[A]$ of the gauge fields only, and then deals with the problems of gauge fixing only in a second step. In particular, this allows to discuss important issues like anomalies already at the level of this effective action \cite{ABanomaly}. Here we will be interested similarly in determining an effective action for two-dimensional gravity and matter systems on spaces of Euclidean signature.

More precisely, the functional integral over the matter field(s) which we denote generically by $\P$, computed with a fixed metric $g$ on a given manifold $\cM$, defines the matter partition function  as $Z_{\rm mat}[g]= \int \cD \P \exp\left( - S_{\rm mat}[g,\P]\right)$. In terms of this, the effective gravitational action  is defined, as usual, as the ratio of $Z_{\rm mat}[g]$ and $Z_{\rm mat}[\wh g]$, where we consider $\wh g$ as a reference metric~:
\be\label{Sgravgen}
\exp\big(-S_{\rm grav}[g,\wh g] \big) = \frac{Z_{\rm mat}[g]}{Z_{\rm mat}[\wh g]}\ .
\ee
In two dimensions, the metric depends on 3 functions, but using the diffeomorphism invariance one can essentially fix two of them, leaving only one geometrically relevant function which can be chosen as an overall conformal factor $e^{2\s(x)}$. Hence, for fixed topology and finitely many fixed (``modular")  parameters $\tau_i$ we can always pick a reference metric $\wh g\equiv \wh g[\tau_i]$ such that any other metric (corresponding to the same $\tau_i$) can be written, up to diffeomorphisms, as\footnote{
This is a global statement. Locally one can go further and choose so-called isothermal coordinates \cite{Chern} which are such that $g_{\m\n}(x)=e^{2\l(x)}\, \dd_{\m\n}$, but obviously this cannot be done globally unless $\int\sqrt{g} \cR=8\pi (1-{\tt g})$ vanishes.
}
\be\label{g-ghattauirel}
g_{\m\n}(x)=e^{2\s(x)}\, \wh g_{\m\n}(x) \ .
\ee
We may then equivalently consider that $S_{\rm grav}$ depends on $\wh g$ and $\s$~: $S_{\rm grav}[g,\wh g] \equiv S_{\rm grav}[\s,\wh g]$.
The definition \eqref{Sgravgen} implies that $S_{\rm grav}$ must satisfy a cocycle identity
\be\label{cocycle}
S_{\rm grav}[g_3,g_2] + S_{\rm grav}[g_2,g_1]=S_{\rm grav}[g_3,g_1] \ .
\ee
The  best-known example is the effective gravitational action for conformal matter coupled to 2D gravity which is the Liouville action  \cite{Liouville1}
\be\label{Liouvaction}
S_{\rm Liouville}[g,\wh g] \equiv S_{\rm Liouville}[\s,\wh g]=\int\d^2 x \sqrt{\wh g} \, \big( \s \wh\D\s + \wh \cR \s\big) 
, \quad g=e^{2\s} \wh g\ ,
\ee
where $\wh\cR$ is the scalar curvature for the metric $\wh g$, and
\be\label{Liouvillegrav}
S_{\rm grav}[g,\wh g]=-\frac{c}{24\pi} S_{\rm Liouville}[g,\wh g] \ ,
\ee
where $c$ is the central charge of the conformal matter system, namely $c=1$ for a massless scalar and $c=\frac{1}{2}$ for a massless Majorana spinor.
An even simpler example satisfying this cocycle identity is the ``cosmological constant action"
\be\label{cosmolconst}
S_{\rm c}[g,\wh g]=\m_0  \int \d^2 x (\sqrt{g} -\sqrt{\wh g})=\m_0 (A-A_0) \ .
\ee
This action typically is also  present as a counterterm, in addition to $S_{\rm Liouville}$, to renormalize certain divergences that might be present.

But these are not the only two-dimensional gravitational actions, satisfying the cocycle condition, that can be constructed and have been studied  in the mathematical literature. Other appropriate functionals are the Mabuchi and Aubin-Yau actions \cite{Mabuchi,AubinYau}, to be defined in the next subsection. These latter functionals   involve not only the conformal factor $\s$ but also  the K\"ahler potential $\F$,  and can also be generalized to higher-dimensional  K\"ahler manifolds, where they  appear in relation with the characterization of constant scalar curvature metrics \cite{AubinYau}.
In the  context of a two-dimensional effective action, they appeared first in ref.~\cite{FKZ} where the matter partition function of non-conformal matter like a massive scalar field was studied.  It was shown there  that  in this case the gravitational action as defined by \eqref{Sgravgen} contains these Mabuchi and Aubin-Yau actions  as first-order in $m^2$ corrections  to the Liouville action. Higher-order corrections in $m^2$ to the effective gravitational action for such massive scalar fields have been obtained\footnote{
Note that at present we call $S_{\rm grav}[g,\wh g]$ what was called $S_{\rm grav}[\wh g,g]$ in refs \cite{FKZ,BL,BL2,BdL2}.
} 
 in \cite{BL2}. The interesting extension to two-dimensional manifolds with boundaries was studied in \cite{BdL2}.

The fixed-area partition function of quantum gravity, with a gravitational action being a sum of the Liouville and Mabuchi actions, has been studied at one loop in \cite{BFK} and at  two and three loops in \cite{BL}.  In a different approach, a rigourous mathematical construction of the functional integral based on the coupling of the Liouville and Mabuchi actions has  been obtained in ref \cite{LRV} by means of probabilistic tools.  

So far, references \cite{FKZ,BL2,BdL2}  considered the effective gravitational action of a two-dimensional massive scalar field. It is natural to try to extend this to the case of two-dimensional massive fermions. The most ``elementary" 2D fermion is a Majorana fermion. As we will discuss below,  this corresponds to a real anti-commuting spinor field with a Majorana-type mass term. Since the corresponding Dirac operator can be chosen purely imaginary, classically, at the level of the Dirac equation, one can impose a Majorana (reality) condition, on the spinor. In the quantum theory however, we need to expand the Majorana field on the eigenfunctions of this purely imaginary Dirac operator, and then these eigenfunctions are necessarily complex. One might ask whether it is then appropriate to talk about Majorana fermions in the quantum theory. However, we can obtain the partition function in terms of the eigenvalues of the (real) square of the Dirac operator  whose eigenfunctions can all be chosen to be real. Moreover, we will show that we can  consistently do the functional integral over a {\it real} anti-commuting spinor field. If one is interested instead in a (complex) Dirac spinor, the corresponding gravitational action is easily seen to be simply twice the one for a Majorana spinor.

In a previous paper \cite{BDE} involving one of the present authors, the study of this effective gravitational action for massive Majorana fermions was initiated.\footnote{
After the present paper was posted on arXiv, the authors of \cite{RSZ} kindly brought their reference to our attention, where they also study the effective gravitational action, up to terms quadratic in the curvature, for two-dimensional massive scalar, Dirac and vector matter fields, along the heat kernel methods as worked out in \cite{Avramidi}. Their emphasis is mostly on  obtaining the beta-functions.
}
While at first this appeared to be a simple generalisation of the case of the scalar field, it actually turned out to be technically quite non-trivial.  Much as for the scalar case, the renormalized Green's functions at coinciding points played an important role. In order  to do a small mass expansion one must be able to define these Green's functions also for the massless theory in which case the contributions of the zero-modes (of the massless Dirac operator) must be subtracted. For spherical topology there are no zero-modes and then this subtlety does not occur. For this reason, \cite{BDE}  established the form of the effective gravitational action only for spherical topology and only up to order $m^2$. Further technical subtleties implied that even in this case the result was more complicated than the one  obtained in \cite{FKZ} or \cite{BL2} for the massive scalar field. Nevertheless, a term characteristic of the Mabuchi action emerged.

In the present paper we will generalise this result  and establish the effective gravitational action for arbitrary genus Riemann surfaces (without boundaries), and in an expansion to all orders in $m^2 A$. We will show that this expansion has a finite radius of convergence !

An important ingredient which enabled us to go beyond spherical topology was the understanding of the structure of the projectors on the zero-modes of the massless Dirac operator. Indeed,  the massless Dirac operator transforms in a  particularly simple way under local conformal transformation (which is neither the case for the massive Dirac operator, nor for the non-zero eigenvalue equation of the Dirac operator, even for zero mass). This implies that the zero-modes of the massless  Dirac operator for metric $g$ are related by a very simple conformal rescaling\footnote{
This result is implicit in \cite{Hitchin, DP} and was spelled out more explicitly in an early draft of \cite{LEL}.
} 
to the zero-modes of the massless Dirac operator for metric $\wh g$, and then a similarly simple relation exists between the zero-mode projectors for metrics $g$ and $\wh g$. This allows us to characterise the zero-mode contributions in general, valid on Riemann surfaces of arbitrary genus. 

The study of the massive and massless Dirac operator $D$ and of its square $D^2$, of their associated Greens's functions and  local zeta-functions  and local heat kernels  occupied a major part in \cite{BDE} and will again occupy an important part of the present paper. However, here we will be able to do a much more systematic study of these quantities and prove several important relations, some of which were only conjectured in \cite{BDE}. As a result, we are able to obtain the effective gravitational action as an all-orders expansion in powers of $m^2$ with each term in the expansion  expressed as an integral over the manifold of these Green's functions, appropriately renormalized, as well as higher Green's functions at coinciding points. As already mentioned, a detailed knowledge of the contributions of the zero-modes (for zero mass) is required to perform the small-mass expansion correctly. 

We will see  that besides the leading Liouville term, we get a new order $m^0$ contribution when  zero-modes of the massless Dirac operator are present. It might seem surprising to get a ``new" order $m^0$ contribution in the massive theory that seemed not to be there in the massless theory. Actually, we will see that this same contribution does arise in the massless theory. However, in the massless theory one could modify or even remove these new terms by appropriately changing the definition of the matter partition function. On the other hand, in the massive theory there is no reason to do such a redefinition and these zero-mode related contributions are genuinely  present.

 At order $m^2$ we get a cosmological constant action, and a local $\int\sqrt{\wh g}\, \s e^{2\s}$ term characteristic of the Mabuchi action,
as well as some further ``non-local terms" involving integrals of the Green's functions computed with the metric $g$. We have been able to rewrite the latter as some multi-local functionals in the conformal factor $\s$ (i.e.~multiple integrals involving the conformal factor at the different integration points) and furthermore only involving the Green's functions computed in the background metric $\wh g$, as well as a finite number of area-like parameters. It would, of course, be desirable to rewrite all terms as purely local (single) integrals expressed in terms of local quantities like the conformal factor, the K\"ahler potential or some generalisation thereof, and the metric and curvature. However, as usual with effective actions, there is no reason that they should be local, and we think at present that our multi-local form (or some equivalent formulation) is the best one can achieve.

The effective gravitational action is of course to be used  if one wants to compute the functional integral over the geometries, which involves in particular a functional integration over the conformal factor ${\cal D}\s$, similarly to what was done in \cite{BFK} at one loop or \cite{BL} for higher loops. The form of the effective gravitational action we have obtained explicitly displays this $\s$-dependence, having it separated from the quantities that only depend on the fixed ``background" metric $\wh g$. This explicit separation should facilitate this integration over the geometries.

We have tried to write the present paper in a hopefully self-contained and pedagogical way. It is organised as follows. In the second part of this introduction we recall a few facts about the Liouville, Mabuchi and Aubin-Yau actions and their stationary points, and  then briefly recall, as a warm-up exercise, how one can show that the effective gravitational action for a massless scalar field is the Liouville action. In the next section we introduce the relevant differential operators : massive Dirac operator $D$, its square $D^2$, and the scalar and spinoral Laplacians. We discuss their eigenvalue problems in some detail and exhibit the eigenvalues and eigenfunctions for the simple example of the flat torus. We work out the transformations of $D$ and of $D^2$  under conformal changes of the metric (both infinitesimal and finite). 
We  also work out exactly how the zero-modes of the (massless) Dirac operator change under the local conformal rescalings, allowing us  to precisely define the zero-mode projectors that are crucial ingredients for the small mass expansion. Indeed, as already mentioned,  these projectors have to be subtracted later-on from the Green's function in order to define the Green's functions of the massless theory around which the small-mass expansion will be done. 

In section 3 we discuss the fermionic functional integral defining the matter partition function and obtain the gravitational action. We show precisely how it is related to the product of the eigenvalues of the Dirac operator $D$ and how this is related to the corresponding zeta-function. We offer also a few remarks about the convergence of these zeta-functions and their definition by analytic continuation. We show how these zeta-functions change under infinitesimal conformal rescalings of the metric.

Then, section~4  contains most of the important technical results~: we embark on a detailed study of the different Green's functions, local zeta-functions $\zeta_\pm(s,x,y)$ and local heat kernels $K_\pm(t,x,y)$. We express precisely how the change  of the zeta function (expressing the gravitational action) under infinitesimal conformal rescalings $\dd\s$ is related to integrals of the local zeta functions times $\dd\s$.
  We prove a precise correspondence between the local zeta function $\zeta_-(s,x,y)$ associated with $D$ and the local zeta function $\zeta_+(s+\frac{1}{2},x,y)$ associated with $D^2$. This relation  allows us to make precise statements\footnote{In particular, this allows us to correct some inexact conjectures made in \cite{BDE} about the small-$t$ asymptotics of $K_-(t,x,x)$, as well as about the poles of $\zeta_-(s,x,x)$ at negative half-integers which turn out not to be there.}
  about the singularity structures of both local zeta functions $\zeta_\pm$, as well as about the small-$t$ asymptotics of the  heat kernel\  $K_-(t,x,y)$, solely from the knowledge of the corresponding asymptotics of $K_+(t,x,y)$ that is worked out in the appendix.
   
All this is put together in section~5 to obtain the variation of the effective gravitational action. We will be able to isolate the zero-mode contributions for arbitrary topology and thus obtain results valid on Riemann surfaces of arbitrary genus. The variation of the gravitational action is obtained as a total variation of a converging  series expansion in powers of $m^2$. ``Integrating" these variations at any order in $m^2$ gives the corresponding gravitational action at any order in terms of integrals over higher and higher Green'a function (computed in the metric $g$) at coinciding points. 
We then work out how the order $m^2$ contribution can be re-expressed in terms of the Mabuchi type term $\int \sqrt{\wh g}\, \s\, e^{2\s}$ and integrals and multiple integrals involving only the conformal factor $\s$ at the different integration points, as well as the Green's function of $D$ for the background metric $\wh g$ and the renormalized Green's function of $D^2$ at coinciding points, also in the background metric $\wh g$.  We expect that a similar rewriting can also be performed on the contributions of  order  $m^4$ and higher.

In appendix A we give the small-$t$ asymptotics of the heat kernel $K_+(t,x,y)$ by solving recursively the associated (heat) differential equation. We do this first quite generally for positive second-order  differential operators and then specialise our results to the squared Dirac operator. No such recursive solution is  available for the corresponding $K_-$ since it does not have a simple initial condition.
 In appendix B we discuss some properties of this $K_-(t,x,y)$ for the flat torus in which case we have explicit eigenvalues and eigenfunctions at our disposal, confirming the general properties obtained in section 4.
 
Results largely overlapping with ours, in particular concerning the zero-mode projectors, have been  obtained independently in \cite{LEL}. Their treatment of the orthonormalisation of the zero-modes allowed us to correct a slight error in the first version of the present paper. 

\subsection{Some (more or less) well-known gravitational actions : Liouville, Mabuchi and Aubin-Yau}

Besides the Liouville action $S_{\rm Liouville}$ which we already defined in \eqref{Liouvaction}, and the somewhat trivial cosmological constant action $S_{\rm c}$ given in \eqref{cosmolconst} there are two other, more or less well-known gravitational actions, namely the Mabuchi and Aubin-Yau actions. 
Let us briefly recall the definitions and basic properties of these  gravitational actions.
While the Liouville action \eqref{Liouvaction} can be written in terms of $\wh g$ and the conformal factor $\s$, the Mabuchi and Aubin-Yau actions are formulated using also the K\"ahler potential $\F$. The conformal factor $\s$ and the K\"ahler potential $\F$ are related by (we may take this as the definition of the K\"ahler potenial $\F$)
\be\label{gg0sig}
g = e^{2\sigma}\wh g \quad , \quad
e^{2\s}= \frac{A}{\wh A}\left(1-\frac{1}{2} \wh A \,\wh\D \F\right)
\quad , \quad
\D=\frac{1}{\sqrt{g}}\del_\m\big(\sqrt{g} g^{\m\n}\del_\n\big) = e^{-2\s} \wh \D \ ,
\ee
where $\wh \D$ denotes the (scalar) Laplacian for the metric $\wh g$ with area $\wh A=\int\d^2 x \sqrt{\wh g}$. 
The second relation in this equation \eqref{gg0sig} shows that the K\"ahler potential describes different metrics of the same total area $A$, and the latter needs to be introduced as an additional variable. This separation becomes particularly clear if we write out how the variations of $\s$ are related to the variations of $\F$ and $A$. It follows from \eqref{gg0sig} that
\be\label{deltasigmadeltaphi}
\dd\s=\frac{\dd A}{2A}-\frac{A}{4} \D\dd\F
\quad \Leftrightarrow \quad
\dd\left(\frac{e^{2\s}}{A}\right)=-\frac{1}{2}\,\wh\D\,\dd\F \ .
\ee

In terms of the K\"ahler potential $\F$, the area $A$ and the conformal factor $\s$, the Mabuchi action on a Riemann surface of genus ${\tt g}$ can then be written as \cite{FKZ}
\be\label{Mab1}
S_{\rm M}[\wh g,g]
=\int\d^2 x \sqrt{\wh g} \left[ 2\pi({\tt g}-1)\F\,\wh\D\,\F+ \Bigl(\frac{8\pi(1-{\tt g})}{\wh A}-\wh \cR\Bigr) \F +\frac{4}{A}\, \s \,e^{2\s} \right] \  ,
\ee
and the Aubin-Yau action as
\be\label{AubY}
S_{\rm AY}[\wh g,g]=-\int\d^2 x \sqrt{\wh g} \left[ \frac{1}{4} \F\,\wh\D\,\F -\frac{\F}{\wh A}\right] \ .
\ee
As already mentioned, they both satisfy a cocycle identity analogous to \eqref{cocycle} and were shown \cite{FKZ} to appear in the effective gravitational action of a (two-dimensional) massive scalar field in the term of first order in an expansion  in $m^2 A$. Note that we can rewrite the Mabuchi action as
\be\label{MabuchiAY}
S_{\rm M}=8\pi(1-{\tt g}) S_{\rm AY}+\int\d^2 x \sqrt{\wh g}\left( \frac{4}{A}\, \s\, e^{2\s} -\wh \cR\F\right) \ .
\ee

Let us now give the variations of  the Liouville, Mabuchi and Aubin-Yau actions under a conformal variation of the metric (equivalently encoded in $\dd\s$ or in $\dd\F$ and $\dd A$, at fixed $\wh g$ and hence fixed $\wh A$). First, the Aubin-Yau action only depends on $\F$, not on $A$, and
\ba\label{AYvar}
\dd S_{\rm AY}[\wh g,g]\hskip-2.mm&=&\hskip-2.mm 
-\int\d^2 x \sqrt{\wh g} \left[ \frac{1}{2} \dd\F\,\wh\D\,\F -\frac{\dd\F}{\wh A}\right] 
=\frac{1}{A}\int\d^2 x \sqrt{\wh g} \left[- \frac{A}{2} \,\wh\D\,\F +\frac{A}{\wh A}\right] \dd\F
= \frac{1}{A}\int\d^2 x \sqrt{\wh g} \, e^{2\s} \, \dd\F  \nonumber\\
&=& \frac{1}{A}\int\d^2 x \sqrt{g}\ \dd\F  \ .
\ea
Next, we have
\ba\label{Mabuchivar1}
&&\dd \ \int\d^2 x \sqrt{\wh g}\left( \frac{4}{A} \s e^{2\s} -\wh \cR\F\right) 
= \int\d^2 x \sqrt{\wh g}\left(\dd\s \, \frac{4}{A} e^{2\s} + 4\s \, \dd\Big( \frac{e^{2\s}}{A}  \Big) -\wh \cR \,\dd\F\right) \nonumber\\
&&= 2 \frac{\dd A}{A} - \int\d^2 x \sqrt{\wh g}\left( 2 \, \wh\D\, \s + \wh \cR  \right)\dd\F \ ,
\ea
where we used the fact that the Laplace operator $\wh \D$ can be freely integrated by parts since we consider only manifolds without boundaries.  Finally, as is well-known, in two dimensions the curvature scalars $\cR$ and $\wh \cR$ are related by 
\be\label{RandRhat}
\cR=e^{-2\s} \Big( \wh \cR + 2 \wh\D\, \s\Big) \ .
\ee
This implies, in particular, that $\sqrt{g}\, \cR =\sqrt{\wh g}\, (\wh \cR + 2 \wh\D\, \s)$.
We can then rewrite \eqref{Mabuchivar1} as
\be\label{Mabuchivar2}
\dd \ \int\d^2 x \sqrt{\wh g}\left( \frac{4}{A} \s e^{2\s} -\wh \cR\F\right) 
=2 \frac{\dd A}{A} - \int\d^2 x \sqrt{g}\, \cR\, \dd\F \ .
\ee
Combining this with \eqref{MabuchiAY} and \eqref{AYvar} gives the variation of the Mabuchi action as
\be\label{Mabuchivar3}
\dd S_{\rm M}[\wh g,g]=
2\frac{\dd A}{A}-\int\d^2 x \sqrt{g}\, \left( \cR-\frac{8\pi(1-{\tt g})}{A}\right) \dd\F \ .
\ee
We see that the stationary points of the Mabuchi action are given by metrics of constant area $A$  and constat curvature $\cR=\frac{8\pi(1-{\tt g})}{A}$. Indeed, any Riemann surface of genus ${\tt g}$ admits a constant curvature metric and then, since $\int\sqrt{g}\,\cR=8\pi (1-{\tt g})$ one necessarily has $\cR=\frac{8\pi(1-{\tt g})}{A}$.

One has similarly for the variation of the Liouville action $S_{\rm Liouville}$ given in \eqref{Liouvaction}.
\be\label{Liouvillevar1}
\dd S_{\rm Liouville}[g,\wh g] =
 \dd\, \int\d^2 x \sqrt{\wh g} \, \big( \s \wh\D\s + \wh \cR \s\big) 
=\int\d^2 x \sqrt{\wh g} \, \big( 2\dd\s \wh\D\s + \wh \cR \dd\s\big) 
=\int\d^2 x \sqrt{g} \, \cR\, \dd\s \ .
\ee
This is the form of the variation which we will use later-on. But we can also re-express $\dd\s$ in terms of $\dd A$ and $\dd\F$ so that
\be\label{Liouvillevar2}
\dd S_{\rm Liouville}[g,\wh g] 
=\int\d^2 x \sqrt{g} \, R\Big( \frac{\dd A}{2A}-\frac{A}{4} \D\dd\F\Big)
=4\pi (1-{\tt g}) \,  \frac{\dd A}{A} - \frac{A}{4} \int\d^2 x \sqrt{g} \, \D \cR\, \dd\F \ ,
\ee
which shows that the stationary points of the Liouville action at constant area $A$ are given by metrics such that $\D \cR=0$, i.e. by constant curvature metrics. As before, then necessarily $\cR=\frac{8\pi(1-{\tt g})}{A}$.

We have seen that the Liouville and Mabuchi actions  admit metrics of constant scalar curvature  as ``solutions of their equations of motion" under the constraint of fixed area. Although not obvious from the previous equation, the variation of the Aubin-Yau action when restricted to the space of Bergmann metrics is similarly related to metrics of constant scalar curvature \cite{AubinYau}.


\subsection{Recap : The gravitational action for a massless scalar field}

Before embarking on the technical complications for the Majorana fermions, it is maybe good to quickly remind the reader of (parts of) the much simpler computation of the gravitational action for a massless (i.e.~conformal) scalar field, leading to the Liouville action.

For a massless scalar field $\f$  on a  2 dimensional (euclidean) manifold without boundary, with metric $g_{\m\n}$, the action is 
\be\label{E1}
S[g,\f]=\int\d^2 x \sqrt{g}\, g^{\m\n} \nabla_\m \f \nabla_\n\f= -\int\d^2 x \sqrt{g}\, \f \D\f \ ,
\ee
where, as in \eqref{gg0sig}, 
$\D = \frac{1}{\sqrt{g}} \del_\n \big( \sqrt{g} \,g^{\m\n} \del_\m\big) \equiv g^{\m\n} \nabla_\n \nabla_\m$
is the scalar Laplace operator.  It is hermitian with respect to the usual scalar product $(f,h)=\int \d^2 x \sqrt{g}\, f^*(x)\, h(x)$. Consider the eigenvalue problem of $-\D$~:
\be\label{E3}
-\D f_n(x)=\l_n f_n(x) \ .
\ee
Since $\D$ is a real operator these eigenfunctions can be chosen to be real. As always they can also  be chosen to be orthonormal:
\be\label{E4}
\int\d^2 x \sqrt{g}\, f_n(x) f_m(x) =\dd_{nm} \ .
\ee
Of course, there are infinitely many eigenfunctions and eigenvalues, so that the index $n$ runs over an infinite set (e.g. over the non-negative integers).
One can then see, since the action $S$ is non-negative, that all eigenvalues are $\l_n\ge 0$. (This is why we consider $-\D$ which is a positive operator, rather than $\D$.) Note that there is always a single zero-mode, i.e.~an eigenfunction $f_0$ with eigenvalue $\l_0=0$. Indeed, the constant function $f_0=\frac{1}{\sqrt{A}}$ is such that $\del_\n f_0=0$ and, hence, $-\D f_0=0$, and moreover $f_0$ is correctly normalised. Conversely, if $\D f_0=0$ one has $\int \sqrt{g}\, g^{\m\n}\del_\m f_0 \del_\n f_0=0$ and then $f_0$ must be constant.

One can then develop any $\f(x)$ on the basis of these eigenfunctions~:
\be\label{E5}
\f(x)=\sum_n c_n f_n(x) \ .
\ee
Inserting this into the action and using the orthonormality of the $f_n$ one gets
\be\label{E6}
S[g,\f]=\int\d^2 x \sqrt{g} \f (-\D \f)
=\sum_{n\ne 0} \l_n c_n^2\ .
\ee
To compute the so-called matter partition function (the field $\f$ is considered to be the matter) for fixed metric $g_{\m\n}$~:
\be\label{E7}
Z_{\rm matter}[g]=\int \cD\f\ e^{-S[g,\f]} \ ,
\ee
we also need to define the functional integral measure $\cD\f$ in such a way that one integrates over all possible field configurations $\f(x)$. This is achieved via the expansion \eqref{E5} by integrating over all the expansion coefficients $c_k$:
\be\label{E8}
\cD\f = \sqrt{A}\int\prod_{k\ne 0} \d c_k \ ,
\ee
where we ``arbitrarily" excluded the integration over the zero-mode coefficient $c_0$ (since it would lead to an infinite factor in \eqref{E7}), and inserted instead a factor $\sqrt{A}$.
Then 
\be\label{E7-1}
\int \cD \f\ e^{-S[g,\f]} =  \sqrt{A}\int\Big(\prod_{k\ne 0} \d c_k\Big) \ e^{-\sum_{n\ne 0} \l_n c_n^2}
=\sqrt{A}\prod_{n\ne 0} \sqrt{\frac{\pi}{\l_n}}
= {\cal N} \, \sqrt{A}\,\big( \Det' (-\D) \big)^{-1/2}
 \ ,
 \ee
 where $\cN$ is some (infinite) normalisation coefficient, and $\Det'$ denotes the determinant computed as the product of all non-zero eigenvalues.  One thus has
 \be\label{E8-1}
 Z_{\rm matter}[g]= {\cal N} \sqrt{A}\big( \Det' (-\D_g) \big)^{-1/2} \ ,
 \ee
 where we added a subscript $g$ on $\D$ to remind us that this is the Laplacian for metric $g$.
 The gravitational action is defined as
 \be\label{E9}
 e^{-S_{\rm grav}[\wh g,g]}=\frac{Z_{\rm matter}[g]}{Z_{\rm matter}[\wh g]} = 
 \sqrt{\frac{A}{\wh A}}\, \Big(\frac{\Det'(-\D_{g})}{\Det'(-\D_{\wh g})}\Big)^{-1/2} \ ,
 \ee
 where the factors ${\cal N}$ now have cancelled. 
 
 One can go on, and rewrite this again in terms of the eigenvalues as
 \ba\label{E10}
 S_{\rm grav}[\wh g,g]= \frac{1}{2} \sum_{n\ne 0} \log \l_n(g) -  \frac{1}{2} \sum_{n\ne 0} \log \l_n(\wh g) 
 -\frac{1}{2}\log \frac{A}{\wh A} \ .
 \ea
 These sums over eigenvalues  typically are divergent but are efficiently regularized (and "renormalized"!) by the well-known technique of zeta-function regularisation \cite{SHzetareg}~:
 One introduces the zeta-function\footnote{Since $n=0$ is excluded from the sum, this rather corresponds to what we would denote as $\wt \zeta_g$ in the main part of this paper.}
 $\zeta_g(s)$ of the operator $-\D_g$ as
 \be\label{E11}
 \zeta_g(s)=\sum_{n\ne 0} \frac{1}{(\l_n(g))^s} 
 \quad \Rightarrow\quad
\zeta_g'(0)=-\sum_{n\ne 0} \log\l_n(g) \ ,
 \ee
 so that
\be\label{E13}
S_{\rm grav}[\wh g,g]= - \frac{1}{2} \zeta_g'(0) + \frac{1}{2} \zeta_{\wh g}'(0) 
-\frac{1}{2}\log \frac{A}{\wh A} \\ .
\ee
Although this is a nice expression, it is not very explicit unless we know all the eigenvalues and are able to compute \eqref{E11}. Instead, one determines the variation of the zeta function under an infinitesimal conformal variation of the metric and then tries to ``integrate" these infinitesimal changes to get the finite difference \eqref{E13}. Obviously, from \eqref{E11}
\be\label{E14}
\dd\zeta_g(s)=-s\sum_{n\ne 0} \frac{\dd\l_n}{\l_n^{s+1}} \ ,
\ee
where $\l_n=\l_n(g)$ and $\dd\l_n$ is its variation when the metric is changed as $g\to e^{2\dd\s} g$.
Under such a variation of the metric, the Laplace operator $\D=e^{-2\s}\wh \D$  changes as $\D\to e^{-2\s-2\dd\s} \wh \D =e^{-2\dd\s}\D=\D-2\dd\s \D$, i.e.~$\dd\D=-2\dd\s \D$. Then there is a corresponding change $\dd\l_n$ of the eigenvalues and change $\dd f_n$ of the eigenfunctions. Comparing the eigenvalue equations for $-\D$ and $-(\D+\dd\D)$ we have
\be\label{E15}
-(\D+\dd\D)(f_n+\dd f_n)=(\l_n+\dd\l_n) (f_n+\dd f_n)
\quad \Rightarrow\quad
-\D \dd f_n -\dd \D f_n = \l_n \dd f_n + \dd\l_n f_n \ .
\ee
We multiply this last equation with $f_n$ and integrate $\int \d^2 x \sqrt{g} ...$. Then the first term gives\break  $-\int d^2 x \sqrt{g} f_n \D \dd f_n = -\int d^2 x \sqrt{g} (\D f_n) \dd f_n=\l_n \int d^2 x \sqrt{g} f_n \dd f_n$ which cancels a corresponding term on the right-hand side. Thus we are left with 
\be\label{E16}
- \int\d^2 x \sqrt{g} f_n(x) \ \dd\D f_n = \dd\l_n \int\d^2 x \sqrt{g} (f_n(x))^2 =\dd\l_n
\ee
Now, recall that $-(\dd\D) f_n=2\dd\s \D f_n =  -2 \dd\s \l_n  f_n$ so that
\be\label{E16-2}
\dd\l_n=-2 \l_n  \int\d^2 x \sqrt{g}\,\dd\s(x)\, ( f_n(x) )^2 \ .
\ee
This is to be inserted into \eqref{E14}~:
\be\label{E16-3}
\dd\zeta_g(s)=2s\sum_{n\ne 0} \frac{\int\d^2 x \sqrt{g}\,\dd\s(x)\, ( f_n(x) )^2 }{\l_n^{s}} 
=2s \int\d^2 x \sqrt{g}\,\dd\s(x)\, \sum_{n\ne 0}\frac{( f_n(x) )^2 }{\l_n^{s}} 
\ee
If we also define the (bi-) local zeta-function as
\be\label{E17}
\zeta_g(s,x,y)=\sum_{n\ne 0} \frac{f_n(x) f_n(y)}{\l_n^s} \ ,
\ee
this can be rewritten as
\be\label{E18}
\dd\zeta_g(s)=2 s \int\d^2 x \sqrt{g}\, \dd\s(x) \, \zeta_g(s,x,x) \ .
\ee
We now take the derivative with respect to $s$. This yields two terms, but at $s=0$ only one of them contributes~:
\be\label{E18-2}
\dd\zeta'_g(0)\equiv\frac{\d}{\d s}\zeta_g(s)\big\vert_{s=0}= 2 \int\d^2 x \sqrt{g}\, \dd\s(x) \, \zeta_g(0,x,x)
\ .
\ee

There is a standard relation between the heat kernel $K(t,x,y)$ for an operator and the corresponding local zeta function $\zeta(s,x,y)$~:
\be\label{E19}
K(t,x,y)=\sum_{n\ne 0} e^{-\l_n t} f_n(x) f_n(y) \quad , \quad
\zeta(s,x,y)=\frac{1}{\G(s)} \int_0^\infty \d t\, t^{s-1}  \,K(t,x,y)\ .
\ee
Possible singularities of the integral arise from the integration region of small $t$ and one can show that the integral defines an analytic function for $\Re s>1$ which can be analytically continued to the whole complex $s$-plane except for poles at $s=1,0,-1,-2,\ldots$. However,  $\G(s)$ also has poles at $s=0,-1,-2,\ldots$ and dividing by $\G(s)$ cancels the poles of the integral, except at $s=1$. We see that $\zeta(s,x,y)$ has a pole at $s=1$ and that its finite values at $s=0,-1,-2,\ldots$ are determined by the singularities of the integral due to the small-$t$ behaviour of the heat kernel.  But the small-$t$ behaviour of the heat kernel is well-known, in particular
\be\label{E20}
K(t,x,x)\sim_{t\to 0} \frac{1}{4\pi t}\Big( 1 + \frac{\cR(x) }{6} \ t + {\cal O}(t^2) \Big)-\frac{1}{A} \ ,\ .
\ee
where the $-\frac{1}{A}$ comes from subtracting the zero-mode $f_0^2=\frac{1}{A}$ which is not included in the sum \eqref{E19}.
From this one gets
\be\label{E21}
\zeta(0,x,x) =\frac{1}{24\pi} \cR(x)-\frac{1}{A} \ ,
\ee
which when inserted in \eqref{E18-2} gives
\be\label{E22}
\dd S_{\rm grav}[\wh g,g]\, =\, -\frac{1}{2}\dd\zeta_g'(0) -\frac{\dd A}{2A}
\, =\, -\frac{1}{24\pi} \int\d^2 x \sqrt{g}\,  \dd\s(x)\, \cR(x)\ .
\ee
We recognise $\dd S_{\rm Liouville}$ and conclude
 \be\label{E23}
 S_{\rm grav}[\wh g,g]=- \frac{1}{24\pi} S_{\rm Liouville}[\wh g,g] \ .
 \ee
This is a famous result, first obtained (using a different method) by Polyakov in 1981 \cite{Liouville1}.

\newpage

\setlength{\baselineskip}{.54cm}

\section{The Dirac operator for Majorana spinors on a curved manifold\label{Dirac}}
\setcounter{equation}{0}

In this section we define the relevant Dirac operator $D$ in 2 Euclidean dimensions, with a Majorana mass term, and relate its square $D^2$ to the spinorial Laplace operator and scalar curvature. We discuss the eigenvalue problem of this Dirac operator and how it changes under (local) conformal transformations. Most importantly, we obtain a sufficient characterisation of the zero-modes of the massless Dirac operator and of the corresponding projectors on these zero-modes, which we will need in section 5.  We first look at flat space, not only because it is simpler but also because it provides the formula valid in any given local orthonormal frame of the curved space. The material of the first part of this section is standard or already contained in \cite{BDE} but we re-derive the relevant formula for a self-contained presentation.

\subsection{Majorana spinors in flat 2D space : $\g$-matrices and Dirac operator}

In two flat Euclidean dimensions we use labels $a=1,2$. Recall that the 3 Pauli matrices satisfy $\s_j \s_k=\dd_{jk}{\bf 1}+ i \e_{jkl}\s_l$. We can then choose two real, hermitian $\g$-matrices as
\be\label{gammachoice}
\g^1=\s_x \ ,  \quad \g^2=\s_z \quad \Rightarrow\quad \{ \g^a,\g^b\}=2\dd^{ab}\, {\bf 1}
 \quad , \quad (\g^a)^\dag=\g^a \quad , \quad (\g^a)^T=\g^a \quad , \quad (\g^a)^*=\g^a 
\ .
\ee 
The chirality matrix (the analogue of $\g_5$ in 4 dimensions)  then is 
\be\label{gamma5}
\g_*=i\g^1\g^2=\s_y \quad \Rightarrow\quad \g_*^\dag=\g_* \quad , \quad \g_*^T=\g_*^*=-\g_* \ ,
\ee
and, of course, $\g_*$ anticommutes with $\g^a$. 
These are all $2\times 2$ matrices and, hence, a spinor $\p$ has two components that are a priori complex.

In 2 Euclidean dimensions the ``Lorentz" group is just the rotation group  SO(2) which has a single generator $J^{12}$.  In the spin-$\frac{1}{2}$ representation as carried by the Dirac spinor this (hermitian) generator is $J^{12}=-\frac{i}{4}[\g^1,\g^2]=-\frac{i}{2}\g^1\g^2=-\frac{1}{2}\g_*$ and the  representation of a finite rotation by an angle $\a$ on a spinor $\p$ is given by the  matrix
$D(\a)=e^{i \a J^{12}}=e^{-i\a\g_*/2}$. More precisely, this means that under such a rotation, acting also on the coordinates\footnote{
Below, on a curved manifold, we will consider separately the notions of covariance under rotations of the spinor and under transformations of the coordinates.
} 
as ${x'}^a=\L^a_{\ b}x^b$, the spinor transforms as 
$\p(x) \to \p'(x') =D(\a)\p(x)$.
Now, with the choice \eqref{gammachoice} of the $\g$-matrices, the matrix $D(\a)$ is real and $\p^*$ transforms exactly as $\p$. Furthermore since $\g_*$ commutes with $D(\a)$,  $\g_*\p$ also transforms as $\p$. Finally, using the relation $D(\a)^{-1} \g^\m D(\a) =\L^\m_{\ \r}\, \g^\r$ one also shows that $(\dsl \p)'(x')= D(\a) \dsl \p(x)$.
Hence, all $\p^*$, $\g_*\p$ and $\dsl\p$ all transform exactly as $\p$. In particular,  it makes sense to impose the Majorana condition $\p^*=\p$, imposing that the two components $\p_1$ and $\p_2$ of $\p$ are both real, or write an equation combining $i\dsl\p$ and $m\g_*\p$.

Note that the spinors are anti-commuting objects obeying $\p_\a\p_\b=-\p_\b\p_\a$, and one has for a complex spinor  $\p^\dag \p=(\p^*)^{\tt T} \p=\p_1^*\p_1+\p_2^*\p_2$. The complex conjugation of such an expression is defined as the hermitian conjugate and involves reversing the order of the terms, i.e. 
\be
(\chi_a \p_\b)^*=\p_\b^* \chi_\a^* \ .
\ee
We then see that $(\p_1^* \p_1)^*=\p_1^* (\p_1^*)^*=\p_1^* \p_1$ and  $\p^\dag \p$ is indeed real, and $\int m\p^\dag \p$ is the standard (real) mass term in the action for a Dirac spinor

For anti-commuting Majorana, i.e.~real spinors, however,  a mass term like  $\int m \p^\dagger  \p =\int m \p^{\tt T}\p=m\int (\p_1\p_1+\p_2\p_2)$ vanishes. We can, nevertheless, introduce a non-vanishing, real mass-term as $\int \p^\dag m\g_*\p=m\int (\p_1 (-i)\p_2 + \p_2 i \p_1) =-2im\int \p_1\p_2$. This is indeed real since $(\p_1\p_2)^*=\p_2\p_1=-\p_1\p_2$. The action for such a Majorano spinor then reads
\be\label{Majaction}
S=\int \p^\dag ( i\dsl + m\g_*)\p \ .
\ee
The corresponding Dirac operator
\be\label{flatDirop}
D=i\dsl + m\g_*
\ee 
is hermitian with respect to the standard inner product on the space of Dirac spinors, 
\be
(\chi, \p)=\int \chi^\dag(x) \p(x)=\int (\chi_1^*\p_1+\chi_2^* \p_2)=(\p,\chi)^* \quad , \quad
(O\chi,\p)=(\chi,O^\dag\p) \ .
\ee 
Indeed, taking the hermitian conjugate of $D$ involves taking the hermitian conjugate of the $\g$ matrices, (which we have chosen to be hermitian), complex conjugation of the $i$ and an integration by parts of $\del_\m$. Hence 
\be
D^\dag=D \ .
\ee  
In particular, we also see that $S=(\p,D\p)=(D\p,\p)=(\p,D\p)^*=S^*$ which confirms that the action is real.
Next, one has
\be
D^2=(i\dsl + m\g_*)(i\dsl + m\g_*)=-\dsl^2 +im(\dsl\g_*+\g_*\dsl)+ m^2=-\dsl^2 + m^2=-\del_\m\del^\m + m^2
\nonumber
\ee 
which incorporates the correct Euclidean continuation of the mass-shell condition $p_\m p^\m + m^2=0$. 

One might ask why we want to study Majorana spinors with the Dirac operator $D$ and not simply   (complex) Dirac spinors  with an ordinary mass term and corresponding action 
$S=\int \p^\dag ( i\dsl + m)\p$. However, the square of the corresponding Dirac operator $\wt D=i\dsl+m$ is $\wt D^2=(i\dsl + m)^2 = -\del_\m\del^\m + m^2+2im \dsl$ and this does {\it not} correspond to anything simple or physical. (It is $(i\dsl-m) (i\dsl +m)$ that instead gives the mass-shell condition.) This is one of the reasons we will focus on the action \eqref{Majaction} and corresponding Dirac operator \eqref{flatDirop}.
Note that this Dirac operator $D$ 
is a purely imaginary. Indeed, the $\g^\m$ are real and $\g_*$ is purely imaginary.  Thus $iD$ is a real operator and  the Dirac equation $D\p=(i\dsl + m\g_*)\p=0$ admits  real solutions $\p$. However, we will be interested in the  corresponding eigenvalue problem  $D\p_n\equiv (i\dsl + m\g_*)\p_n=\l_n\p_n$.
Since $D$ is hermitian, the eigenvalues $\l_n$ are real and then the $\p_n$ cannot be real, but must be complex.  We will discuss this eigenvalue problem more generally below, on curved space.

\newpage
\subsection{Curved space : $\g$-matrices, covariant derivatives, Dirac operator and spinorial Laplacian}

To define  spinors on a curved space (-time) one needs to introduce locally orthonormal frames which is done by introducing the so-called viel-bein $e^a_\m$. To define the Dirac operator one needs to define the covariant derivative $\nabla_\m$ which involves the spin connection $\o^{ab}_\m$ which is related to the $e^a_\m$ by the zero-torsion condition. In this subsection we recall how this is done and we also establish a few useful relations. Of course, the equations in this subsection are well-known to supergravity practitioners,  but some are still somewhat non-trivial, and we find it useful to re-derive them here for further reference. Since most of the discussion is not specific to two dimensions we first discuss arbitrary $n$-dimensional spaces (with Euclidean signature\footnote{
The whole discussion can also be repeated for Lorentzian signature, but one would need to pay attention to the upper or lower positions of the flat indices $a,b,\ldots$ also.
}) 
and only set $n=2$ in the end.  Our detailed discussion also allows us to make some  statements about possible zero-modes of the Dirac operator that will be useful in the sequel.

\subsubsection{Local orthonormal frames and spin connection (in $n$ dimensions)}

Starting from the metric tensor written using the coordinate one-forms $\d x^\m$, one introduces the one-forms $e^a$ of the local orthonormal frame as 
\be\label{metric}
g_{\m\n}(x)\, \d x^\m \otimes \d x^\n = \dd_{ab}\,  e^a(x) \otimes e^b(x) \ .
\ee
Both sets of indices $\m,\n,\ldots$ and $a,b,\ldots$ take  values $1, 2,\ldots n$.
We can always express the $e^a$ in terms of the $\d x^\m$ and vice versa:
\be\label{vielbeins}
e^a(x)=e^a_\m(x)\,  \d x^\m \quad , \quad \d x^\m=E^\m_b(x) \,e^b(x) \quad , \quad e^a_\m(x) \, E^\m_b(x)=\dd^a_b \quad , \quad 
E^\m_b(x) e^b_\n(x)=\dd^\m_\n \ .
\ee
As indicated here, the viel-bein $e^a_\m$ and inverse viel-bein $E^\m_a$ depend on the point $x$ on the manifold.  Combining \eqref{vielbeins} and \eqref{metric} results in the relations
\be
g_{\m\n}\,E^\m_a E^\n_b=\dd_{ab}\quad , \quad \dd_{ab}\, e^a_\m e^b_\n=g_{\m\n} 
\quad \text{and}\quad\
\dd^{ab} E^\m_a E^\n_b=g^{\m\n} \quad , \quad g^{\m\n}e^a_\m e^b_\n=\dd^{ab}\ .
\ee
The $\dd_{ab}$ is to be interpreted as the metric in the local orthonormal frames. The $e^a_\m$ and $E^\m_a$ implement the change of basis between these orthonormal local frames and the coordinate frames but, of course, in general they cannot be interpreted as  Jacobian matrices for a change of coordinates. Since the metric in the orthonormal frame is $\dd_{ab}$ and the inverse metric $\dd^{ab}=\dd_{ab}$, raising and lowering the corresponding indices $a,b,\ldots$ has no effect and we write indifferently $T^{ab}\equiv T^a_{\ b}$ etc. Of course this does not apply to the ``coordinate indices" $\m,\n,\ldots$.

The relevant connection in the local orthonormal frame is the spin-connection $\o^{ab}_\m$. It is defined in terms of the connection coefficients $\G^c_{ab}$ in the local orthonormal frame
as $\o^a_{\ b\m}=\G^a_{cb}e^c_\m$.  The $\G^c_{ab}$ can be obtained from the $\G^{\r}_{\n\m}$ in the coordinate frame through the (inhomogeneous) transformation rules for the connection coefficients as $\G^c_{ab}=E_a^\m E_b^\n\, \G^\r_{\m\n} e^c_\r + E_a^\m \del_\m E_b^\r e^c_\r$. However, since we will only work on Riemannian manifolds, where the torsion vanishes and the metric is covariantly constant, the spin-connection can be more easily obtained through the zero-torsion condition which is most compactly written as an equation for the spin-connection one-form $\o^{ab}=\o^{ab}_\m \d x^\m$ using the exterior derivative $\d$~:
\be\label{zero-torsion}
\d e^a+ \o^{ab}\wedge e^b =0 
\quad \Leftrightarrow\quad \del_\m e^a_\n - \del_\n e^a_\m  
+ \o^{ab}_\m e^b_\n - \o^{ab}_\n e^b_\m=0.
\ee
Requiring the metric to be covariantly constant implies that the spin-connection is antisymmetric in $a$ and $b$:
\be\label{metriccovder3}
\nabla_\m g_{\n\r}=0 \quad \Leftrightarrow \quad
\nabla_\m \dd_{ab}=0 \quad \Leftrightarrow \quad
 \o_{ab\m}=-\o_{ba\m} \ .
\ee
Then there are $n \times \frac{n(n-1)}{2}$ independent components $\o^{ab}_\m$, while \eqref{zero-torsion} provides exactly the same number of equation, hence determinig the spin-connection completely in terms of the $e^a_\m$. It is not too difficult to derive  the following useful relations
\ba\label{covdereaEa}
\nabla_\m e^a_\n &\equiv& \del_\m e^a_\n +\o^{ab}_\m e^b_\n -\G^\r_{\m\n}e^a_\r =0 \ , \nonumber\\
\nabla_\m E_a^\n &\equiv&\del_\m E_a^\n + \G^\n_{\m\r} E^\r_a - E^\n_b \o^{ba}_\m=0 \ .
\ea
Finally we note that we obtain the Riemann curvature tensor and curvature 2-form from the spin connection as
\be\label{curvature2form}
R^{ab}\equiv \frac{1}{2}R^{ab}_{\ \ \m\n} \d x^\m\wedge \d x^\n=\d \o^{ab}+\o^{ac}\wedge \o^{cb} 
\quad , \quad \
R^a_{\ b\m\n} =e^a_\l E_b^\r R^\l_{\ \r\m\n} .
\ee


Let us now discuss the Dirac matrices.
The  Dirac-matrices $\g^a$ of the previous subsection satisfied the Clifford algebra $\{\g^a,\g^b\}=2\dd^{ab} {\bf 1}$ as appropriate in flat space. 
This means that they are also the appropriate Dirac matrices in the local orthonormal frames.  To avoid confusion, we temporarily denote these Dirac-matrices as $\tilde\g^a$ so that
\be
\{\tilde\g^a,\tilde\g^b\}=2\dd^{ab} {\bf 1}
\ee 
One then defines the curved-space Dirac matrices as 
\be\label{curvedgammas}
\g^\m=E^\m_a \,\tilde\g^a \quad \Rightarrow\quad \{\g^\m,\g^\n\}=E^\m_a E^\n_b \,\{\tilde\g^a,\tilde\g^b\}
=E^\m_a E^\n_b \,2 \dd^{ab} {\bf 1} = 2 g^{\m\n} {\bf 1}\ .
\ee
We assume again that all $n$ matrices $\tilde\g^a$ are chosen to be hermitian so that also
\be\label{gammamuherm}
(\g^\m)^\dag=\g^\m \ .
\ee
To simplify the notation, we will usually not put the tilde over the flat-space $\g$-matrices with the convention that $\g^\m, \g^\n, \g^\r$ etc are always curved-space $\g$-matrices and $\g^a, \g^b, \g^c$ etc are always flat-space $\g$-matrices. The only potential source of  confusion comes when the indices take numerical values, but we will always use the convention that $\g^1,\ \g^2,\ldots $ are flat-space $\g$-matrices.

\subsubsection{Covariant derivatives (in $n$ dimensions)}

Let us now recall how to construct the appropriate covariant derivative acting a the spinor $\p$.  A spinor transforms (locally) in the spin representation of the local  Lorentz group $SO(n-1,1)$ or at present - since we are in Euclidean signature - the local rotation group $SO(n)$ as
\be\label{localspinortransf}
\p(x)\to \p'(x) = L_{\rm Dirac}(x) \p(x) \equiv \exp\Big( \frac{i}{2} \t^{cd}(x) J^{\rm Dirac}_{cd}\Big) \p(x) \ ,
\ee
where the $ \t^{cd}(x)=-\t^{dc}(x)$ are the (local) rotation angles in the $cd$ plane, and the $J^{\rm Dirac}_{cd}=-J^{\rm Dirac}_{dc}$ are the corresponding hermitian $SO(n)$ generators in the spin / Dirac representation
\be\label{SOnDiracrepres}
J^{\rm Dirac}_{cd}=-\frac{i}{2}\g_{cd}\equiv -\frac{i}{4}[\g_c,\g_d] \quad , \quad 
(J^{\rm Dirac}_{cd})^\dag=J^{\rm Dirac}_{cd}
\ee
It follows from the Clifford algebra satisfied by the $\g^a$ that the $J^{\rm Dirac}_{cd}$ indeed satisfy  the $SO(n)$ algebra 
\be\label{SONalg}
[J_{cd},J_{ef}]=-i \big( \dd_{de} J_{cf} - \dd_{df}J_{ce}-\dd_{ce}J_{df}+\dd_{cf}J_{de}\big) \ .
\ee
Note that the local $SO(n)$ transformations \eqref{localspinortransf} do {\it not} act on the coordinates~: we have the same $x$ before and after the transformation.
The corresponding local $SO(n)$ transformation of the basis forms $e^a$ is
\be\label{SOnbasistrans}
e^a(x) \to {e'}^a(x)=\big(L_{\rm vect}(x)\big)^{ab} \ e^b(x)=
\Big[\exp\Big( \frac{i}{2} \t^{cd}(x) J^{\rm vect}_{cd}\Big) \Big]^{ab} \ e^b(x) \ ,
\ee
where
\be
(J^{\rm vect}_{cd})^{ab} =-i \big(\dd^a_c\dd^b_d-\dd^b_c\dd^a_d\big) \ .
\ee
are the purely imaginary and anti-symmetric, hence hermitian, generators of the vector representation of $SO(n)$. Of course, they also satisfy the $SO(n)$-algebra \eqref{SONalg}.

How does the spin-connection transform under such a local $SO(n)$ transformation? The torsion being a tensor, its vanishing cannot be affected by a (local) change of basis. Hence, in the new frame we must still have
$\d {e'}^a+{\o'}^{ab} {e'}^b=0$.
Using  $\d {e'}^a=\d \big(L^{ab} e^b)= \d L^{ab}  e^b+ L^{ab} \d e^b=\big(\d L^{ac} - L^{ab} \o^{bc}\big) e^c$, we get 
\be\label{spinconntransf}
{\o'}^{ab}=L^{ac}\o^{cd} (L^{-1})^{db} -(\d L^{ac}) (L^{-1})^{cb} = L^{ac}\o^{cd} (L^{-1})^{db} +L^{ac}(\d L^{-1})^{cb} \ ,
\ee
showing that the $\o^{ab}$ transform exactly as a gauge field in the adjoint representation of the local rotation group. Indeed, the matrices $L$ appearing in \eqref{spinconntransf}
are what we above called $L_{\rm vect}$ and we have 
$\o^{ab}=\frac{i}{2} \o^{cd} (J^{\rm vect}_{cd})^{ab}$. But the transformation of the $\o^{cd}_\m$ cannot not depend on the representation we choose, so that we also have
 \be\label{spinconntransf4}
 \frac{i}{2} {\o'}^{cd}(x)  J^{\rm Dirac}_{cd}
 = \frac{i}{2} \o^{cd}(x) \, L_{\rm Dirac}(x)\,  J^{\rm Dirac}_{cd} \, L^{-1}_{\rm Dirac}(x) 
 + L_{\rm Dirac}(x) ) \d L^{-1}_{\rm Dirac}(x) \ .
 \ee

The covariant derivative $\nabla_\m$ of a spinor $\p(x)$ should be defined, as usual,  such that under a local $SO(n)$ rotation $\nabla_\m\p(x)$ transforms exactly as $\p(x)$. 
If we define
\be\label{spinorcovder}
\nabla_\m\p=\del_\m \p + \frac{i}{2} \o^{cd}_\m  J^{\rm Dirac}_{cd} \p 
\quad \Leftrightarrow\quad
\nabla_\m\p_\a=\del_\m \p_\a + \frac{i}{2} \o^{cd}_\m  (J^{\rm Dirac}_{cd})_{\a\b} \p_\b
\ee
it is straightforward to check, using \eqref{localspinortransf} and \eqref{spinconntransf4} that we have indeed
\be\label{psidertransf}
\nabla'_\m \p'(x)\equiv \big( \del_\m + \frac{i}{2} {\o'}^{cd}_\m(x)  J^{\rm Dirac}_{cd} \big) \big( L_{\rm Dirac}(x) \p(x) \big) =  L_{\rm Dirac}(x) \nabla_\m \p(x) \ ,
\ee
where $\nabla'_\m$ involves ${\o'}^{cd}_\m$. Note again that $x$ is unchanged so that $\nabla'_\m$ involves $\del_\m$ and not some $\del'_\m$.
Later-on we will also encounter covariant derivatives acting on other objects transforming in different representations and, accordingly, these covariant derivatives will include other connection terms.  If we want to insist that we are dealing exactly with the covariant derivative acting on the spinor as just defined we will write $\nabla_\m^{\rm sp}$ so that
\be\label{spincovder3}
\nabla_\m^{\rm sp} \equiv \del_\m  + \frac{i}{2} \o^{cd}_\m  J^{\rm Dirac}_{cd} 
= \del_\m  + \frac{1}{4} \o^{cd}_\m  \g_{cd} 
\quad , \qquad \nabla_\m \p= \nabla_\m^{\rm sp}\p \ .
\ee

As a general rule, a covariant derivative $\nabla_\m$ acting on some quantity $F_{(i)}$ transforming in a representation ${\cal R}_i$ of the local rotation group is alway defined in such a way that $\nabla_\m F_{(i)}$ transforms in the same representation ${\cal R}_i$. Just as we wrote $\nabla^{\rm sp}$ to emphasise this fact we could write $\nabla_\m^{(i)}$, but one usually doesn't. It follows from this remark, that the covariant derivative obeys the Leibnitz rule, 
\be\label{Leibnitz}
\nabla_\m (F_{(i)} F_{(j)})= (\nabla_\m F_{(i)}) F_{(j)} +  F_{(i)} \nabla_\m F_{(j)} \ .
\ee 
It similarly follows that in an integral of a quantity that is a scalar under the local rotation group, the covariant derivatives can be freely integrated by parts (possibly up to a boundary term).
One application of these rules is the following identity we will need later-on
\be\label{covderintbyparts}
-\int\sqrt{g}\, (\nabla_\m f) \p^\dag \g^\m \chi 
= \int\sqrt{g}\, f\, \big( (\nabla_\m^{\rm sp} \p)^\dag \g^\m \chi +\p^\dag \g^\m \nabla_\m^{\rm sp}\chi \big) \ ,
\ee
where $f=f(x)$ is a scalar (so that $\nabla_\m f = \del_\m f$) and $\p=\p(x)$ and $\chi=\chi(x)$ are Dirac  (or Majorana) spinors. It is relatively straightforward to check this identity explicitly by integrating by parts, using $\frac{\del_\m\sqrt{g}}{\sqrt{g}} =\G_{\m\s}^\s$, $\g^\m=E^\m_a \g^a$ and the relations
\eqref{covdereaEa}, as well as the hermiticity of the $J_{cd}^{\rm Dirac}$ and the commutation relation of the latter with the $\g^a$.


Let us insist that we have discussed local, i.e.~$x$-dependent frame transformations by some $L(x)$ in the appropriate representation of $SO(n)$. We have not made any coordinate transformation. In the present formalism, local $SO(n)$ transformations and coordinate transformations are totally decoupled. One may, of course, also do a coordinate transformation $x^\m\to {x'}^\m= f^\m(x^\n)$. Under such a coordinate transformation the spinor $\p$, as well as all frame vectors like $v^a$ are inert, i.e.~behave as scalar quantities. For a vector e.g.~one goes from the $v^a$ to the $v^\m$ by the appropriate change of basis, i.e.~$v^\m=E^\m_a v^a$ and it is the transformation  of the $E^\m_a$ under coordinate transformations, namely ${E'}^\m_a=\frac{\del {x'}^\m}{\del x^\n} E^\n_a$,  that implies that the $v^\m$ will indeed transform as the components of a coordinate vector, i.e.~${v'}^\m=\frac{\del {x'}^\m}{\del x^\n} v^\n$.
 
Obviously, both $\del_\m$ and $\o^{ab}_\m$ transform under coordinate transformations as the covariant components of a coordinate vector and, hence, so does $\nabla_\m \p$. Let us temporarily denote the quantities in the $x'$ coordinates by a tilde (rather than a prime which we used for the local $SO(n)$ transformations). Since $\p$ behaves as a scalar quantity under these coordinate transformation we then have $\tilde\p(x')=\p(x)$ and thus indeed
\be
\tilde\nabla_\m \tilde\p(x')=\frac{\del x^\n}{\del {x'}^\m} \nabla_\n \p(x) \ .
\ee
 
Next, the Dirac operator should be the obvious curved space generalisation of \eqref{flatDirop}, i.e.\break  $D=i\Nsl +m\g_*$ where we define the operator $\Nsl$ as
\be
\Nsl = \g^\m\, \nabla_\m=\g^a E_a^\m\, \nabla_\m
\quad \Rightarrow\quad
\Nsl\,\p\equiv \Nsl^{\rm \; sp}\p= \g^\m\, \nabla_\m^{\rm sp} \p=\g^a E_a^\m\, \nabla_\m^{\rm sp} \p  
\ .
\ee
We have just seen that under coordinate transformations $\nabla_\m^{\rm sp} \p$ transforms as the covariant components of a coordinate vector, while $E^\m_a$ transforms as the contravariant components. Of course, the $\g^a$ are just numerical matrices that do not change. Hence, $\Nsl\,\p$ is a scalar under coordinate transformations, just as $\p$.
Next, under a local Lorentz transformation $L$, we have seen that $\nabla_\m^{\rm sp}\p$ transforms with $L_{\rm Dirac}$ while ${E'}^\m_a=(L_{\rm vect})_{ab} E^\m_b$.  Hence\footnote{
Note that $L_{\rm Dirac}$, just as $\g^a$, are matrices while $(L_{\rm vect})_{ab}$ are the components of the matrix $L_{\rm vect}$. Hence, $(L_{\rm vect})_{ab}$ commutes with $L_{\rm Dirac}$ and with $\g^a$. Also, just as in flat space, we have
$L_{\rm Dirac}^{-1}  \g^a L_{\rm Dirac}=(L_{\rm vect})_{ac} \g^c$.
}
\ba\label{Nsltrans}
&& \Nsl\, \p(x) = \g^a E^\m_a(x) \nabla_\m^{\rm sp} \p(x) \ \  \to \ \ 
\g^a {E'}^\m_a(x) \nabla^{'\rm sp}_\m\p'(x)
= \g^a \, \big((L_{\rm vect})_{ab} E^\m_b\big) \,\big( L_{\rm Dirac}  \nabla_\m^{\rm sp} \p(x) \big)\nonumber\\
&&=  (L_{\rm vect})_{ab} \ L_{\rm Dirac} \ \big( L_{\rm Dirac}^{-1}\g^a L_{\rm Dirac} \big)\,E^\m_b\, \nabla_\m^{\rm sp} \p(x) 
=  (L_{\rm vect})_{ab} \ L_{\rm Dirac} \ (L_{\rm vect})_{ac} \g^c  \,E^\m_b\, \nabla_\m^{\rm sp} \p(x) \nonumber\\
&& = L_{\rm Dirac} \  \g^b E^\m_b\, \nabla_\m^{\rm sp} \p(x) 
= L_{\rm Dirac} \  \Nsl\, \p(x) \ .
\ea
This means that under both, coordinate transformations and local Lorentz transformations, $\Nsl^{\rm \; sp}\p$ transforms exactly as $\p$. In particular, this implies that the appropriate covariant derivative $\nabla_\n$ acting on $\Nsl\,\p$ is again $\nabla^{\rm sp}_\n$~:
\be
\nabla_\n \Nsl\,\p=\nabla^{\rm sp}_\n \Nsl\,\p=\nabla^{\rm sp}_\n \Nsl^{\rm \; sp} \p
\quad \Rightarrow \quad 
\Nsl\, \Nsl\, \p = \Nsl^{\rm \; sp} \Nsl^{\rm \; sp}\, \p
= \g^\m \nabla^{\rm sp}_\m \g^\n  \nabla^{\rm sp}_\n\, \p \ .
\ee
However, $\Nsl\, \Nsl\, \p$ is {\it not} $\g^\m  \g^\n \nabla_\m^{\rm sp}  \nabla_\n^{\rm sp}\p$. Instead one has, of course
$\Nsl\, \Nsl\, \p= \g^\m  \nabla_\m \big(\g^\n  \nabla_\n^{\rm sp} \p\big)$ and the question is how to ``commute" the $\nabla_\m$ with the $\g^\n$. On the one hand,
from the usual rules about how the covariant derivative should act on coordinate and frame indices we have
\be\label{gammacovder}
\nabla_\m \g^a=\del_\m \g^a +\o_\m^{ab}\g^b= \o_\m^{ab}\g^b
\quad \Leftrightarrow \quad
\nabla_\m\g^\n=\del_\m\g^\n +\G^\n_{\m\r}\g^\r\ ,
\ee
where the equivalence of both relations is due to $E^\m_a$ being covariantly constant, cf \eqref{covdereaEa}. On the other hand, $\nabla_\m$ also is a matrix, just like the $\g^\n$, and one must also take into account the non-vanishing commutator. This then leads to
\be\label{gammanablacom}
\nabla_\m (\g^\n \nabla_\n \p)=\g^\n \nabla_\m  \nabla_\n \p \ ,
\ee
where on the right-hand side the covariant derivatives are, of course, $\nabla_\m  \nabla_\n \p=\nabla_\m^{\rm sp}\nabla_\n^{\rm sp}\p - \G_{\m\n}^\r\nabla_\r^{\rm sp}\p$. Formally, the $\nabla_\m$ just passes through the $\g^\n$. One could say that the contribution from the commutator cancels the contribution from $(\nabla_\m\g^\n)$~! In any case, we have
\be\label{Nslsquare}
\Nsl\, \Nsl\, \p \equiv (\g^\m \nabla_\m) (\g^\n \nabla_\n) \p = \g^\m \g^\n \nabla_\m \nabla_\n\p
=  \g^\m \g^\n \Big( \nabla_\m^{\rm sp} \nabla_\n^{\rm sp}  -\G_{\m\n}^\r \nabla_\r^{\rm sp}\Big)\p\ .
\ee
Next, using $\g^\m\g^\n=g^{\m\n}+\g^{\m\n}$ (where, of course, 
$\g^{\m\n}=\frac{1}{2}[\g^\m,\g^\n]$) we have
\be\label{Nslsquare2}
\Nsl\, \Nsl\, \p = \g^\m\g^\n\nabla_\m\nabla_\n\p=g^{\m\n}\nabla_\m\nabla_\n\p + \g^{\m\n}\nabla_\m\nabla_\n\p
=g^{\m\n}\nabla_\m\nabla_\n\p+\frac{1}{2} \g^{\m\n} [\nabla_\m,\nabla_\n]\p \ .
\ee
The commutator of two covariant derivatives is related to the curvature tensor as
\be
[\nabla_\m,\nabla_\n]=\frac{i}{2} R_{\m\n}^{\ \ cd} J_{cd} \ ,
\ee
where $J_{cd}$ are the $SO(n)$ generators in the appropriate representation.\footnote{
For a vector e.g. one has
$[\nabla_\m,\nabla_\n] v^a=\frac{i}{2} R_{\m\n}^{\ \ cd} (J_{cd}^{\rm vect})^{ab} v^b
=\frac{1}{2} R_{\m\n}^{\ \ cd} (\dd_c^a \dd_d^b-\dd_d^a\dd_c^b) v^b
=R_{\m\n}^{\ \ ab} v^b$
which is indeed the familiar formula.
} 
When acting on $\p$ the generators are those in the Dirac representation and
\be
[\nabla_\m,\nabla_\n] \p=\frac{i}{2} R_{\m\n}^{\ \ cd} (J_{cd}^{\rm Dirac}) \p
=\frac{1}{4} R_{\m\n}^{\ \ cd} \g_{cd} \p \ .
\ee
Inserting this into \eqref{Nslsquare2} we get
\be\label{Nslsquare3}
\Nsl\, \Nsl \,\p = \D_{\rm sp}\, \p
+\frac{1}{8} \g^{\m\n} R_{\m\n}^{\ \ cd} \g_{cd} \p \ .
\ee
where we defined the spinoral Laplace operator as
\beb\label{spinorlapl}
\D_{\rm sp}=g^{\m\n} \Big( \nabla_\m^{\rm sp} \nabla_\n^{\rm sp}  -\G_{\m\n}^\r \nabla_\r^{\rm sp}\Big)\ .
\eeb

Finally, in any even dimension $n$, there is an appropriate chirality matrix $\g_*\sim \frac{1}{n!}\e_{\m_1\ldots\m_n} \g^{\m_1}\ldots \g^{\m_n} = \frac{1}{n!} \e_{a_1\ldots a_n} \g^{a_1}\ldots \g^{a_n}$. This shows that this chirality matrix is the same as one defines in flat space. We have $\{\g_*,\g^a\}=\{\g_*,\g^\m\}=0$. Including if necessary\footnote{
In Euclidean signature one has $\g_*=i\g^1\ldots \g^n$ for $n=4k+2$ and $\g_*=\g^1\ldots \g^n$ for $n=4k$.
} 
a factor of $i$ we have $\g_*^2={\bf 1}$ and $\g_*^\dag=\g_*$ (provided one has chosen hermitian $\g^a$ as we did).  One can then define the hermitian, massive Dirac operator as the curved-space generalisation of \eqref{flatDirop}~:
\be\label{Dopcurved2n}
D=i\Nsl+m\g_* \ .
\ee
Since $J_{cd}^{\rm Dirac}$ commutes with $\g_*$ we see that
$\nabla_\m^{\rm sp}$ commutes with $\g_*$ and the operator $\Nsl=\g^\m \nabla_\m^{\rm sp}$ anticommutes with $\g_*$.
We then have
\beb\label{Dsquared}
D^2=-\Nsl\, \Nsl + m^2 = -\D_{\rm sp} -\frac{1}{8} \g^{\m\n} R_{\m\n}^{\ \ cd} \g_{cd} + m^2 \ .
\eeb

\subsubsection{Back to two dimensions}

In two dimensions the general formula simplify drastically. The local rotation group $SO(2)$ is abelian and only has a single generator $J_{12}$. In its Dirac representation there is only $J_{12}^{\rm Dirac}=-\frac{i}{2}\g_{12}=-\frac{i}{2}\g_1\g_2=-\frac{1}{2}\g_*$ as already discussed in subsection 2.1.
 Similarly, the spin-connection $\o^{ab}$ also is just a single one-form  $\o^{12}=-\o^{21}$. We then have
\be
\o^{cd}_\m \g_{cd}= 2 \o^{12}_\m \g_{12}
=-2 i \o^{12}_\m \g_*\ .
\ee
We find it convenient to define
$\o_\m=2\o^{12}_\m$,
so that
\beb\label{covderspinor2d}
\nabla_\m^{\rm sp} = \del_\m - \frac{i}{4} \o_\m \g_* \quad , \quad \o_\m=2\o^{12}_\m \ .
\eeb
Note that since $\g_*=\s_y$ is purely imaginary, $\nabla_\m^{\rm sp}$ is a real (anti-hermitian) differential operator. Obviously, it is a $2\times 2$-matrix, and so is $\Nsl = \g^\m\, \nabla_\m^{\rm sp}=\g^a E_a^\m\, \nabla_\m^{\rm sp}$.

Moreover, in two dimensions $\g^{\m\n}R_{\m\n}^{\ \ cd}\g_{cd}$ simplifies as
\be
\g^{\m\n}R_{\m\n}^{\ \ cd}\g_{cd}=\g^{ab}R_{ab}^{\ \ cd}\g_{cd}
=4\g^{12} R_{12}^{\ \ 12}\g_{12}=4 (-i\g_*)^2 R_{12}^{\ \ 12}=-4 R_{12}^{\ \ 12}=-2 \cR\ .
\ee
where $\cR=R_{12}^{\ \ 12}+R_{21}^{\ \ 21}=2R_{12}^{\ \ 12}$ is the scalar Ricci curvature. 
Thus
\beb\label{Dsquared-2d}
D^2=-\Nsl\, \Nsl + m^2 = -\D_{\rm sp} +\frac{1}{4} \cR + m^2 \ ,
\eeb
 where the spinorial Laplacian $ \D_{\rm sp}$ was defined in \eqref{spinorlapl}.
Just as $\nabla_\m^{\rm sp}$, the spinorial Laplacian  is a $2\times 2$-matrix differential operator. More precisely, it has a piece proportional to the identity matrix and involving the scalar Laplacian 
and a piece proportional to $\g_*$:
\be\label{spinorLapl2}
 \D_{\rm sp} = {\bf 1}_{2\times 2}\Big( \D_{\rm scalar} -\frac{1}{16}\o^\m\o_\m \Big)
 -\frac{i}{4}\, \g_*\, \Big( (\nabla_\m \o^\m) + 2 \o^\m \del_\m \Big) \ ,  
\ee
where the scalar Laplacian is defined by  \eqref{gg0sig}, or equivalently by
\be\label{scalarlapl}
\D_{\rm scalar}=g^{\m\n}(\del_\m\del_\n-\G^\r_{\m\n}\del_\r)\ .
 \ee

The hermitian, massive Dirac operator $D=i\Nsl+m\g_*$ was already defined in \eqref{Dopcurved2n}. With our 2-dimensional real $\g^a$ it is moreover purely imaginary.
For later reference, let us note that with our conventions  it is very explicitly given by
\be\label{matrixstrDstructure}
D=i\s_x \cD_1+i \s_z \cD_2+ m \s_y\quad , \quad
\cD_1=E_1^\m\del_\m -\frac{1}{4} E_2^\m\o_\m \quad,\quad \cD_2=E_2^\m\del_\m +\frac{1}{4} E_1^\m\o_\m\ .
\ee
The eigenvalue problem of $D$ is
\be\label{evproblem}
D \p_n\equiv (i\Nsl + m\g_*)\p_n=\l_n\p_n \ .
\ee
Since $D$ is hermitian, its eigenvalues $\l_n$ are real. On the other hand,  $D$ is purely imaginary and then the eigenfunctions $\p_n$ cannot be real (unless $\l_n=0$). Taking the complex conjugate of \eqref{evproblem} we see that $\p_n^*$ is also an eigenfunction but with eigenvalue $-\l_n$~:
\be\label{evproblem2}
D \p_n^*=-\l_n \p_n^* \ .
\ee
For all $\l_n\ne 0$, within any couple $(\p_n,\p_n^*)$ we decide to call $\p_n$ the eigenfunction with the positive eigenvalue. With this convention we always have 
\be\label{lambdannonnegative}
\l_n\ge 0\ .
\ee
We let
\be\label{psichiphi}
\p_n=\frac{1}{\sqrt{2}}(\chi_n+i \f_n)\ ,
\ee
with real $\chi_n$ and $\f_n$
and then taking the real and imaginary parts of \eqref{evproblem} gives
\beb\label{evphichi}
D \chi_n=i \l_n\f_n
\quad , \quad
D \f_n=-i \l_n\chi_n \ .
\eeb
Of course, the $\chi_n$, $\ \f_n$, $\ \p_n$ and $\p_n^*$ are all eigenfunctions of $D^2$ with eigenvalue $\l_n^2$~:
\be\label{ev2phichi}
D^2\chi_n=\l^2_n\chi_n
\quad , \quad
D^2\f_n= \l^2_n\f_n 
\quad , \quad
D^2\p_n=\l^2_n\p_n 
\quad , \quad
D^2\p^*_n=\l^2_n\p^*_n\ .
\ee

The Dirac operator $D$, as well as $i\Nsl$, are hermitian with respect to the inner product
\be\label{innerprod}
(\P_1,\P_2)=\int\d^2 x \sqrt{g}\, \P_1^\dag \P_2 \ .
\ee
As usual, the eigenfunctions $\p_n$ corresponding to different $\l_n$ are orthogonal. Similarly, for the $\p_n^*$. Also all $\p_n^*$ are orthogonal to all $\p_k$ (as long as $\l_n\ne 0$). Equivalently, assuming the eigenfunctions to be also normalised, we have for the real and imaginary parts
\be\label{orthonomality}
(\chi_n,\chi_k)=\dd_{nk} \quad , \quad
(\f_n,\f_k)=\dd_{nk} \quad , \quad
(\chi_n,\f_k)=0 \quad (\l_n\ne 0)\ .
\ee
Note that the eigenfunctions of $D$ (the $\p_n$ and $\p_n^*$) are automatically eigenfunctions of $D^2$, but the converse is not necessarily true as is examplified by the $\chi_n$ and $\f_n$. What is true is that within any eigenspace of $D^2$ with eigenvalue $\L_n=\l_n^2$ one can find linear combinations (corresponding precisely to the $\p_n$ and $\p_n^*$) that are eigenfunctions of $D$ with eigenvalues $+\l_n$ and $-\l_n$.


\subsection{The example of the flat torus with arbitrary periods}\label{flattorus-sec2}

It will be helpful to be able to check our general statements in the sequel of this paper against at least one example where we explicitly know the eigenvalues and eigenfunctions. The simplest example is the flat torus. Then the metric is just $g_{\m\n}=\dd_{\m\n}$ and $\Nsl=\dsl$. Of course, being flat, we will not be able to appreciate any effects of curvature.
 The relevant Dirac operator then is
\be\label{flatD}
D=i\dsl+m\g_* = i\s_x\del_1 + i \s_z \del_2 + \s_y m =i\, \begin{pmatrix} \del_2 & \del_1 - m\\ \del_1 + m & - \del_2 \end{pmatrix} 
\quad , \quad 
D^2=-(\del_1^2+\del_2^2) + m^2
\ .
\ee
There are still different tori corresponding to different periods.\footnote{
For fermions we could also consider different spin structures, i.e. periodic or anti-periodic boundary conditions around one or the other circle of the torus, leading to four different spin structures. Here we will only investigate the doubly periodic boundary conditions, which are the only ones for which zero-modes are present.}

\vskip3.mm
\noindent
{\bf The flat ``square" torus~: } We will shortly consider the flat torus with arbitrary periods, but we begin by looking
at the ``square" torus corresponding to a square of length $2\pi$ with opposite sides identified. 
The eigenvalue problem for $D^2$, i.e. $D^2 \p_n=\l_n^2\p_n$ yields already
\be\label{lambdatorus}
\l_n^2\equiv \l_{\vec{n}}^2=n_1^2 + n_2^2 + m^2 \quad , \quad \p_n\sim e^{i n_1 x^1 + i n_2 x^2}\ .
\ee
We denote by $\l_{\vec{n}}$ the positive square root of $\l_{\vec{n}}^2$\,. Inserting
  the ansatz
$\p_{\vec{n}}(x^1,x^2)=\begin{pmatrix} a\\ b \end{pmatrix} e^{i n_1 x^1 + i n_2 x^2}$
into $D \p_{\vec{n}}=\l_{\vec{n}}\p_{\vec{n}}$ gives the two equivalent equations
$-(n_1+im) b=(\l_{\vec{n}}+n_2)a$ and $-(n_1-im) a=(\l_{\vec{n}}-n_2)b$.
This is easily solved, up to a common normalisation, which we choose such that $\int \d^2x\,  \p_{\vec{n}}^\dag \p_{\vec{n}}=1$~:  
\be\label{eigentorussol}
\p_{\vec{n}}(\vec{x})
=\frac{1}{2\pi \sqrt{2\l_{\vec{n}}(\l_{\vec{n}}+n_2)}} 
\begin{pmatrix} n_1+im \\ -\l_{\vec{n}}-n_2 \end{pmatrix} 
e^{i n_1 x^1 + i n_2 x^2} \ .
\ee
Since $\l_{\vec{n}}$ only depends on $n_1^2+n_2^2$, each eigenvalue is (at least) fourfold degenerate.\footnote{
In the present example where both circles have the same radius, there is a further degeneracy under the exchange of $n_1$ and $n_2$.
} 
Also
\be\label{psinstar}
\p_{\vec{n}}^*= -\p_{-\vec{n}}\big\vert_{\l_{\vec{n}}\to -\l_{\vec{n}}} \ ,
\ee
which shows that $\p_{\vec{n}}^*$ is indeed an eigenfunction of $D$ with eigenvalue $ -\l_{\vec{n}}$. Thus the complete set of eigenfunctions is the set of all $\p_{\vec{n}}$ and all $\p_{\vec{n}}^*$.  

\vskip3.mm
\noindent{\bf Flat torus with arbitrary periods~:}
A general flat two-dimensional torus is obtained as the quotient of ${\bf R}^2$ by a lattice $\G$ where $\G$ is generated by two linearly independent vectors $\vec{\o}_1$ and $\vec{\o}_2$, i.e.~we identify any two points of ${\bf R}^2$ that differ by $p\vec{\o}_1+q\vec{\o}_2$ with integer $p, q$. With respect to an orthonormal basis $\vec{e}_1,\ \vec{e}_2$ we can write $\vec{x}=x^1\vec{e}_1+x^2\vec{e}_2$, $\ \vec{\o}_1=\o_{11}\vec{e}_1+\o_{12}\vec{e}_2$ and $\vec{\o}_2=\o_{21}\vec{e}_1+\o_{22}\vec{e}_2$, so that one identifies $x^1\simeq x^1+p\o_{11}+q\o_{21}$ and $x^2\simeq x^2 +p\o_{12}+q\o_{22}$. The formula simplify, if one chooses the basis $\vec{e}_1,\ \vec{e}_2$ such that $\vec{e}_1$ is aligned with $\vec{\o}_1$ which means $\o_{12}=0$. Also, by scaling the coordinates appropriately one may assume that $\o_{11}=2\pi$.  With this choice one writes
\be\label{omegaperiods}
\vec{\o}_1=2\pi\,\vec{e}_1 \quad , \quad \vec{\o}_2=2\pi \big(\tau_1 \vec{e}_1+\tau_2 \vec{e}_2 \big)
\quad , \quad
x^1\simeq x^1+2\pi (p + \tau_1 q) \quad , \quad x^2\simeq x^2 + 2\pi \tau_2 q \ ,
\ee
with integer $p$ and $q$. 
If we use a complex coordinate $z=x^1+ix^2$ instead, the periodicity is
\be\label{zperiods}
z\simeq z+2\pi \big( p+\tau q) \quad , \quad \tau=\tau_1+i\tau_2 \quad , \quad \tau_2>0 \ \ .
\ee
The area of the torus now is given by the area of the fundamental cell which is a parallelogram with sides $\vec{\o}_1$ and $\vec{\o}_2$. Hence the area is $A=2\pi \times 2\pi \tau_2$. 
Note that the case $\tau_1=0$ corresponds to a ``rectangular" torus.
The  ``wave-vectors" $\vec{k}$ are such that $\vec{k}\cdot \big(p\vec{\o}_1+q\vec{\o}_2\big)\in 2\pi{\bf Z}$ for all integer $p,q$. They form the ``dual lattice" and are given by
\be\label{duallatticetorus}
\vec{k}=n_1 \vec{e}_1 + \frac{n_2-\tau_1 n_1}{\tau_2}\vec{e}_2 \quad , \quad n_1,n_2\in {\bf Z} \ ,
\ee
as one can easily check.
This means that ($x^1=\frac{z+\zb}{2},\ x^2=\frac{z-\zb}{2i}$)
\be\label{periodicfunctionbasis}
\exp\big(i\vec{k}_{\vec{n}}\cdot\vec{x}\big)=\exp\big(i n_1 x^1 +  i\frac{n_2-\tau_1 n_1}{\tau_2} x^2\big)
=\exp\Big(-\frac{1}{2\tau_2}\big( (n\bar\tau - m)z-(n\tau-m)\zb\big)\Big) \ .
\ee
has the correct periodicity properties \eqref{omegaperiods}, resp.~\eqref{zperiods}.

The Dirac operator $D$ is still the same as in \eqref{flatD} since the present torus is still flat, only the periodic boundary conditions are different. Then also $D^2=-(\del_1^2+\del_2^2) + m^2$ as before and, hence, $D^2 e^{i\vec{k}_{\vec{n}}\cdot \vec{x}}=\l^2_{\vec{n}} e^{i\vec{k}_{\vec{n}}\cdot \vec{x}}$ with
\be\label{lambdasquaretorustau}
\l^2_{\vec{n}} =\vec{k}^2_{\vec{n}}+m^2=n_1^2 + \big( \frac{n_2-\tau_1 n_1}{\tau_2}\big)^2 + m^2 \ .
\ee
If we denote $k_1=n_1$ and $k_2= \frac{n_2-\tau_1 n_1}{\tau_2}$ then almost exactly as before\footnote{
The only difference is the appearance of the additional normalisation factor $\frac{1}{\sqrt{\tau_2}}$ since now the area of the torus is $2\pi \times 2\pi \tau_2$ instead of $(2\pi)^2$.
} 
we get for the eigenfunction of $D$
\be\label{eigentorustausol}
\p_{\vec{n}}(\vec{x})
=\frac{1}{2\pi \sqrt{2\tau_2 \l_{\vec{n}}(\l_{\vec{n}}+k_2)}} 
\begin{pmatrix} k_1+im \\ -\l_{\vec{n}}-k_2 \end{pmatrix} 
e^{i k_1 x^1 + i k_2 x^2} \ ,
\ee
together with the complex conjugate functions which still satisfy
\be\label{psinstartau}
\p_{\vec{n}}^*= -\p_{-\vec{n}}\big\vert_{\l_{\vec{n}}\to -\l_{\vec{n}}} \ .
\ee

\vskip3.mm
\noindent
{\bf Zero-modes of $D$~:} As discussed in general above, for $m\ne 0$ there are no zero-modes of $D$, while for $m=0$ the zero-modes correspond to $\vec{k}_{\vec n}=0$, i.e.~$n_1=n_2=0$. However, in this case
 \eqref{eigentorussol} and \eqref{eigentorustausol} are indeterminate. We may instead first take to $n_1=n_2=0$ and then consider the limit $m\to 0$ which gives the two zero-modes of definite chirality~:
\be\label{toruszeromodes}
\p_{\vec{0}}=\frac{1}{2\pi\sqrt{2\tau_2}} \begin{pmatrix} i \\ -1 \end{pmatrix} \quad , \quad
\p_{\vec{0}}^*=\frac{1}{2\pi\sqrt{2\tau_2}} \begin{pmatrix} -i \\ -1 \end{pmatrix}  
\quad , \quad
\g_* \p_{\vec{0}}= \p_{\vec{0}} \quad , \quad \g_* \p_{\vec{0}}^*= -  \p_{\vec{0}} \ .
\ee
Below, we will  define the ``projector" on the zero-modes in general as $P_0(x,y)=\sum_{i} \p_{0,i}(x) \p_{0,i}^\dag(y)$. At present, for the general flat torus  this simply gives
\be\label{zeromodeprojtorustau}
P_0=\p_{\vec{0}}\, \p_{\vec{0}}^\dag + \p_{\vec{0}}^*\, {\p_{\vec{0}}^*}^\dag=\frac{1}{(2\pi)^2} \frac{1}{2\tau_2} \begin{pmatrix} 1&-i\\i&1\end{pmatrix} +{\rm c.c} 
=\frac{1}{(2\pi)^2\tau_2} \ {\bf 1}_{2\times 2}  =\frac{1}{A} \ {\bf 1}_{2\times 2} \ ,
\ee
where $A=(2\pi)^2\tau_2$ is  the area of the torus.


\subsection{Conformal transformation of the Dirac operator}\label{confsec}

As discussed in the introduction, we will consider two metrics $g_{\m\n}$ and $\wh g_{\m\n}$ that are conformally equivalent, i.e.~that are simply related by a local ($x$-dependent) conformal factor which we write as $e^{2\s}\equiv e^{2\s(x)}$~:
\beb\label{confrescg}
g_{\m\n}=e^{2\s}\, \wh g_{\m\n} \ .
\eeb
We will consider $\wh g_{\m\n}$ as a reference metric (also referred to as the background metric) and $g_{\m\n}$ as a ``general" metric obtained by varying the conformal factor $e^{2\s(x)}$. This view is particularly useful in two dimensions where, by a change of coordinates, any metric can be brought into the form  $e^{2\s(x)} \wh g_{\m\n}$ with a fixed reference metric $\wh g_{\m\n}$. In particular, in two dimensions we have the relation \eqref{RandRhat} between the curvature scalar $\cR$ for the metric $g$ and the one $\wh \cR$ for the metric $\wh g$, namely $\cR=e^{-2\s} \big( \wh \cR + 2 \wh\D\, \s\big)$, as well as 
\be\label{gghatdeterminant}
\sqrt{g}=\sqrt{\wh g} \, e^{2\s}\ .
\ee
Since $\sqrt{\wh g}\, \wh\D_{\rm scalar}=\del_\m \big( \wh g^{\m\n} \sqrt{\wh g} \, \del_\n\big)$ we see that $\sqrt{g} \,\cR = \sqrt{\wh g}\,  \wh\cR + \text{total derivative}$, which is the reason why the integral $\int \sqrt{g} \,\cR$ is independent of the conformal deformation and it is a topological invariant : its value only depends on the topology of the two-dimensional manifold $\cM$. Indeed, it equals $4\pi \chi(\cM)$ where $\chi(\cM)$ is the Euler characteristic of $\cM$. For a manifold without boundaries the latter is given in terms of the genus ${\tt g}$ as $\chi(\cM)=2(1-{\tt g})$, so that $\int_{\cM} \sqrt{g} \cR=8\pi (1-{\tt g})$.
We already noted that the Laplace operators $\D$ and $\wh\D$ are simply related by $\D=e^{-2\s}\wh\D$. We now want to establish how the operators $\Nsl$ and $\wh\Nsl$ are related.

\subsubsection{Dirac operator in $n$ dimension }

It is not difficult to work this out in general dimension $n$. Below, we will then specialise to $n=2$.
Writing the metric in terms of the local frame forms $e^a$ as in \eqref{metric} we see that there is also a corresponding conformal rescaling of the frame forms as $e^a=e^\s \hat e^a$ and correspondingly for the vielbeins $e^a_\m$ and their inverses $E^\m_a$ as
\be\label{confresce}
e^a_\m=e^\s \hat e^a_\m \quad , \quad E_a^\m=e^{-\s} \hat E_a^\m \ ,
\ee
How are the spin-connections $\hat\o^{ab}$ and $\o^{ab}$ related ? 
Writing out the  zero-torsion condition for $e^a=(e^\s \hat e^a)$ and $\o^{ab}$, and using the one for 
 $\hat e^a$ and $\hat\o^{ab}$, gives
\be\label{zerrotorsionconf}
0= \d e^a+\o^{ab} e^b =e^\s \Big[ \d\s\, \hat e^a-\hat\o^{ab} \hat e^b +  \o^{ab}  \hat e^b\Big] \ ,
\ee
which is solved as
\be\label{omegaresc}
\o^{ab}_\m=\hat\o^{ab}_\m +(\hat e^a_\m \hat E^{b\l} - \hat e^b_\m \hat E^{a\l})\del_\l\s \ .
\ee
It follows that
\be\label{spinorderconfgeneraldim}
\nabla_\m^{\rm sp} =\del_\m+\frac{1}{4}\o^{cd}_\m \g_{cd}
=\ \hat \nabla_\m^{\rm sp}  +\frac{1}{2}\hat e^c_\m \hat E^{d\l} \del_\l\s\,\g_{cd} 
\ee
Using $\g^\m=E^\m_a\g^a=e^{-\s} \hat E^\m_a\g^a=e^{-\s}\hat\g^\m$, we arrive at
\be\label{spinorderconfgeneraldim-2}
\Nsl 
= e^{-\s} \big( \wh \Nsl +\frac{1}{2}\g^c \hat E^{d\l} \del_\l\s\,\g_{cd} \big)
=  e^{-\s} \Big( \wh \Nsl +\frac{n-1}{2} \,(\hat \dsl\, \s) \Big)\ ,
\ee
where we used $\g^c\g_{cd}=(n-1)\g_d$. This can also be written as
\be\label{spinorderconfgeneraldim-4}
\Nsl = e^{-\frac{n+1}{2}\s} \wh \Nsl\  e^{\frac{n-1}{2}\s} \ .
\ee
For an infinitesimal variation $\dd\s$  of the conformal factor this yields
\be\label{Nslvar}
\dd \Nsl= -\dd\s \Nsl + \frac{n-1}{2} \g^\l\del_\l \dd\s 
\equiv  -\dd\s \Nsl + \frac{n-1}{2}\, \big(\dsl\, \dd\s \big)\ .
\ee

Consider now the Dirac operator $D=i\Nsl + m \g_*$ as given in \eqref{Dopcurved2n}.
Obviously, under a conformal rescaling the term $m\g_*$ does not change, so that 
\be\label{Dvarconf}
D=i\Nsl + m \g_*=  e^{-\frac{n+1}{2}\s} \, i \wh \Nsl\  e^{\frac{n-1}{2}\s} + m \g_* 
\quad  \Rightarrow\quad
\dd D= i \dd \Nsl  = -\dd\s \, i\Nsl + \frac{n-1}{2}\, i (\dsl\, \dd\s)
\ .
\ee
One can also work out the variation of $D^2$ which requires some care, with the result 
\be\label{deltaD2-3}
\dd D^2 = -2\dd\s\,(D^2-m^2)+(\dsl\dd\s)\Nsl 
- (n-1) g^{\m\n} (\del_\m\dd\s) \nabla_\n^{\rm sp} -\frac{n-1}{2} (\D_{\rm scalar} \dd\s) \ .
\ee
Actually, we will not need this formula. The relation \eqref{spinorderconfgeneraldim-4} and its direct consequences \eqref{Nslvar} and \eqref{Dvarconf} is all we will need.

\subsubsection{Dirac operator in two dimensions}

Let us now specialise to 2 dimensions i.e. $n=2$ in the previous formula. Then
\beb\label{spinorderconf2dim}
\Nsl = e^{-\frac{3}{2}\s} \wh \Nsl\  e^{\frac{1}{2}\s} \ ,
\eeb
as well as
\be\label{deltaD2dim}
\dd D=-\dd\s (D-m\g_*) +  \frac{i}{2}\,  (\dsl\, \dd\s)
\ee
\be\label{deltaD2-2d-2}
\dd D^2 = -2\dd\s\,(D^2-m^2) + \g^{\m\n}(\del_\m\dd\s) \nabla_\n^{\rm sp} 
 -\frac{1}{2} (\D_{\rm scalar} \dd\s)  \ ,
\ee
where $\g^{\m\n}=\e^{\m\n} i\g_*$. 
Let us check that the factors $e^{-\frac{3}{2}\s}$ and $e^{\frac{1}{2}\s}$ in \eqref{spinorderconf2dim} are exactly what is needed for $i\Nsl$ to be hermitian~:
Indeed, $ \wh\Nsl$ is hermitian for the metric $\wh g$. Then
\ba\label{hermitiancheck}
(\p_1,i\Nsl\, \p_2)&\equiv& \int \d^2 z \sqrt{g}\, \p_1^\dag i\Nsl\, \p_2
=\int \d^2 z \sqrt{\wh g}\, e^{2\s}\, \p_1^\dag\,  e^{-\frac{3}{2}\s} i\wh\Nsl\, \big( e^{\frac{1}{2}\s} \p_2\big)\nonumber\\
&=&\int \d^2 z \sqrt{\wh g}\,  \,  \big(i\wh\Nsl\, ( e^{\frac{1}{2}\s}\p_1)\big)^\dag  \, \big( e^{\frac{1}{2}\s} \p_2\big)
=\int \d^2 z \sqrt{g}\, \,  \big(i\Nsl \p_1\big)^\dag  \, \p_2 
\equiv (i\Nsl\, \p_1,\p_2)\ ,
\ea
so that $i\Nsl$ is hermitian for the metric $g$. Actually, the same argument applies in $n$ dimensions with \eqref{spinorderconfgeneraldim-4} and $\sqrt{g}=e^{n\s}\sqrt{\wh g}$.
We see that we coud have derived the formula \eqref{spinorderconfgeneraldim-4} and \eqref{spinorderconf2dim} solely from the requirement that $i\Nsl\,$ is hermitian for the metric $g$ if $i\wh\Nsl\,$ is hermitian for the metric $\wh g$.

\setlength{\baselineskip}{.52cm}

\subsection{Zero-modes\label{zeromodesec}}

We will now collect a few results and remarks one can rather easily derive about the zero-modes of the {\it massless} Dirac operator $i\Nsl$ on a two-dimensional manifold without boundaries.

We will easily re-derive that there are no zero-modes of $i\Nsl\,$ for spherical topology, and we explicitly obtain the zero-modes for toroidal topology (with doubly periodic spin structure). For genus ${\tt g}\ge 2$ we do not have explicit formulae for the zero-modes, but we can nevertheless sufficiently characterise their dependences on the conformal factor, as needed later-on. It is actually known \cite{DP, Hitchin} that for genus ${\tt g}\ge 3$, even the number of zero-modes depends on the conformal class of the metric and no simple formula for their number can exist.

\subsubsection{General remarks about zero-modes}

First recall that $i\Nsl\ $ is purely imaginary so that if $i\Nsl\, \p_0=0$ then also $i\Nsl\, \p^*_0=0$. Hence, if $\p_0$ is a complex zero-mode, then $\p_0^*$ is another zero-mode.

As already noticed, $i\Nsl$ anticommutes with $\g_*$. Then, for $m\ne 0$, $D=i\Nsl + m\g_*$ neither commutes nor anti-commutes with $\g_*$ and we cannot have eigenfunctions of $D$ of definite chirality (i.e.~being also eigenfunctions of $\g_*$). However, for $m=0$, $D=i\Nsl\ $ anti-commutes with $\g_*$, so that $\g_*\p_n$ is eigenfunction of $D$ with eigenvalue $-\l_n$. Thus for $\l_n\ne 0$ (and $m=0$), $\p_n$ and $\g_*\p_n$ necessarily are orthogonal. For $\l_n=0$ (and $m=0$), however, this is not the case. Instead one can always choose a basis of definite chirality eigenfunctions. Indeed,  if $\p_0$ is a zero-mode of $i\Nsl$ without being an eigenstate of $\g_*$ then $\g_*\p_0$  and $\p_0$ are two  linearly independent zero-modes and one can form the two linear combinations $\p_{0,\pm}=\frac{1}{2} (1\pm\g_*)\p_0$. Obviously, the $\p_{0,\pm}$ now are zero-modes of definite chirality. As is well known, the difference of the number of positive and negative chirality zero-modes of $i\Nsl\,$ is called its index. Since $\g_*=\s_y=\begin{pmatrix}0&-i\\i&0\end{pmatrix}$, the positive chirality eigenfunctions must be $\sim\begin{pmatrix}1\\i\end{pmatrix}$ and the negative chirality eigenfunctions must be  $\sim\begin{pmatrix}1\\-i\end{pmatrix}$. We see that they are intrinsically complex, and complex conjugation switches the chirality. Hence there are as many positive chirality as negative chirality zero-modes, i.e.~the index of $i\Nsl$ is zero. Indeed, this index is related to the Pontryagin invariant (see e.g.~\cite{ABanomaly}) which vanishes in $n=4k+2$ dimensions, and in particular in 2 dimensions.

On the other hand, $D^2$  commutes with $\g_*$, and one can then take all the eigenfunctions (zero and non-zero modes) of $D^2$ to have definite chirality. From the discussion of the previous paragraph it is then clear that in general these definite chirality eigenfunctions of $D^2$ are {\it not} eigenfunctions of $D$.

There are some simple standard arguments about when zero-modes can exist. First of all, assuming the eigenfunctions to be normalised,
\be\label{positivity}
\l_n^2-m^2=(\p_n,(D^2-m^2)\p_n)=(\p_n,(i\Nsl)^2\p_n)=(i\Nsl\p_n,i\Nsl\p_n) \ge 0 \ .
\ee
We see that 
\beb\label{Lambdanconstrqint}
\l_n^2 > m^2\ , \quad  \text{unless $i\Nsl\, \p_0=0$ in which case $\l_0^2=m^2$} \ . 
\eeb
In particular, if $m\ne 0$ we always have $\l_n^2\ge m^2>0$ and there cannot be any zero-modes of the massive Dirac operator. 
Furthermore, if $i\Nsl\,$ admits a zero-mode, then the smallest eigenvalue is $\l_0=m$.

Next, we have 
\be\label{Spinorlaplcond}
(\p_n,(-\D_{\rm sp})\p_n)=-\int\d^2 x\sqrt{g}\, \p_n^*\, g^{\m\n}\nabla_\m \nabla_\n\p_n
=\int\d^2 x\sqrt{g} (\nabla_\m^{\rm sp}\p_n)^* g^{\m\n} \nabla_\n^{\rm sp}\p_n \equiv \a^2 \ge 0 \ .
\ee
Since the metric $g^{\m\n}$ is positive definite, $\a=0$ means that $\nabla_\m^{\rm sp}\p_n=0$ for both $\m=1$ and $\m=2$, i.e.~$\p_n$ must be covariantly constant. However, since $[\nabla_1, \nabla_2]\p \sim \cR \g_*\p$, we see that a covariantly constant spinor can only exist on a manifold with vanishing curvature, $\cR=0$ everywhere, i.e.~only on the flat torus. For all other two-dimensional manifolds without boundary (and metrics) one has $\a^2>0$.
Next, it follows   from \eqref{Dsquared-2d}   that 
\be\label{meancurvrel}
\l_n^2-m^2 = \a^2 +\frac{1}{4} \int\sqrt{g}\, \cR\, \p_n^\dag \p_n =\a^2 +\frac{1}{4} \langle \cR \rangle_n \ ,
\ee
where $\langle \cR \rangle_n$ denotes the ``mean curvature in the state $\p_n$"
If we choose a constant curvature metric, $\cR=\frac{8\pi}{A} (1-{\tt g})$
where ${\tt g}$ is the genus of the Riemann surface, this becomes
\be\label{meancurvrel-2}
\l_n^2-m^2 = \a^2 + \frac{2\pi}{A} (1-{\tt g})  \ .
\ee
We will show soon that there is a simple relation between the zero-modes of the metric $\wh g$ and the zero-modes of the metric $g=e^{2\s}\wh g$. In particular, we may choose $\wh g$ to be the constant curvature metric and then use \eqref{meancurvrel-2} to argue about the existence of the corresponding zero-modes or not. In particular,
we see that for spherical topology (${\tt g}=0$) the right-hand side is $\a^2+\frac{2\pi}{A}>0$ so that $\l_n^2 > m^2$ and the Dirac operator cannot have any zero-mode even for $m=0$. For the flat torus we have $\l_n^2= m^2 + \a^2$ and for $m=0$ one can have zero-modes provided $\a=0$, meaning that the zero-modes must be covariantly constant spinors that precisely exist (only) on the flat torus as we have just seen. Then there are also corresponding zero-modes on an arbitrary torus.\footnote{
Since the latter cannot be covarianty constant and hence must have $\a^2>0$ they necessarily have a negative {\it mean} curvature~: $\langle \cR \rangle_0=-4\a^2$.
}
Finally, for genus two and larger, the right-hand side of \eqref{meancurvrel-2} can vanish and there are zero-modes of the massless Dirac operator. 

\vskip3.mm
\noindent
\underline{Remarks on comparing eigenvalues for vanishing and non-vanishing mass~:}
Let us denote by $\pm\l_n^{(0)}$ the positive and negative eigenvalues of $D$ for $m=0$, i.e.~of $i\Nsl\ $~: $i\Nsl\, \p^{(0)}_{\pm n, i}=\pm \l_n^{(0)} \p^{(0)}_{\pm n, i}$, where the additional index $i$ for the corresponding eigenfunctions takes into account that the eigenvalues $\pm\l_n^{(0)}$ can be degenerate. Now, for $\pm\l_n^{(0)}\ne 0$ these eigenfunctions are in general not eigenstates of $\g_*$ and, hence, the $\p^{(0)}_{\pm n, i}$ are {\it not} eigenfunctions of $D=i\Nsl+m\g_*$. Obviously then, there cannot be any linear relation between the $\l_n$ and the $\l_n^{(0)}$. However, we have $D^2 \p^{(0)}_{\pm n, i}=\big ( (i\Nsl)^2+m^2\big) \p^{(0)}_{\pm n, i}=\big( 
(\pm\l_n^{(0)})^2+m^2\big) \p^{(0)}_{\pm n, i}$, so that the $\p^{(0)}_{\pm n, i}$ are eigenfiunctions of $D^2$ with eigenvalues 
\be\label{lambdam=0andm}
\l_n^2=(\l_n^{(0)})^2+m^2 \ .
\ee 
This means that the eigenfunctions $\p_n$ corresponding to $\l_n$ can be identified with some linear combination of the $ \p^{(0)}_{\pm n, i}$, and similarly for the $\p_{-n}=\p_n^*$ corresponding to $-\l_n$.

\subsubsection{Zero-modes on a general torus\label{gentoruszeromodessec}}

Before discussing the case of arbitrary genus, let us first consider the zero-modes on the torus where we can give very explicit results. The general torus is defined by a) the periodicity conditions of the flat torus and b) a metric given by $g_{\m\n}=e^{2\s(x)}\dd_{\m\n}$ (i.e.~$\wh g_{\m\n}=\dd_{\m\n}$), where $\s(x)\equiv \s(x^1,x^2)$ must also satisfy the appropriate periodicity conditions.  Thus from \eqref{spinorderconf2dim} we get
\be
D=ie^{-3\s(x)/2} \dsl \, e^{\s(x)/2} + m \g_*
= e^{-3\s(x)/2} \big( i\s_x\del_1 + i\s_z \del_2 \big) e^{\s(x)/2} + m  \s_y \ .
\ee
The derivative and the mass terms behave differently under the conformal transformations. However, we see that {\it for zero mass}, the zero-modes $\p_{0,i}$ are simply determined by
\be\label{Nslpsi0eq}
D^{m=0} \p_{0,i}(x) =i\Nsl\, \p_{0,i}(x)= 0
\quad \Rightarrow\quad i\dsl \big(e^{\s(x)/2}  \p_{0,i}(x)\big)=0 \ ,
\ee
and one identifies $e^{\s(x)/2}  \p_{0}(x)$ with the zero-modes $\wh \p_{0,i}$ of $i\dsl$, i.e.~the zero-modes of the flat torus as determined above in \eqref{toruszeromodes}. Actually, this argument generalises to arbitrary genus as we will discuss shortly.  What is special for the torus is that we know explicitly the $\wh \p_{0,i}$. Hence the two zero-modes of $i\Nsl\, $ are
\be\label{gentoruszeromodesunnorm}
\p_{0,1}(x)= c\,e^{-\s(x)/2} \,\wh \p_{0,1}(x)= c\ \frac{e^{-\s(x)/2}  }{2\pi\sqrt{2\tau_2}} \begin{pmatrix} i \\ -1 \end{pmatrix} 
\quad , \quad 
\p_{0,2}(x)= c\, e^{-\s(x)/2} \,\wh \p_{0,2}(x)=    c\ \frac{e^{-\s(x)/2}}{2\pi\sqrt{2\tau_2}} \begin{pmatrix} -i \\ -1 \end{pmatrix}  \ ,
\ee
where $c$ is a normalisation constant to be determined.
Indeed, the normalisation now is defined with respect to the metric $g=e^{2\s}\wh g$ as (no sum over $i$)
\be\label{normalisation}
1=\int\d^2 x \sqrt{g(x)}\, \p_{0,i}^\dag(x)\p_{0,i}(x)
=\int\d^2 x \sqrt{\wh g}\, e^{2\s(x)} \frac{|c|^2}{(2\pi)^2 \tau_2} e^{-\s(x)}
= \frac{|c|^2}{(2\pi)^2 \tau_2}\int\d^2 x \sqrt{\wh g}\, e^{\s(x)} \ .
\ee
Of course, the integral in the last expression is {\it not} the area which would be $A=\int\d^2 x \sqrt{\wh g}\, e^{2\s(x)}$. Also note that $(2\pi)^2 \tau_2$ is the area of the torus with metric $\wh g$, i.e. for $\s=0$. We now define the ``area-like" parameters
\be\label{AhatAn}
A_{(n)}=\int\d^2 x \sqrt{\wh g}\, e^{n\s(x)} \quad \Rightarrow \quad A_{(2)}=A \ , \quad A_{(0)}=\wh A = (2\pi)^2 \tau_2\ .
\ee
Hence, up to a choice of phase, $c=\sqrt{\frac{\wh A}{A_{(1)}}}=2\pi \sqrt{\frac{\tau_2}{A_{(1)}}}$, and
\be\label{gentoruszeromodes}
\p_{0,1}(x)=  \frac{e^{-\s(x)/2}  }{\sqrt{2 A_{(1)}}} \begin{pmatrix} i \\ -1 \end{pmatrix} 
\quad , \quad 
\p_{0,2}(x)=  \frac{e^{-\s(x)/2}  }{\sqrt{2 A_{(1)}}}  \begin{pmatrix} -i \\ -1 \end{pmatrix}  \ .
\ee
It follows that the projector on the zero-modes  is
\be\label{zeromodeprojgentorus}
P_0(x,y)=\sum_{i=1}^2 \p_{0,i}(x)\, \p_{0,i}^\dag(y) 
=\frac{e^{-(\s(x)+\s(y))/2}}{A_{(1)}} \ {\bf 1}_{2\times 2}  =\frac{e^{-(\s(x)+\s(y))/2}}{\int\d^2 z \sqrt{\wh g}\,e^\s} \ {\bf 1}_{2\times 2} \ ,
\ee
which now depends on $x$ and $y$, but  is still proportional to the unit matrix (as for the flat torus).

The zero-modes $\p_{0,1}$ and $\p_{0,2}$ as defined in \eqref{gentoruszeromodes} have definite chirality and then also are eigenfunctions of $D$ with non-zero mass. Indeed, we have $\g_* \p_{0,1}=\p_{0,1}$ and $\g_* \p_{0,2}=-\p_{0,2}$ and then
\be
D\p_{0,1}=m\p_{0,1} \quad , \quad D\p_{0,2}=-m\p_{0,2} \ .
\ee
Thus, the above projector $P_0$ also is the projector on the eigen-modes of $D$, that become zero-modes when the mass is taken to zero.

\subsubsection{Zero-modes of $i\Nsl\,$ for arbitrary genus ${\tt g}\ge 1$}

We have seen before that for genus zero (spherical topology) there are no zero-modes of $i\Nsl\,$, and we have just discussed the two zero-modes for a general genus-one manifold. Now we want to establish some further general results valid for arbitrary genus.  For higher genus surfaces there are more zero-modes, but in any case there are only finitely many of them. As already mentioned, their number depends \cite{DP, Hitchin} on the spin structure and, for genus ${\tt g}\ge 3$ also on the conformal class of the metric, i.e.~on the $\wh g(\tau_i)$. Of course, within any conformal class this number is fixed, and we have explicitly related the $\p_{0,i}$ to the $\wh\p_{0,i}$.
As usual, we suppose that the zero-modes are all ortho-normalised~:
\be\label{zeromodesorthonormal}
i\Nsl \, \p_{0,i}=0 
\quad , \quad
\int\d^2 x \sqrt{g(x)} \,\p^\dag_{0,i}(x) \p_{0,j}(x)=\dd_{ij} \ .
\ee
As already discussed, these zero-modes can be chosen to have definite chirality, so that even for non-vanishing mass they continue to be eigenstates of the Dirac operator~:
$D\p_{0,i}=m\g_* \p_{0.i}=\pm m\,\p_{0,i}$.
Recall that the zero-modes of definite chirality necessarily are complex and $\p_{0,i}$ and $\p^*_{0,i}$ have opposite chirality and, hence,  that there must be an even number $2n_0$ of zero-modes, $n_0$ with positive chirality and $n_0$ with negative chirality, resulting in a vanishing index of $i\Nsl$. Let us then choose the labelling of the zero-modes such that $\p_{0,i}, \ i=1,\ldots n_0$ have positive chirality and $\p_{0,i}, \ i=n_0+1,\ldots 2n_0$ have negative chirality, and moreover such that $\p_{0,i}^*=\p_{0,i+n_0},\ i=1,\ldots n_0$.

The projector on these zero-modes then is
\be\label{zeromodeproj}
P_0(x,y)=\sum_{i=1}^{2n_0} \p_{0,i}(x)\, \p_{0,i}^\dag (y) 
=\sum_{i=1}^{n_0} \big(\p_{0,i}(x)\, \p_{0,i}^\dag (y)+ \p^*_{0,i}(x)\, {\p^*_{0,i}}^\dag (y)\big) \ .
\ee
This is indeed a projector in the sense that we have, by \eqref{zeromodesorthonormal}, the relation
\be\label{projproperty}
\int\d^2 z \sqrt{g(z)} P_0(x,z) P_0(z,y) = P_0(x,y)\ .
\ee
Since the zero-modes were chosen to have definite chirality $\pm 1$ it follows that
\be\label{gammastarPzero}
\g_* P_0(x,y) \g_*=P_0(x,y)  \quad \Leftrightarrow\quad \big[ \g_*,P_0(x,y)\big]=0\ .
\ee
Now, the only matrices that commute with $\g_*$ are  ${\bf 1}_{2\times 2}$ and $\g_*$ itself.
It follows that $P_0$ has the following matrix structure
\be\label{P0matrixstructure}
P_0(x,y)= p_{0,0}(x,y) \, {\bf 1}_{2\times 2} + p_{0,*}(x,y) \, \g_* \ .
\ee
One can also argue, by choosing a basis of real zero-modes instead, that  $P_0(x,x)$ is real and hermitian, excluding the $\g_*$-piece for coinciding points. However, this argument does not exclude a $\g_*$-piece for $x\ne y$.
For an arbitrary torus (with modular parameter $\tau$ and conformal factor $\s$), we could explicitly show that the matrix-structure is indeed $P_0(x,y)\sim {\bf 1}_{2\times 2}$, and it would be nice to show this for arbitrary genus.

It will also be useful to define similarly
\be\label{zeromodeQ}
Q_0(x,y)=\sum_{i=1}^{n_0} \big(\p_{0,i}(x)\, \p_{0,i}^\dag (y)- \p^*_{0,i}(x)\, {\p^*_{0,i}}^\dag (y)\big) \ .
\ee
Obviously
\be\label{PQrel}
Q_0(x,y)=\g_* P_0(x,y)=P_0(x,y)\g_* 
=p_{0,0}(x,y) \, \g_*+ p_{0,*}(x,y) \, {\bf 1}_{2\times 2} \ ,
\ee
so that
\be\label{projpropertyQ}
\int\d^2 z \sqrt{g(z)} Q_0(x,z) Q_0(z,y) 
=\int\d^2 z \sqrt{g(z)} P_0(x,z) P_0(z,y) = P_0(x,y)\ .
\ee

\subsubsection{Conformal variation of the zero-modes and of the zero-mode projector\label{confvarzeroproj}}

We will need the variation of the zero-modes  under conformal transformations. Fortunately this can be easily obtained exactly, generalising the above discussion for the torus. Recall that under  $g=e^{2\s}\wh g$ we have ${\Nsl}=e^{-3\s/2}\wh\Nsl\, e^{\s/2}$. 
Then $\Nsl\, \p_{0,i}=0$ is equivalent to $\wh\Nsl\, (e^{\s/2}\p_{0,i})=0$ so that $e^{\s/2}\p_{0,i}$ is a zero-mode of $\wh\Nsl\,$ and we can identify $e^{\s(x)/2}\p_{0,i}(x)\sim \wh\p_{0,i}(x)$,
i.e.~the  zero-modes of $\Nsl$ are $\p_{0,i}(x)\sim e^{-\s(x)/2}\wh\p_{0,i}(x)$. However, a priori, they are not ortho-normalised. Instead, we can consider the linear combinations of the $n_0$ positive chirality zero-modes
\be\label{Nslzeromodesgeneral}
\p_{0,k}(x)=\sum_{j=1}^{n_0}\,e^{-\s(x)/2}\, \wh\p_{0,i}(x)\, c_{ik}\ ,
\ee 
and the corresponding complex conjugate relation for the negative chirality zero-modes. Imposing the ortho-normality of the $\p_{0,k}$ leads to
\be\label{CcalPrel}
\dd_{kl}=\sum_{i,j} c_{ik}^* {\cal P}_{0,ij} c_{jl} \ ,
\ee
where we have defined the (constant) matrix ${\cal P}_0$ as
\be\label{P0ijdef}
{\cal P}_{0,ij}=\int \d^2 x \sqrt{\wh g(x)}\, e^{\s(x)}\ \wh\p_{0,i}^{\,\dag}(x) \wh\p_{0,j}(x) \ .
\ee
Although the
${\cal P}_{0,ij}$ are constants, they are actually  functionals of the conformal factor $\s$. 
In matrix notation \eqref{CcalPrel} reads ${\bf 1}=c^\dag {\cal P}_0\, c$ so that ${\cal P}_0=(c^{-1})^\dag c^{-1}$ and ${\cal P}_0^{-1}= c\, c^\dag$ which shows that ${\cal P}_0$ is hermitian.\footnote{
Of course these relations only determine the matrix $c$ up to right multiplication with any unitary matrix, consistent with the fact that this just leads to another orthonormal basis of the positive chirality zero-modes $\p_{0,k}$.
} In components this reads $c_{ki} c_{li}^*=({\cal P}_0)^{-1}_{kl}$, and
inserting \eqref{Nslzeromodesgeneral}  into the above definition of the zero-mode projector $P_0$ we find
\be\label{P0sigma}
P_0(x,y) =\sum_{i=1}^{n_0}  \p_{0,i}(x)\p_{0,i}^\dag(y) + {\rm c.c.}
=e^{-(\s(x)+\s(y))/2} \, \sum_{i,j=1}^{n_0} \big( ({\cal P}_0^{-1})_{ij} \wh\p_{0,i}(x) \wh\p_{0,j}^{\,\dag}(y)\,+\, {\rm c.c.}\big) \ .
\ee
Since $Q_0=\g_* P_0=P_0 \g_*$, $\ \ Q_0$ obeys a completely analogous relation except that the negative chirality modes contribute with a minus sign to the sum. Hence, it will be enough if we further discuss $P_0$ only.
For our later computations it will  also be useful to introduce the following notations
\ba\label{Ptdef}
P_0(x,y)&=&e^{-\s(x)/2} \Pt_0(x,y) e^{-\s(y)/2}
\quad , \quad 
\Pt_0(x,y)=\sum_{i,j=1}^{n_0} \big( ({\cal P}_0^{-1})_{ij} \wh\p_{0,i}(x) \wh\p_{0,j}^{\,\dag}(y)\,+\, {\rm c.c.}\big) \ , \\
\label{P0hatdef}
\wh P_0(x,y)&=&\sum_i \big(\wh\p_{0,i}(x)\wh\p^\dag_{0,i}(y)+{\rm c.c.}\big) \ .
\ea
One must keep in mind that the quantity $\Pt_0(x,y)$ still depends on $\s$
 due to the appearance of $e^\s$ in the ortho-normalisation matrix ${\cal P}_0$ of the zero-modes. Hence, the overall $\s$ dependences of the projectors are rather non-trivial, but perfectly explicit.
Under an infinitesimal conformal variation with $\dd\s$ we have
\be\label{deltaP0}
\dd P_0(x,y)=-\frac{1}{2}\big( \dd\s(x)+\dd\s(y)\big) P_0(x,y) 
- \Big[\sum_{i=1}^{n_0} ( c^\dag \dd {\cal P}_0 c)_{ij} \ \p_{0,i}(x) \p_{0,j}^{\, \dag}(y) \,+\, {\rm c.c.} \Big] \ .
\ee
A somewhat simpler quantity is
\be\label{P0confvar2}
\int\d^2 y \sqrt{g}\, \dd\s(y) \tr P_0(y,y)
= \sum_{i,j=1}^{n_0} ({\cal P}_0^{-1})_{ij}  \dd {\cal P}_{0,ji}\,+\, {\rm c.c.} =   2\, \dd\, \Tr \log {\cal P}_0
=2\, \dd\, \log\det {\cal P}_0\ ,
\ee
where here $\Tr$ and $\det$ denote the trace and determinant of the $n_0\times n_0$ matrices and the factor 2 occurs since ${\cal P}_0$ is hermitian and hence the trace is real.

How shall we interpret the quantities ${\cal P}_{0,ij}$~? We have seen that for
the example of a torus, $n_0=1$ and  $\wh\p_{0,i}^{\,\dag}(z) \wh\p_{0,i}(z)=\frac{1}{\wh A}$, where $\wh A=\int\d^2 z \sqrt{\wh g}$. Hence there is a single ${\cal P}_{0}$ which equals ${\cal P}_{0}=\frac{A_{(1)}}{\wh A}$ with $A_{(1)}=\int\d^2 z \sqrt{\wh g} \, e^\s$ an area-like constant. Similarly, on a general genus Riemann surface we will interpret the ${\cal P}_{0,ij}$ as a finite number of area-like parameters. Of course, the appearance of a term $({\cal P}^{-1}_{0})_{ij}$ is non-local in $\s$ but it is no more  non-local than a factor $\frac{1}{A}$.

\newpage
\setlength{\baselineskip}{.60cm}
\section{Fermionic matter partition function and gravitational action}

\subsection{Fermionic functional integral for Majorana spinors}

We define the matter partition function for fermionic matter with action $S=\int\d^2 x \sqrt{g}\,\p^\dag D\p$, where $D=i\Nsl+m\g_*$, on a two-dimensional manifold with metric $g$ as the functional integral
\be\label{Zg}
Z_{\rm mat}[g]= \int \cD \P \exp\left( - S[g,\P]\right) \quad , \quad S[g,\P]=\int\d^2 x \sqrt{g}\, \P^\dag D_g \P\ ,
\ee
where we wrote $D_g$ to insist that this is the Dirac operator $D$ for the metric $g$ (and corresponding vielbein $e$ and spin connection $\o$).  We want to consider here only {\it Majorana spinors} $\P$, which means that $\P(x)$ is an arbitrary anti-commuting {\it real} spinor and $\P^\dag\equiv\P^{\rm T}$. The Dirac operator is purely imaginary.  Nevertheless, our action is real, as it should. Indeed,  recall that for  anti-commuting objects $a$ and $b$,  complex conjugation is defined as hermitian conjugation and reverses the order, so that $(a b)^*=b^* a^*$. Now  $(D_g\P)^*=-D_g\P$. Then
$(S[g,\P])^*=\int \big(\P^{\rm T} D_g \P\big)^*
=\int  ((D_g \P)^*)^{\rm T} \P =-\int  (D_g \P)^{\rm T} \P 
=+\int  \P^{\rm T}  D_g \P= S[g,\P]$, where we used the anti-commutativity in the last step.

Before going on, let us show that if we considered instead a Dirac spinor, i.e.~an anti-commuting {\it complex}  spinor $\P_{\rm Dirac}=\P+i \wt \P$, where $\P$ and $\wt\P$ are two anti-commuting {\it real} (Majorana) spinors, the action for the Dirac spinor (with the same Dirac operator $D$) would simply be the sum of the actions for the real and imaginary parts, $\P$ and $\wt\P$, separately without any cross-term, namely
\be\label{SDSM1}
S[g,\P_{\rm Dirac}]=S[g,\P]+S[g,\wt \P] \ .
\ee
Indeed,
\be\label{SDSM2}
S[g,\P_{\rm Dirac}]=\int \P_{\rm Dirac}^\dag D \P_{\rm Dirac} 
= \int (\P+i \wt \P)^\dag D (\P+i \wt \P)
=S[g,\P]+S[g,\wt \P]  - i \int \big( \wt\P^\dag D \P - \P^\dag D \wt\P\big) \ ,
\ee
and, using the hermiticity of $D$ and the fact that $\P$ and $\wt\P$ are real and anti-commuting and that $D$ is purely imaginary, one easily shows that $\int \big( \wt\P^\dag D \P - \P^\dag D \wt\P\big) =0$.
It then follows rather straightforwardly that one has for the corresponding partition functions
\be\label{DiracmatterZ}
Z^{\rm Dirac}_{\rm mat}[g]= \int \cD \P_{\rm Dirac} \exp\left( - S[g,\P_{\rm Dirac}]\right)
= \int \cD \P \cD \wt\P  \exp\left( - S[g,\P]- S[g,\wt\P]\right)=\Big( Z_{\rm mat}[g]\Big)^2\ .
\ee
Thus the matter partition function for a Dirac spinor is simply the square of the matter partition function for a Majorana spinor, and we may thus consider the latter as the more ``fundamental" object to study. Hence, for the remainder of this paper we will consider Majorana spinors only.

We can then expand the anti-commuting real (Majorana) spinor $\P$ on a complete set of real eigenfunctions of $D_g$:
\be\label{eigenexp}
\P(x)=\frac{1}{\sqrt{\m}}\Big( \sum_n (b_n \chi_n(x) + c_n \f_n(x)) +\sum_i d_i \p_{0,i}(x) \Big) \ .
\ee
Here and in the following it is understood that the $\f_n$ and $\chi_n$ are the real  eigenfunctions of $D_g^2$, satisfying $D_g\f_n=-i\l_n\chi_n$ and $D_g\chi_n=i\l_n\f_n$ (cf \eqref{evphichi}) for strictly positive eigenvalues $\l_n>0$. In particular, they are orthonormalised according to
\eqref{orthonomality}.
The orthonormalized zero-modes, if present, are denoted $\p_{0,i}$.
Recall that zero-modes are present only for $m=0$ and genus ${\tt g}\ge 1$ and then there are only finitely many of them. Of course, they are also orthogonal to the $\f_n$ and $\chi_n$. 

The eigenfunctions $\chi_n,\ \f_n$ and $\p_{0,i}$ are real, {\it commuting} functions, providing an orthonormalized  basis. Since $\p$ is anticommuting,  the expansion coefficients $b_n, c_n, d_i$  must be real and {\it anti}-commuting. Then
\be\label{Psidaggerexp}
\P^\dag(x)=\frac{1}{\sqrt{\m}}\Big( \sum_n (b_n \chi^\dag_n(x) + c_n \f^\dag_n(x)) +\sum_i d_i \p^\dag_{0,i}(x) \Big) \ ,
\ee
were actually $\chi^\dag_n=\chi^{\rm T}_n$, $\f^\dag_n=\f^{\rm T}_n$, $\p_{0,i}^\dag=\p_{0,i}^{\rm T}$.
We have also introduced  an arbitrary mass scale $\m$ so that these coefficients $b_n,  c_n, d_i$ are dimensionless. Indeed,  from the normalisation condition of the eigenmodes one sees that the $\chi_n$, $\f_n$ and $\p_{0,i}$ have engineering dimension one, i.e. $\chi_n\sim\f_n\sim \p_{0,i}\sim \m$, and
since $\P$ must have dimension $\frac{1}{2}$ so that the action 
$ \int\d^2 x \sqrt{g}\, \P^\dag D_g \P$ is dimensionless, we see from \eqref{eigenexp} that  $b_n, \ c_n, \ d_i$ are indeed dimensionless thanks to this explicit factor $\frac{1}{\sqrt{\m}}$.

Inserting the expansions \eqref{eigenexp} and \eqref{Psidaggerexp} into the action $S$ as given in \eqref{Zg} we  get, thanks to the orthonomality of the eigenfunctions and the anticommutativity of the expansion coefficients $b_n$ and $c_n$
\be\label{actionexp}
S[g,\P]= \int\d^2 x \sqrt{g}\, \P^\dag D_g \P= 2 i \sum_n \frac{\l_n}{\m} c_n b_n \ ,
 \ee
 where of course only the non-zero modes contribute. Note that, despite the explicit appearance of the factor $i$, this expression is real as we have shown before. Indeed, $(ic_n b_n)^*=-i b_n c_n=+i c_n b_n$.
 
The functional integral measure $\cD\P$ is defined in terms of grassmann integrals over these coefficients $b_n$ and $c_n$ as $\cD\P=\int \prod_l \d b_l \d c_l$, but not including the $d_i$.  For grassmannian variables the integrals are non-vanishing only if each $b_n$ and $c_n$ appears exactly once in the integrand. Expanding the exponential then gives
\be\label{Zg2}
Z_{\rm mat}[g]= \int \prod_l \d b_l\, \d c_l \exp\left( -2i \sum_n \frac{\l_n}{\m} c_n b_n  \right) 
= {\cal N} \prod_n \frac{\l_n}{\m}\ ,
\ee
where ${\cal N}$ is some overall (infinite) ``normalisation" constant that will drop out in the end.
The product is only over all strictly positive eigenvalues $\l_n$. In principle one could have also done the integral over the zero-mode coefficients $d_i$ but then one needs to insert an extra factor $\prod_i d_i$ inside the functional integral  to get a non-vanishing result. This result would have been the same as  in \eqref{Zg2}, where we excluded the zero-mode integration from the beginning. Since there is a certain arbitrariness in dealing with the zero-modes, if present, one can also define $Z_{\rm mat}$ with some zero-mode related factor included (like e.g.~$\sqrt{A}$ in the case of a scalar matter field, see \cite{BL2}). We will come back to this point later.

The gravitational action was defined in \eqref{Sgravgen} as 
$Z_{\rm mat}[g]=e^{-S_{\rm grav}[g,\wh g]}Z_{\rm mat}[\wh g]$, or equivalently\break
$S_{\rm grav}[g,\wh g]=-\log \frac{Z_{\rm mat}[g]}{Z_{\rm mat}[\wh g]} \ ,$
so that
\be\label{gravac2}
S_{\rm grav}[g,\wh g]=-\log \prod_n \frac{\l_n[g]}{\m}+ \log  \prod_n \frac{\l_n[\wh g]}{\m} \ .
\ee
Since the product over $n$ only involves the non-zero eigenvalues this is sometimes denoted as $\prod_n'$, but in our convention $n$ runs only over the non-zero eigenvalues anyway, so the prime is not necessary.
We may also rewrite this in terms of the determinant of $D^2$, which has
eigenvalues $\l_n^2$ for the $\chi_n$ and $\l_n^2$ for the $\f_n$, and possibly $0$ for the $\p_{0,i}$. The determinant of $D^2$ computed on the space of non-zero-modes is conventionally denoted by $\Det'$ so that here we are interested in $\Det' D^2$~:
\be\label{D2det}
\Det' D^2 =\big( \prod_n \l_n^2 \big)^2\ ,
\ee
and 
\beb\label{gravac3}
S_{\rm grav}[g,\wh g]=-\frac{1}{4}\log \Det' D^2[g] + \frac{1}{4}\log \Det' D^2 [\wh g] \ .
\eeb

\subsection{Zeta function regularization}

All these determinants and products of eigenvalues are, of course, ill-defined since the infinite products are divergent, and have to be regularized. We will use the standard tool of regularization via the corresponding zeta-functions \cite{SHzetareg}. 


\subsubsection{Zeta-functions of positive operators}

The zeta-function of an operator ${\cal O}$ with eigenvalues $\L_n>0$ is defined as
\be\label{zetaOdef}
\zeta_{\cal O}(s) =\sum_n \L_n^{-s}\ .
\ee
Proceeding formally, i.e.~disregarding whether or not the series converges, the determinant of ${\cal O}$ is related to the derivative of this zeta function at $s=0$ as follows
\be\label{zetaOderiv}
\zeta'_{\cal O}(s) \equiv \frac{\d}{\d s} \zeta_{\cal O}(s) = -\sum_n (\log\L_n) \L_n^{-s}
\quad \Rightarrow \quad
\zeta'_{\cal O}(0) = -\sum_n (\log\L_n)=-\log \Det {\cal O} \ .
\ee
For the operator $D^2$ we are interested in, the corresponding zeta-function is 
\be\label{ZetaD2}
\zeta_{D^2}(s)=2\sum_n (\l_n^2)^{-s} \ ,
\ee
since each eigenvalue $\l_n^2>0$ occurs twice, once with eigenfunction $\chi_n$ and once with $\f_n$. Again, also, the sum obviously does not include any zero eigenvalue in case there are zero-modes.

The zeta-function regularization is a standard technique to regulate and possibly assign finite values to otherwise diverging sums. The zeta-functions of operators   are generalisations of Riemann's zeta function $\zeta_{\rm R}$,  defined as
\be\label{Riemannzeta}
\zeta_{\rm R}(s)=\sum_{n=1}^\infty n^{-s}\ .
\ee


\subsubsection{A short discussion of Riemann's zeta function}

It will be useful to briefly discuss some of the well-known properties of Riemann's zeta-function, in view of the generalisation of these properties to the zeta-functions of operators. Riemann's zeta function as defined by the sum \eqref{Riemannzeta} should be
considered as a function of a complex argument $s\in {\bf C}$.
Then the sum is absolutely convergent for $\Re s>1$ ($\Re s$ designates the real part of $s$). Indeed, for complex $s$ write $s=s_1+ i s_2$ so that $n^{-s}=n^{-s_1} n^{-i s_2}=n^{-s_1}  e^{-is_2 \log n}$ and $|n^{-s}|=n^{-s_1}$, which shows that the sum indeed converges absolutely for $s_1>1$, i.e. $\Re s>1$.
This then defines an analytic function $\zeta_{\rm R}(s)$ for all $\Re s>1$.   This function can  be analytically continued to $\Re s \le 1$.
The function that results from analytically continuing $\zeta_{\rm R}$ will still be still called $\zeta_{\rm R}$. Riemann has shown that this analytic continuation defines a meromorphic function on the whole complex plane with a single simple pole at $s=1$ with unit residue. This means that $\zeta_{\rm R}(s)$ is an analytic function on all of ${\bf C}\backslash \{1\}$ and that $\lim_{s\to 1}(s-1)\zeta_{\rm R}(s)=1$. In particular, $\zeta_{\rm R}(0)=-\frac{1}{2}$ is finite.

An important ingredient  in this analytic continuation is Euler's Gamma function, defined as
\be\label{Gamma}
\G(s)=\int_0^\infty \d t\, t^{s-1} e^{-t} \quad , \quad \Re s>0\ .
\ee
Integrating by parts, it is trivial to show that, for $\Re s>0$, $\G(s)$ satisfies the functional relation
$\G(s+1)=s\G(s)$.
This relation can then be used to define $\G(s)$ for $\Re s \le 0$, except at the isolated points $s= 0, -1, -2, \ldots$. Indeed, the functional relation shows that $\G(s)$ has simple poles at all negative integers and 0, and is otherwise analytic.
Similarly, for $\Re s>1$, the sum \eqref{Riemannzeta} admits a simple integral representation
\be\label{zetaintegral}
\sum_{n=1}^\infty n^{-s} = \frac{1}{\G(s)}\int_0^\infty \d t\, t^{s-1} \frac{1}{e^t-1}\quad  , \quad \Re s>1 \ ,
\ee
as one sees by expanding $(e^t-1)^{-1}=\sum_{n=1}^\infty e^{-nt}$,
and interchanging the sum and the integral, which is justified when both sums converge absolutely, i.e.~for $\Re s>1$.
For later reference let us denote 
\be\label{Riemannheatkernel}
K_{\rm R}(t)=\sum_{n=1}^\infty e^{-t n}=\frac{1}{e^t-1} \ ,
\ee
and refer to it as the ``Riemann heat kernel". Then the previous equation can be summarised as
\be\label{zetaintegral3}
\zeta_{\rm R}(s) = \frac{1}{\G(s)}\int_0^\infty \d t\, t^{s-1} K_{\rm R}(t) \ .
\ee
We will heavily use a generalisation of this relation later-on.

The basic functional relation used to perform the analytic continuation of  $\zeta_{\rm R}$ is the  identity proven by Riemann: the function $\xi(s)= \pi^{-s/2} \G(s/2) \zeta_{\rm R}(s)$ satisfies $\xi(s)=\xi(1-s)$. The proof of this identity uses various standard techniques including Poisson resummation and Mellin transforms  of theta functions (similar to the integral representation \eqref{zetaintegral}). We will not go into details here. Instead, we will give a heuristic argument why the integral representation \eqref{zetaintegral} or \eqref{zetaintegral3} of $\zeta_{\rm R}$ can be analytically continued to yield a meromorphic function on the whole complex plane with a single simple pole at $s=1$ with residue 1. 
Again, this argument will be generalised below to study the zeta-function associated with our operator $D^2$.

First, observe that any singularity of the integral $\int_0^\infty \d t\, t^{s-1} \frac{1}{e^t-1} $ can only come from the region $t\to 0$ where the integrand behaves as $t^{s-2}$, while for $t\to\infty$ the exponential decrease of the integrand ensures convergence.  So let us isolate this small-$t$ region by choosing  some small (but finite) positive real $\e>0$ and split the integral  \eqref{zetaintegral} into 2 parts:
\be\label{zetaint2}
\frac{1}{\G(s)}\Big( \int_0^{\e} \d t\, t^{s-1} \frac{1}{e^t-1}+\int_\e^\infty \d t\, t^{s-1} \frac{1}{e^t-1}\Big)\ .
\ee
In the first integral we can replace $\frac{1}{e^t-1}=\frac{1}{t+t^2/2+\ldots}$ by $\frac{1}{t}$ up to terms that contribute at most an order $\e$ correction. Then the first integral is elementary and yields 
\be
\int_0^\e \d t\, t^{s-2}=\frac{t^{s-1}}{s-1}\Big\vert^\e_0=\frac{\e^{s-1}-0}{s-1}\quad \text{since}\ \Re s> 1 \ .
\ee 
This term exhibits the pole at $s=1$. Indeed, letting $s\to 1$ (for fixed $\e$) one has $\e^{s-1}=e^{(s-1)\log\e}= 1 +(s-1)\log\e +\ldots$ and we get for the first integral $\frac{1}{s-1} +\log\e +{\cal O}(s-1)$. On the other hand, the second integral in \eqref{zetaint2} is regular for all $s\in {\bf C}$ and defines some analytic function for all $s\in {\bf C}$.
We see that this allows to define $\zeta_{\rm R}$ as a meromorphic function on ${\bf C}$ with a single simple pole at $s=1$ with residue 1. One might still worry about the order $\e$-corrections to the first integral as they seem to lead to $\int_0^\e \d t\, t^{s-1}=\frac{t^{s}}{s}\Big\vert^\e_0=\frac{\e^{s}-0}{s}$, i.e.~to a pole at $s=0$. However, this pole is cancelled by the zero of $\frac{1}{\G(s)}$ at $s=0$. The same cancellation occurs for all would-be poles at all negative integers. Hence there is only a single pole at $s=1$.

\subsubsection{Expressing the gravitational action in terms of $\zeta_{D^2}$}

Let us now come back to the zeta-function \eqref{ZetaD2} of our operator $D^2$. 
In the next section, we will discuss the convergence properties of the sum, as well as the properties of its analytic continuation, and see that just as $\zeta_{\rm R}$, our $\zeta_{D^2}(s)$ is analytic, except for a single simple pole at $s=1$. Hence, formally  taking the derivative of the sum with respect to $s$ and setting $s=0$ gives as before
\be\label{zetader}
\zeta'_{D^2}(0)=-2\sum_n (\log\l_n^2) =-2 \log \prod_n \l_n^2 \ .
\ee
Of course, the expression on the left-hand side is computed from the analytic continuation and thus provides a way to make sense of the otherwise diverging sum on the right-hand side.
Similarly, $(\log\m^2)\, \zeta_{D^2}(0)=2 \sum_n \log\m^2 = -2\log\prod_n \frac{1}{\m^2}$.
These $\zeta_{D^2}$ depend on the metric $g$ via the dependence of $D^2$ on $g$. Thus, to indicate the dependence on $g$ we should denote them as $\zeta_{D^2_g}$ but for notational convenience we will simple denote them as $\zeta_g$. 
Then
\be
-2\log \prod_n \frac{\l_n^2}{\m^2} =\zeta'_{D^2_g}(0)+(\log \m^2 )\,\zeta_{D^2_g}(0) 
\equiv \zeta'_{g}(0)+(\log \m^2 )\,\zeta_{g}(0) \ .
\ee
It follows that the (regularized) gravitational action is
\beb\label{gravac4}
S_{\rm grav}[g,\wh g]=\frac{1}{4} \Big( \zeta_{g}'(0) +(\log\m^2)\, \zeta_{g}(0)\Big) 
- \frac{1}{4} \Big( \zeta_{\wh g}'(0) +(\log\m^2)\, \zeta_{\wh g}'(0)\Big) \ .
\eeb

We want to determine this gravitational action for $g_{\m\n}(x)=e^{2\s(x)} \wh g_{\m\n}(x)$. Our strategy will be to first determine $\dd S_{\rm grav}$ for infinitesimal $\dd \s$ (around an arbitrary metric $g$) and then ``integrate" this variation to obtain $S_{\rm grav}[g,\wh g]$. Obviously, the variation of the gravitational action is given in terms of the variation of the zeta-function $\zeta_g(s)$ (and of its derivative $\zeta'_g(s)$) around $s=0$. To obtain this variation, we need to study the variations of the eigenvalues $\l_n$ under a corresponding variation of the Dirac operator $D$, which we will do next. To fully exploit these results we will need to study in some detail Green's functions, local zeta functions and local heat kernels and their variations  - which will be the subject of  section 4.

\subsection{Variation of the eigenvalues : perturbation theory}

We want to study how the eigenvalues $\l_n$ (or $\l_n^2$) change under local conformal rescalings of the metric with some $\dd\s(x)$. The variation $\dd D=i\dd\Nsl$ of the Dirac operator $D$  has been obtained in sect.~\ref{confsec}, see eqs \eqref{Nslvar} and \eqref{deltaD2dim}.
Recall that $D\chi_n=i\l_n\f_n$ and $D\f_n=-i\l_n\chi_n$. Then
under $D\to D+\dd D$ we have $\l_n\to \l_n+\dd \l_n$, as well as $\chi_n\to \chi_n+\dd\chi_n$ and $\f_n\to \f_n+\dd \f_n$:
\be\label{pert1}
(D+\dd D)(\chi_n+\dd\chi_n)=i(\l_n+\dd\l_n)(\f_n+\dd\f_n) \quad \Rightarrow \quad\
\dd D\chi_n + D \dd\chi_n = i \dd\l_n \f_n+i \l_n\dd\f_n\ .
\ee
Taking the inner product with $\f_n$ and using the hermiticity of $D$ one gets
\be\label{pert2a}
i\dd\l_n (\f_n,\f_n)=(\f_n,\dd D \chi_n)+(\f_n,D \dd\chi_n) -i\l_n (\f_n,\dd\f_n)
=(\f_n,\dd D \chi_n)+i\l_n (\chi_n, \dd\chi_n) -i\l_n (\f_n,\dd\f_n)
\ee
so that
\be\label{pert2}
\dd\l_n=-i(\f_n,\dd D\chi_n) + \l_n \Big( (\chi_n,\dd\chi_n)-(\f_n,\dd\f_n)\Big) \ .
\ee
In ``ordinary" first order perturbation theory in quantum mechanics one has a relation analogous to $(\chi_n,\dd\chi_n)=(\f_n,\dd\f_n)=0$ that translates that both $\chi_n$ and $\chi_n+\dd\chi_n$ are normalised to one, and similarly for the $\f_n$. Here, however,  one must take into account that the $\chi_n$ and $\f_n$ are normalised with the metric $g$, while $\chi_n+ \dd\chi_n$ and $\f_n+\dd\f_n$  are normalised with the metric $e^{2\dd\s} g$. Let us write out what this  implies (to first order in the perturbation)~:
\be\label{pertnormdetailed}
1=\int \d^2 x \,  \sqrt{g}\, e^{2\dd\s}\, (\chi_n+\dd\chi_n)^\dag (\chi_n+\dd\chi_n)
=\int \d^2 x \,  \sqrt{g}\,  \big( \chi_n^\dag \chi_n + 2\dd\s \chi_n^\dag \chi_n +\dd\chi_n^\dag \chi_n + \chi_n^\dag \dd\chi_n\big)
\ee
The first term $\int \d^2 x \,  \sqrt{g}\, \chi_n^\dag \chi_n$ is just $(\chi_n,\chi_n)=1$, while $\dd\chi_n^\dag \chi_n=\dd\chi_n^T\chi_n=\chi_n^T\dd\chi_n=\chi_n^\dag \dd\chi_n$ and we get
\be\label{pertnorm}
(\chi_n,\dd\chi_n)=-\int\d^2 x \sqrt{g}\, \dd\s(x) \chi_n^\dag(x) \chi_n(x) 
\equiv -\int \dd\s \, \chi_n^\dag\chi_n\ ,
\ee
where  $\int \ldots$ is short-hand for $\int \d^2 x \sqrt{g} \ldots $.
Similarly, $(\f_n,\dd\f_n)=-\int\dd\s \f_n^\dag \f_n$. Replacing  $\dd D=i \dd\Nsl$,  we can then rewrite \eqref{pert2} as
\be\label{pert3}
\dd\l_n=(\f_n,\dd \Nsl\,\chi_n) + \l_n \int \dd\s\big( \f_n^\dag \f_n-\chi_n^\dag \chi_n  \big) \ .
\ee
Next, we insert $\dd\Nsl=-\dd\s \Nsl + \frac{1}{2}\dsl\dd\s$, cf.~\eqref{deltaD2dim},  and integrate by parts (cf \eqref{covderintbyparts}), so that
\ba\label{perrt4}
(\f_n,\dd \Nsl\,\chi_n)&=&-\int\dd\s \f_n^\dag \Nsl\,\chi_n +\frac{1}{2}\int  \del_\l\dd\s \f_n^\dag \g^\l \chi_n
=-\int\dd\s\Big( \f_n^\dag \Nsl\,\chi_n +\frac{1}{2} (\Nsl\,\f_n)^\dag \chi_n +\frac{1}{2}\f_n\Nsl\,\chi_n\Big)\nonumber\\
&=&-\int\dd\s \Big( -i\f_n^\dag (D-m\g_*)\chi_n +\frac{i}{2} \big((D-m\g_*)\f_n\big)^\dag\chi_n -\frac{i}{2} \f_n^\dag (D-m\g_*)\chi_n\Big)\nonumber\\
&=&-\int\dd\s\Big(\frac{3}{2}\l_n\f_n^\dag\f_n -\frac{1}{2}\l_n\chi_n^\dag\chi_n +im\f_n^\dag\g_*\chi_n\Big) \ ,
\ea
and then
\be\label{pert5}
\dd\l_n=-\int \dd\s\Big( \frac{\l_n}{2} (\f_n^\dag\f_n+\chi_n^\dag\chi_n) + i m \f_n^\dag \g_* \chi_n\Big)\ .
\ee
Since $\f_n$, $\chi_n$ and $i\g_*$ are real, we have $\f_n^\dag i\g_* \chi_n=(\f_n^\dag i\g_* \chi_n)^\dag=\chi_n^\dag (-i) \g_* \f_n$, and we can also rewrite this as
\be\label{pert5-2}
\dd\l_n=-\frac{1}{2}\int \dd\s\Big( \l_n (\f_n^\dag\f_n+\chi_n^\dag\chi_n) + i m (\f_n^\dag \g_* \chi_n-\chi_n^\dag \g_* \f_n)\Big)\ .
\ee
This allows us to express the variation of $\zeta(s)$ as
\be\label{zetavar}
\dd\zeta(s)=2\sum_n \dd \big(\l_n^{-2s}\big)= -4s\sum_n \frac{\dd\l_n}{\l_n^{2s+1}}
=2s\hskip-1.mm \int\dd\s\sum_n \Big( \frac{1}{\l_n^{2s}}  (\f_n^\dag\f_n+\chi_n^\dag\chi_n) + im \frac{1}{\l_n^{2s+1}} (\f_n^\dag \g_*\chi_n-\chi_n^\dag \g_*\f_n) \Big)  .
\ee

Let us make a remark about zero-modes if they are present. By zero-modes we here mean solutions $\p_{0,i}$ of $i\Nsl \, \p_{0,i}=0$, or in terms of the real and imaginary parts $i\Nsl\, \f_{0,i}=i\Nsl\,  \chi_{0,i}=0$. Then $i\l_0\f_{0,i}=D\chi_{0,i}=m\g_*\chi_{0,i}$ so that necessarily $\l_0=m$ and $\g_*\chi_{0,i}=i\f_{0,i}$. Similarly $\g_*\f_{0,i}=-i\chi_{0,i}$. We have seen that under a conformal transformation the zero-modes change in a rather simple way so as to remain zero-modes of the conformally transformed  $i\Nsl$. In particular then the new eigenvalue is still $\l_0=m$, so that
\be\label{deltalambda0}
\dd\l_0=0 \ .
\ee 
Let us check this from \eqref{pert5}. One has $\f_{0,i}^\dag \f_{0,i}= -i\f_{0,i}^\dag\g_* \chi_{0,i}$, and $\chi_{0,i}^\dag \chi_{0,i}=i\chi_{0,i}^\dag \g_* \f_{0,i}=\big(i\chi_{0,i}^\dag \g_* \f_{0,i}\big)^\dag$ $=-i\f_{0,i}^\dag\g_* \chi_{0,i}$. It follows that the bracket on the right-hand side of \eqref{pert5} vanishes for $n=0$, confirming $\dd\l_0=0$. It follows that in \eqref{zetavar} we can just as well exclude the contributions of the zero-modes and replace $\sum_n\to \sum_{n\ne 0}$.

\newpage
\setlength{\baselineskip}{.58cm}
\section{The tool box : heat kernels, zeta-function and Green's functions}


In this section we set up and discuss the technical tools we need to obtain the gravitational action. This includes in particular local heat kernels, local zeta functions and Green's functions. As we will see, they are all related through various relations. An important detailed computation about one of the two the heat kernels we introduce, namely $K_+$, is deferred to appendix A. Some partial results about the other heat kernel ($K_-$) are presented in appendix B. General references on the heat kernel are \cite{Gilkey, Avramidi, Vassil}.


\subsection{The basic tools : definitions, relations and properties\label{defandbasicheat}}

Throughout this section we assume the mass $m$ is non-vanishing, so that $D=i\Nsl+m\g_*$ has no zero-modes. As discussed above, $i\Nsl$ has zero-modes for manifolds of genus ${\tt g}\ge 1$, but the non-vanishing mass implies that $\l_n\ge m>0$. The absence of zero-modes of $D$ is an important assumption. It means in particular that all zeta-functions $\zeta$, Green's functions $G$, determinants $\Det$, etc can be straightforwardly defined in terms of sums or products of {\it all} eigenvalues / eigenfunctions. On the other hand, in the zero-mass limit the zero-modes of $i\Nsl$ become zero-modes of $D$. One can then still define corresponding quantities $\wt\zeta$, $\wt G$, $\Det'$ etc by excluding the zero-modes from the sum / product.
However, these modified quantities  satisfy different relations from those satisfied by $\zeta$, $G$ and $\Det$. For example, the Green's function $G(x,y)$ of $D^2$  satisfies $D^2_x G(x,y)=\dd^{(2)}(x-y) {\bf 1}_{2\times 2}$ while in the presence of  zero-modes, $\wt G(x,y)$ satisfies a similar relation, $D^2_x \wt G(x,y)=\dd^{(2)}(x-y) {\bf 1}_{2\times 2}-P_0(x,y)$  where on the right-hand side one must subtract  $P_0(x,y)$, the projector on the zero-modes, which was discussed in subsection \ref{zeromodesec}. Since we are interested in the effective gravitational action for massive matter, one might think that this complication just does not occur. However, for massive bosonic (scalar) matter a fruitful method was to do a small mass expansion around  zero-mass, and the same will be true here. One must then express everything in terms of quantities that have a smooth limit ass $m\to 0$, and these are the $\wt G$, $\wt\zeta$, etc.  On the torus we could compute the zero-mode projector exactly, and on higher genus Riemann surfaces we could still sufficiently characterise its structure. This will  allow us in the next section to obtain the variation of the gravitational action in an expansion in $m^2$ to all orders, and on arbitrary genus Riemann surfaces. This is quite some progress with respect to \cite{BDE}~!

Having said this, we emphasize again that throughout this section we assume that $D$ has no zero-modes (i.e.~that either $m\ne 0$ or that we deal with spherical topology). Recall that we had defined the complex eigenfunctions $\p_n$ and $\p_n^*$ (and hence also the real eigenfunctions $\chi_n$ and $\f_n$) such that $\l_n > 0$  and
\be\label{eigeneqrecall}
D\p_n=\l_n\p_n \quad , \quad D\p_n^*=-\l_n \p_n^* \quad , \quad
D\chi_n=i\l_n\f_n \quad,\quad D\f_n=-i\l_n\chi_n \ .
\ee
In terms of these eigenfunctions
the completeness relation  reads
\be\label{completeness}
\sum_n \big(\chi_n(x)\chi_n^\dag(y)+\f_n(x)\f_n^\dag(y)\big) = \sum_n \big(\p_n(x)\p_n^\dag(y)+\p^*_n(x)(\p_n^*(y))^\dag\big)  = \frac{\dd^{(2)}(x-y)}{\sqrt{g}}\ {\bf 1}_{2\times 2}\ .
\ee

\subsubsection{Local zeta-functions}

We define two local zeta-functions $\zeta_+(s,x,y)$ and $\zeta_-(s,x,y)$ as
\ba\label{Zetadefs}
\zeta_+(s,x,y)&=&\sum_n \l_n^{-2s} \big(\chi_n(x)\chi_n^\dag(y)+\f_n(x)\f_n^\dag(y)\big) \ ,\nonumber \\
\zeta_-(s,x,y)&=&\sum_n \l_n^{-2s} \big(\chi_n(x)\f_n^\dag(y)-\f_n(x)\chi_n^\dag(y)\big) \ ,
\ea
where $x=(x^1,x^2)$ and $y=(y^1,y^2)$ denote points on the manifold.  Note that these local zeta-functions are real $2\times 2$-matrices.
They can also be rewritten in terms of the $\p_n$ and $\p_n^*$ as
\ba\label{Zetadefs2}
\zeta_+(s,x,y)&=&\sum_n \l_n^{-2s} \big(\p_n(x)\p_n^\dag(y)+\p^*_n(x)(\p_n^*)^\dag(y)\big) \ ,\nonumber \\
\zeta_-(s,x,y)&=&i\ \sum_n \l_n^{-2s} \big(\p_n(x)\p_n^\dag(y)-\p^*_n(x)(\p_n^*)^\dag(y)\big)  \ .
\ea
As we will see below, $\zeta_+(1,x,y)$ is the Green's function of $D^2$, while $-i\zeta_-(\frac{1}{2},x,y)$ is the Green's function of $D$. It is for this reason that we need to introduce the ``strange-looking" $\zeta_-(s,x,y)$.

As mentioned above, we will also encounter the local  zeta-functions $\wt\zeta_\pm(s,x,y)$ with the zero-modes of $i\Nsl\,$ (if present) excluded from the sums. The contributions of these zero-modes is precisely given in terms of the zero-mode projectors $P_0$ defined in \eqref{zeromodeproj} and $Q_0$ defined in \eqref{zeromodeQ}. Hence we obviously have
\be\label{zetatilde-zeta}
\wt\zeta_+(s,x,y)=\zeta_+(s,x,y) - \frac{P_0(x,y)}{m^{2s}}
\quad , \quad
\wt\zeta_-(s,x,y)=\zeta_+(s,x,y) - i\,\frac{Q_0(x,y) }{m^{2s}}\ .
\ee

There is an important relation between $\zeta_+$ and $\zeta_-$ which generalises the relation between the corresponding Green's functions. 
It follows immediately from \eqref{eigeneqrecall} and the definitions \eqref{Zetadefs} that
\ba\label{zetapmrel}
D_x\, \zeta_+(s,x,y)=-i \,\zeta_-(s-\frac{1}{2},x,y) \quad &,& \quad
D_x\, \zeta_-(s,x,y)=i \,\zeta_+(s-\frac{1}{2},x,y) \ ,\nonumber\\
\nonumber\\
D_x^2\, \zeta_\pm(s,x,y)&=&\zeta_\pm(s-1,x,y) \ ,
\ea
where the subscript $x$ on $D$ indicates that the derivatives are with respect to $x$. Since $D_x P_0(x,y)=m\g_* P_0(x,y) = m Q_0(x,y)$ and $D_x Q_0(x,y)=m\g_* Q_0(x,y) = m P_0(x,y)$, these relations \eqref{zetapmrel} immediately also carry over to the $\wt\zeta_\pm(s,x,y)$.

Denoting the  trace over the $2\times 2$ matrices by $\tr$, and using the orthonormality of the $\f_n$ and $\chi_n$ we have \be\label{Zetaint}
\int \d^2 x \sqrt{g}\, \tr\, \zeta_+(s,x,x) =
2 \sum_n \l_n^{-2s} \equiv \zeta(s) \quad , \quad
\int \d^2 x \sqrt{g}\, \tr \zeta_-(s,x,x)=0 \ ,
\ee
where $\zeta(s)$ is  the zeta function of $D^2$.

The following (anti-) hermiticity relations follow directly from the definitions \eqref{Zetadefs}~:
\ba\label{zetaherm}
\big( \zeta_+(s,x,y)\big)^\dag = \zeta_+(s,y,x) \quad &\Rightarrow&\quad \text{$\zeta_+(s,x,x)$ is real and hermitian}\ , \nonumber\\
\big( \zeta_-(s,x,y)\big)^\dag = -\zeta_-(s,y,x) \quad &\Rightarrow&\quad \text{$\zeta_-(s,x,x)$ is real and anti-hermitian} \ .
\ea
There are 3 possible hermitian and real matrices, namely $\s_x,\ \s_z$ and ${\bf 1}_{2\times 2}$, while there is only a single real and anti-hermitian matrix, namely $i\g_*$. Hence, we can already assert that
\be\label{zetamoinsxxmatrix}
\zeta_-(s,x,x)\sim i\g_* \  .
\ee
On the other hand, for $\zeta_+(s,x,x)$ the present argument would allow 3 possible matrix structures, while later-on we will find that $\zeta_+(s,x,x)\sim{\bf 1}_{2\times 2}$ \  .

The convergence properties of these local zeta-functions depend essentially on the large-$n$ behaviour of the eigenvalues $\l_n^2$. Indeed,  equation \eqref{Zetaint} more or less suggests that multiplying the $\l_n^{-2s}$ by $\chi_n^\dag(x) \chi_n(y)$ and $\f_n^\dag(x)\f_n(y)$ does not change the convergence properties.\footnote{
This argument is certainly true for the {\it convergence} properties of the sum. For large eigenvalues, we expect the corresponding $\p_n(x)\p_n^\dag(y)$ to be rapidly oscillating functions of $x$ and $y$, leading to cancellations between adjacent terms in the sum, and thus to a possible improvement of the convergence of the sum. But this rapidly oscillating behaviour and the corresponding cancellations also mean that one has to examine more carefully the usual argument  made for the heat kernel, see below, that for small $t$ the sum is dominated by the large eigenvalues.} 
Of course,  discussing the convergence of a sum of functions (or even functions of 2 arguments) needs somewhat more care and requires specifying an appropriate norm in which the convergence is to be appreciated, but without going into these details it is clear that we get a good deal of information about the convergence of the local zeta-functions by studying the behaviour of the large eigenvalues, which amounts to studying the convergence of the (integrated) zeta-function $\zeta(s)$.

The behaviour of the large eigenvalues of $D^2$ is dictated by the leading 2-derivative term in $D^2$ which, by \eqref{spinorLapl2}, is the same as the one of the scalar Laplacian, and which is the same as in flat space (with periodic boundary conditions). This amounts to saying that the large eigenvalue behaviour (``UV-behaviour") reflects the short-distance properties of the manifold, and at short distances all manifolds are locally flat. Thus the leading large $n$ behaviour of the $\l_n$ must be the same as in the case of the flat torus where the eigenvalues depend on two integers $n_1$ and $n_2$. 
We have seen before that for general periods of the torus the eigenvalues are $\l_{\vec n}^2=n_1^2+\big(\frac{n_2+\tau_1 n_1}{\tau_2}\big)^2+m^2$.
Thus, for large eigenvalues we will always have $\l_n^2\equiv \l^2_{\vec{n}}\simeq a^2 \big[ n_1^2+\big(\frac{n_2+\tau_1 n_1}{\tau_2}\big)^2\big]$ for some $a$ which sets the scale. This is essentially the statement of Weyl's law \cite{Weylslaw} about the large eigenvalues of the Laplace operator.\footnote{
We see that the behaviour of the large eigenvalues does not depend much on the details of the geometry. It depends to some extend on the global structure like boundary conditions.  And it does, of course, crucially  depend on the dimensionality of the manifold.
}
Then for large $|n_1|$ and large $|n_2|$ we approximately have
\be
\sum_{(n_1,n_2)\ne (0,0)}\l_n^{-2s} 
\simeq a^{-2s} \sum_{(n_1,n_2)\ne (0,0)} \big[ n_1^2+\big(\frac{n_2+\tau_1 n_1}{\tau_2}\big)^2\big]^{-s}
\simeq a^{-2s}  \int_{\vec{k}\notin {\cal D}} \d^2 k \,\big[k_1^2+\big(\frac{k_2+\tau_1 k_1}{\tau_2}\big)^2\big]^{-s} \ ,
\ee
where ${\cal D}$ is some domain ``of unit radius" around the origin. Assuming convergence, we change variables to $p_1=k_1, \ p_2=\frac{k_2+\tau_1 k_1}{\tau_2}$ and go to polar coordinates so that
\be
\sum_{n_1,n_2}\l_n^{-2s}\simeq  2\pi a^{-2s} \tau_2 \int_1^\infty \d p\, p\, (p^2)^{-s}=\pi a^{-2s}   \tau_2\int_1^\infty \d \xi\, \xi^{-s}
\simeq \pi a^{-2s} \tau_2 \sum_{n=1}^\infty n^{-s}=\pi a^{-2s}  \tau_2\, \zeta_{\rm R}(s)\ .
\ee
We see that the convergence properties of the zeta-functions of Laplace-like operators in two dimensions are exactly the same as for Riemann's zeta-function !
It follows that, just as for the sum defining Riemann's zeta function $\zeta_{\rm R}(s)$, the zeta function $\zeta(s)\equiv \zeta_{D^2}(s)$ is defined by a convergent sum  for $\Re s>1$. As  explained before, we then also expect that the sums defining the local zeta-functions $\zeta_\pm(s,x,y)$  are convergent expressions for $\Re s>1$ and are otherwise defined by analytical continuation. Below, we will indeed confirm this expectation and show that, for $x\ne y, $ $\zeta_\pm(s,x,y)$  can be defined as  analytic functions for all $s\in {\bf C}$, and that $\zeta_+(s,x,x)$ has a single pole at $s=1$ while $\zeta_-(s,x,x)$ has a single pole at $s=\frac{1}{2}$.

\subsubsection{Local heat kernels}

We similarly define local heat kernels that are again $2\times 2$-matrices :
\ba\label{Kdefs}
K_+(t,x,y)&=&\sum_n e^{-\l_n^2 t} \big(\chi_n(x)\chi_n^\dag(y)+\f_n(x)\f_n^\dag(y)\big)
= \sum_n e^{-\l_n^2 t} \big(\p_n(x)\p_n^\dag(y)+\p^*_n(x)(\p_n^*)^\dag(y)\big)  \ , \nonumber\\
K_-(t,x,y)&=&\sum_n e^{-\l_n^2 t}  \big(\chi_n(x)\f_n^\dag(y)-\f_n(x)\chi_n^\dag(y)\big) 
= i\ \sum_n  e^{-\l_n^2 t}  \big(\p_n(x)\p_n^\dag(y)-\p^*_n(x)(\p_n^*)^\dag(y)\big)\ .  \nonumber\\
\ea
Again, if necessary, one can also define the corresponding local heat kernels with the zero-modes of $i\Nsl\,$ excluded as
\be\label{K-Ktilde}
\wt K_+(t,x,y)=K_+(t,x,y)-e^{-m^2 t}P_0(x,y)
\quad , \quad 
\wt K_-(t,x,y)=K_-(t,x,y)-ie^{-m^2 t}Q_0(x,y) \ .
\ee
One could also define
\be\label{KKdef}
{\cal K}(t,x,y)=\frac{1}{2} K_+(t,x,y) - \frac{i}{2} K_- (t,x,y) = \sum_n e^{-\l_n^2 t} \p_n(x)\p_n^\dag(y) \ .
\ee
Then the real functions $K_\pm$ constitute (2 times) the real and ($-2$ times) the imaginary parts of $2{\cal K}$.

It follows from \eqref{Kdefs} that the large-$t$ behaviour of the heat kernels is controlled by the smallest eigenvalue. Since we assumed that there is no zero-mode, we have $\l_n^2 >0$ and for $t\to\infty$ all our heat kernels vanish (at least) as $e^{-\l_{\rm min}^2 t}$, where $\l_{\rm min}$ denotes the smallest eigenvalue~:
\be\label{heatatinfty}
K_\pm(t,x,y) \sim_{t\to\infty} {\cal O}(e^{-\l_{\rm min}^2 t}) \ .
\ee

From the definitions \eqref{Kdefs}  one sees that  $K_+$ and $K_-$ (and hence also ${\cal K}$) satisfy the ``generalised heat equation"
\be\label{heateq}
\left(\frac{\d}{\d t} + D_x^2\right) K_{\pm}(t,x,y) = 0 \ ,
\ee
where the subscript $x$ on $D^2$ indicates that the derivatives are with respect to $x$. 
In appendix A we explain how one can obtain the asymptotic small-$t$ expansion of $K_+(t,x,y)$ from this differential equation and an appropriate ``initial" condition for $K_+$. Indeed, one sees from the definition \eqref{Kdefs} and the closure relation \eqref{completeness} that for $t\to 0$ we have 
\be\label{Kplusinitial}
K_+(t,x,y)\sim \frac{ \dd^{(2)}(x-y)}{\sqrt{g}}\, {\bf 1}_{2\times 2} \ ,
\ee 
which is the ``usual" initial condition for a heat kernel, as also assumed in the appendix. However, no such simple initial condition holds for $K_-$. This is why the method of  appendix A yields $K_+$ and not $K_-$.
In particular, in the appendix we work out the relevant first few terms of the small-$t$ expansion of $K_+$ which we will need below.

Another way to see the differences between $K_+$ and $K_-$ is  to note that $K_+$ contains the full sum of all eigenfunctions of $D$, namely the $\p_n$ and the $\p_n^*$. Thus if one uses a different basis for these eigenfunctions one still gets the same $K_+$. Moreover, the $\p_n$ and the $\p_n^*$ appear symmetrically. We can write $K_+(t,x,y)=\bra{x} \sum_{\l_n^2} e^{-\l_n^2 t} P(\l_n^2) \ket{y}$ where $P(\l_n^2)$ is the projector on the eigenspace of $D^2$ with eigenvalue $\l_n^2$. This makes clear that one could use any basis of eigenfunctions of $D^2$. This also explains why one is able (in principle) to obtain $K_+$  uniquely by solving the heat equation \eqref{heateq} for the operator $D^2$ with the appropriately prescribed initial condition. However, this is not true for the imaginary part $K_-$ of ${\cal K}$. We see from \eqref{Kdefs} that the definition of $K_-$ is not simply a sum over all eigenfunctions of $D^2$, but that we made a certain distinction between the eigenfunctions of $D$ with eigenvalues $\l_n>0$ and those with eigenvalues $-\l_n<0$. (Recall that we always take $\l_n>0$.) Clearly, the operator $D^2$ does not make this distinction, and hence, one cannot simply get the $K_-$ by solving \eqref{heateq}. 

What can be said about the matrix structure of the heat kernel $K_+(t,x,y)$ ? As just explained, $K_+$ can be obtained by solving the heat equation \eqref{heateq} with initial condition \eqref{Kplusinitial}. Now
it is clear from \eqref{Dsquared} and \eqref{spinorLapl2}, that the matrix structure of $D^2$ is ${\bf 1} (\ldots) + \g_* (\ldots)$. It then follows that  $K_+(t,x,y)$ must have the same  matrix structure:
\be\label{matrixstrK}
K_+(t,x,y)= K_+^0(t,x,y) \, {\bf 1}+ K_+^*(t,x,y) \, \g_*\ .
\ee

What can be said about $K_-$~? Define an auxiliary quantity $L(t,x,y)$ as 
\be\label{Ldef}
L(t,x,y)=i \sum_n \frac{e^{-\l_n^2 t} }{\l_n} \big(\chi_n(x)\chi_n^\dag(y)+\f_n(x)\f_n^\dag(y)\big) 
=  i \sum_n \frac{e^{-\l^2_n t} }{\l_n}   \big(\p_n(x)\p_n^\dag(y)+\p^*_n(x)(\p_n^*)^\dag(y)\big) \ ,
\ee
which is now constructed from the eigenfunctions and eigenvalues of $D^2$ only, without any distinction between positive and negative  eigenvalues. We then have
\be\label{K-Lrel}
K_-(t,x,y)=D_x L(t,x,y) \ ,
\ee
Obviously, $L(t,x,y)$ also satisfies the heat equation \eqref{heateq}. Its initial condition is
\be\label{Linniital}
L(t,x,y)\sim_{t\to 0} \ 
i \sum_n \frac{1}{\l_n} \big(\chi_n(x)\chi_n^\dag(y)+\f_n(x)\f_n^\dag(y)\big) \ .
\ee
However, contrary to the completeness relation, we do not recognise here any known function (or distribution). This prevents us from obtaining $L$ in analogy to what is done for $K_+$ in appendix A.

Let us nevertheless see what we can say about $K_-$ and $K_+$ on general grounds. 
In analogy with the discussion for the local heat kernels, it follows directly from the definitions \eqref{Kdefs} of $K_+$ and $K_-$ in terms of the eigenfunctions $\p_n$ and $\p_n^*$ that
\be\label{Khermiticity}
(K_+(t,x,y))^\dag=K_+(t,y,x) \quad , \quad (K_-(t,x,y))^\dag=-K_-(t,y,x) \ ,
\ee
so that for $x=y$ we simply have
\be\label{Kequalhermiticity}
(K_+(t,x,x))^\dag=K_+(t,x,x) \quad , \quad (K_-(t,x,x))^\dag=-K_-(t,x,x) \ .
\ee
This means that $K_+(t,x,x)$ is hermitian and $K_-(t,x,x)$ is anti-hermitian. Moreover, they are both  real. The only real and anti-hermitian matrix is $i\g_*$ and it follows that~:
\beb\label{Kxxmatrix}
K_+(t,x,x)\sim {\bf 1}_{2\times 2}
\quad , \quad
K_-(t,x,x) \sim \g_* \ .
\eeb


\subsubsection{Relating zeta functions and heat kernels}

There is a well-known relation between zeta-functions and heat kernels which generalises the relation \eqref{zetaintegral3} for Riemann's zeta function. This relation also generalises to the local zeta-functions and local heat kernels we have defined~:
\beb\label{zetaK}
\zeta_\pm(s,x,y)= \frac{1}{\G(s)} \int_0^\infty \d t\, t^{s-1}\, K_\pm (t,x,y) \ .
\eeb
Indeed, these relations \eqref{zetaK} follow trivially (for $\Re s>1$) from inserting the definitions of $K_\pm$ as a sum over the eigenvalues and eigenfunction, interchanging sum and integration and doing the integrations term per term as $(\G(s))^{-1}\int_0^\infty  \d t\, t^{s-1}\,  e^{-t \l^2_n} = \l_n^{-2s} (\G(s))^{-1}\int_0^\infty  \d t'\, {t'}^{s-1}\,  e^{-t'}= \l_n^{-2s}$. Then, the relation \eqref{zetaK} can be obtained on all of ${\bf C}$ as an identity between meromorphic functions  by analytic continuation. 

The relation \eqref{zetaK} is a special case of the Mellin transform. The Mellin transform $\cM (f)$ of a (continuous) function $f$ is
\be\label{Mellin}
\vf(s)\equiv(\cM(f))(s) = \int_0^\infty \d t\, f(t)\, t^{s-1} \ .
\ee
If this integral is absolutely convergent in a strip $a< \Re s <b$, then $\vf(s)$ is analytic in this strip, and $f$ can be recovered by the inverse Mellin transform defined as
\be\label{inverseMellin}
f(t)=(\cM^{-1}(\vf))(t) = \frac{1}{2\pi i} \int_{c-i\infty}^{c+i\infty} \d s\, \vf(s)\, t^{-s}  \ , \quad a<c<b\ .
\ee

We can then interpret equation \eqref{zetaK} by saying that the $\G(s)\, \zeta_\pm(s,x,y)$ are the Mellin transforms of the heat kernels $K_\pm(t,x,y)$. The inverse Mellin transform then allows us to get the heat kernels back from the zeta functions as
\be\label{heatfromzeta}
K_\pm(t,x,y) = \frac{1}{2\pi i}  \int_{c-i\infty}^{c+i\infty} \d s\, t^{-s}\, \G(s) \, \zeta_\pm(s,x,y) \ ,
\ee
where the real $c$ must be chosen such that $\G(c+i\s)\, \zeta_\pm(c+i\s,x,y)$ is analytic for all real $\s$ and tends uniformly to zero  as $\s\to \pm\infty$. One can show that any $c>1$ is a convenient choice. This is discussed in more detail below.

The relation \eqref{zetaK} translates \eqref{Kxxmatrix} into the corresponding relations for the zeta-functions at coinciding points, confirming the structure of $\zeta_-(s,x,x)$ already obtained earlier, and providing the corresponding result for $\zeta_+(s,x,x)$~:
\beb\label{zetaxxmatrix}
\zeta_+(s,x,x)\sim {\bf 1}_{2\times 2}
\quad , \quad
\zeta_-(s,x,x) \sim \g_* \ .
\eeb
As already discussed, one can get quite some information  about $K_+$ from the differential equation it satisfies, which is not the case of $K_-$.  However, much of the information about $K_+$ translates via \eqref{zetaK}  into corresponding information about $\zeta_+$. We can then use the relations 
\eqref{zetapmrel} to obtain $\zeta_-$ from $\zeta_+$ and obtain the corresponding information about $\zeta_-$. Finally one can use the inverse Mellin transformation to deduce statements about $K_-$.

To begin with note that, since $\zeta_+$  can be obtained from $K_+$ by \eqref{zetaK}, its matrix structure must be the same (cf \eqref{matrixstrK})~:
\be\label{zetaxymatrixstr}
\zeta_+(s,x,y)= \zeta_+^0(s,x,y) {\bf 1}_{2\times 2} + \zeta_+^*(s,x,y) \g_* \ .
\ee
Next,  recall from  \eqref{matrixstrDstructure} that
$i\Nsl=i\s_x \cD_1+i \s_z \cD_2$, where $\cD_1$ and $\cD_2$ are some differential operators not involving any matrix. It follows that $i\Nsl\,\g_*=-i\g_*\,\Nsl=-\s_z \cD_1+\s_x\cD_2$ and
\be\label{matrixtraces}
\tr i\Nsl=\tr \g_* i\Nsl=\tr i\Nsl\, \g_*=0 \ .
\ee
Recall the first relation \eqref{zetapmrel} which we rewrite as  $\g_*\zeta_-(s,x,y)=i\g_*(i\Nsl_x+m\g_*) \zeta_+(s+\frac{1}{2},x,y)$. Then taking the Dirac trace of this relation,  only the part $\sim m$ survives~:
\be\label{zeta-g*trace}
\tr \g_* \zeta_-(s,x,y)= i m \tr \zeta_+(s+\frac{1}{2},x,y)\ .
\ee
As we will see in the next subsection, the variation of the gravitational action involves only\break $\tr \zeta_+(s,x,x)$ and $\tr \g_* \zeta_-(s,x,x)$, and the previous relation shows that the knowledge of\break $\tr \zeta_+(s,x,x)$ should be all we need to know. But we can also easily establish this same relation without the traces~:  we have seen in \eqref{zetaxxmatrix}  that $ \zeta_-(s,x,x)\sim\g_*$ and $ \zeta_+(s,x,x)\sim {\bf 1}_{2\times 2}$, so that indeed
\beb\label{zetaxx-g*}
\zeta_-(s,x,x)= i m \g_* \ \zeta_+(s+\frac{1}{2},x,x)\  .
\eeb
We will soon show that $\zeta_+(s,x,x)$ is a meromorphic function with a single pole at $s=1$. The previous relation should be understood as an identity between meromorphic functions and, in particular, $\zeta_-(s,x,x)$ then has a single pole at $s=\frac{1}{2}$.
Let us insist that one should not conclude from this relation that $\zeta_-(s,x,x)$ vanishes for $m=0$, since our whole discussion assumed that there are no zero-modes of the Dirac operator. It is only for spherical topology where this conclusion would be correct, while for genus larger or equal to one, we expect $\zeta_+(s,x,x)$ to contain a contribution $\sim \frac{1}{m^s}$, so there is a possible zero-mode contribution  to $\zeta_-(s,x,x)$ even as $m\to 0$.

\subsubsection{Small-$t$ asymptotics  and poles of zeta functions at coinciding points}

Let us now study the possible singularities of the integrals \eqref{zetaK} in order to establish the analyticity properties of $\zeta_\pm (s,x,y)$.
Since $K_\pm$ vanish exponentially for large $t$, cf \eqref{heatatinfty}, the integral always converges at $t\to\infty$.  Hence any divergences of the integral must come from the region $t\to 0$ where the $K_\pm$ are singular. Thus any singularities (poles) of the zeta-functions are related to the small-$t$ behaviour of the heat kernel. Furthermore, since $\G(s)$ has poles at  $s=0,-1,-2,\ldots$, one sees that $\frac{1}{\G(s)}$ vanishes at $s=0,-1,-2,\ldots$, resulting in finite values of the zeta-functions at $s=0,-1,-2,\ldots$. In particular, these finite values are also determined by the divergences of the integral due to the small-$t$ behaviour of $K$.  We will explain this in further detail below. There is quite some abundant literature on the small-$t$ asymptotics of heat kernels associated to various differential operators, in particular at coinciding points $x=y$. 

The general expectation is that, due to the presence of $e^{-\l^2_n t}$, the small-$t$ behaviour of the heat kernel is related to the behaviour of the large eigenvalues of order $\l^2_n\sim \frac{1}{t}$, and as  discussed above, this is related to the leading-derivative terms of the differential operator $D^2$ and the small-scale structure of the manifold. However, while this expectation turns out to be correct for our heat kernels $K_\pm(t,x,x)$ at coinciding points, as well as for $K_+(t,x,y)$ at $x\ne y$, it will {\it not} be true for $K_-(t,x,y)$ at $x\ne y$. Indeed, for $x\ne y$, the local heat kernels not only involve the $e^{-\l^2_n t}$ but also the $\p_n(x)\p_n^\dag(y)$ and the latter typically oscillate rapidly for large $n$ (large $\l^2_n$). This can potentially lead to important cancellations and correspondingly the sum no longer is dominated by the eigenvalues of order $\l^2_n\sim \frac{1}{t}$ but by much smaller eigenvalues already. Obviously, this will change the small-$t$ aymptotics of $K_-(t,x,y)$. We will see this very explicitly in appendix B for the $K_-(t,x,y)$ on the flat torus.

As explained above, our heat kernel $K_+$ can, in principle, be entirely determined by solving the heat equation. In particular,  its small-$t$ behaviour can be determined in an asymptotic expansion from the differential equation \eqref{heateq}, with the leading behaviour being the same as in flat space, and the subleading terms being given in terms of the local curvature and derivatives of the curvature. This is studied in some detail in appendix A. Of particular interest will be the expansion at coinciding points $x=y$. It is shown in  appendix A eq~\eqref{F01coinc2d} that\footnote{
In the appendix, we have determined the first few coefficients of the heat kernel $K(t,x,y)$ as follow from the differential equation. As argued above, this actually only determines $K_+(t,x,y)$, so eq~\eqref{F01coinc2d} translates into an equation for $K_+(t,y,y)$.
Note   that, solely from the differential equation, we could also have expected a contributions $\sim\g_*$. Indeed, such contributions do appear  in the intermediate steps of the computation, in particular for $K_+(t,x,y)$ with $x\ne y$, but they drop out when setting $x=y$ to obtain $K_+(t,x,x)$, in agreement with the general result \eqref{Kxxmatrix}.
}, 
\be\label{K+shortdist}
K_+(t,x,x)\sim_{t\to 0}\frac{1}{4\pi t}\Big( 1 -\big(\frac{\cR(x)}{12}+m^2\big) \, t + {\cal O}(t^2) \Big)\, {\bf 1}_{2\times 2} \ .
\ee
A heuristic way to understand this result  is to note that the heat kernel, at coinciding points, for the two-dimensional scalar Laplacian $\D_{\rm scalar}$ is well-known (cf \eqref{E20}) to be $K_{\D_{\rm scalar}}(t,x,x)$ $=\frac{1}{4\pi t}\big( 1 + \frac{\cR}{6}t+ \ldots\big)$. Now $K_+$ is the heat kernel for $D^2$, which involves the spinorial Laplacian $\D_{\rm sp}$ which, of course, differs from $\D_{\rm scalar}$ but has the same leading derivative terms. One might then expect that it also gives the same  result to this order\footnote{
This assumption  is not entirely true as one sees from the detailed computations in the appendix. Indeed, the subleading single derivative term gives important contributions to the first subleading term in the small-$t$ expansion. Since this first-derivative terms differs between $\D_{\rm scalar}$ and $\D_{\rm sp}$ by $-\frac{i}{2}\g_* \o^\m\del_\m$ one might have expected a different contribution to the first subleading terms in the small-$t$ expansion of $K_+$. However, at this order, these additional contributions cancel.
}. 
Moreover, one expects that the additional $\frac{\cR}{4}+m^2$ in $D^2$ simply gives an extra $e^{-(\cR/4+m^2)t}\simeq 1 -(\frac{\cR}{4} + m^2)t$, to this order, resulting in $\frac{1}{4\pi t}\big( 1 + \frac{\cR}{6}t+ \ldots\big) \times \big( 1 -\frac{\cR}{4} t- m^2 t+\ldots\big) = \frac{1}{4\pi t}\big( 1 -\frac{\cR}{12} t -m^2 t +\ldots\big)$, in agreement with \eqref{K+shortdist}.

Inserting the result \eqref{K+shortdist} for $K_+$ into \eqref{zetaK} allows us  to get much information about  $\zeta_+(s,x,x)$ as we will explain next. More generally, the small-$t$ expansion is (cf appendix A) 
\be\label{Kxxrexp}
K_+(t,x,x)\sim_{t\to 0}\frac{{\bf 1}_{2\times 2}}{4\pi t}\ \sum_{r=0}^\infty F_r(x,x) t^r \ \quad , \quad F_0(x,x)=1\ .
\ee
This is an asymptotic expansion for small $t$ which means it disregards any exponentially small terms ${\cal O}(e^{-a/t})$ that can be present but are invisible in such an expansion in powers of $t$. As discussed above, we also know that $K_+(t,x,x)$ vanishes exponentially as $e^{-\l^2_{\rm min} t}$ as $t\to\infty$. It follows that upon
evaluating the integral \eqref{zetaK} any possible divergences can only come from the small-$t$ region\footnote{
We have introduced an arbitrary large mass scale $\m$ so that $\m^{-2}$ is small enough for the asymptotic expansion to be applicable. This is similar to the small but finite $\e$ we had introduced when discussing Riemann's zeta function.
} 
$\int_0^{\m^{-2}} \ldots$ where the asymptotic expansion \eqref{Kxxrexp} is valid. Inserting this expansion into this integral, and interchanging the integral and the sum, the $r^{\rm th}$ term in the sum is (this is to be understood as evaluated for $\Re s>1$ and then analytically continued)
\be\label{Frpolestr}
\frac{F_r(x,x)}{4\pi} \frac{1}{\G(s)}\int_0^{\m^{-2}} \d t\, t^{s+r-2}
=\frac{F_r(x,x)}{4\pi} \frac{1}{\G(s)}\ \frac{(\m^{-2})^{s+r-1}}{s+r-1}\ , \quad r=0,1,2,\ldots
\ee
Let us insist that, since the integral is evaluated for $\Re s >1$, the contribution from the lower boundary is $0^{s+r-1}=0$.
The expression on the right-hand side of \eqref{Frpolestr} can then be analytically continued~: it is regular, except possibly as $s\to 1-r$. Indeed, Euler's $\G$-function has no zeros so that $\frac{1}{\G(s)}$ is regular. On the other hand, $\G(s)$ has poles whenever $s\to 0,-1,-2,\ldots$ which corresponds to $r=1,2,\ldots$, but not $r=0$. Hence, we must distinguish the cases $r=0$ and $r=1,2,\ldots$.
For $r=0$, i.e. $s\to 1$, we have $\G(s)\sim  1 -\g(s-1)$, where $\g\simeq 0.57$ is Euler's constant, and then
\be\label{F0polestr}
\frac{F_0(x,x)}{4\pi} \frac{1}{\G(s)}\int_0^{\m^{-2}} \d t\, t^{s-2}\sim_{s\to 1} \frac{F_0(x,x)}{4\pi} \Big( \frac{1}{s-1}+\g+\log\m^{-2} + {\cal O}(s-1)\Big) \ .
\ee
For $r=1,2,\ldots$, we have  $\frac{1}{\G(s)}\sim_{s\to 1-r} (-)^{r-1} (r-1)!\, (s-1+r)$ so that the ``would-be pole" of the integral is cancelled by this zero and one gets
\be
\frac{F_r(x,x)}{4\pi} \frac{1}{\G(s)}\int_0^{\m^{-2}} \d t\, t^{s+r-2}
\sim_{s\to 1-r} \frac{F_r(x,x)}{4\pi} \Big( (-)^{r-1} (r-1)!\,+ {\cal O}(s+r-1)\Big) \ , \quad r=1,2,\ldots
\ee
We conclude that
\beb\label{zetaplusatnegativeint}
\zeta_+(s,x,x)\sim_{s\to 1} \Big(\frac{1}{4\pi(s-1)} + \text{regular}\Big)\  {\bf 1}_{2\times 2} 
\ , \quad
\zeta_+(k,x,x)=\frac{(-)^k}{4\pi} k!\,F_{1-k}(x,x) \  {\bf 1}_{2\times 2} \ , \  k=0,-1, \ldots 
\eeb
In particular for $k=0$, we read from \eqref{K+shortdist} that $F_1(x,x)=-\big(\frac{\cR(x)}{12}+m^2\big)\, {\bf 1}_{2\times 2}$ which yields
\be\label{zetapluszero}
\zeta_+(0,x,x)=-\frac{1}{4\pi}  \big(\frac{\cR(x)}{12}+m^2\big)\, {\bf 1}_{2\times 2} \ .
\ee

What about $\zeta_-(s,x,x)$~? We know from \eqref{zetaxx-g*} that 
$\zeta_-(s,x,x)= i m \g_* \ \zeta_+(s+\frac{1}{2},x,x)$.
This means that $\zeta_-(s,x,x)$ must be analytic in $s$ except for a simple pole at $s=\frac{1}{2}$. More precisely,  \eqref{zetaplusatnegativeint} translates into
\be\label{zetaminusatnegativeint}
\zeta_-(s,x,x)\sim_{s\to 1/2} \Big( \frac{im}{4\pi(s-\frac{1}{2})}\, + \text{regular}\Big) \g_* 
\ , \ 
\zeta_-(k-\frac{1}{2},x,x)=(-)^k\frac{ i m}{4\pi} F_{1-k}(x,x) \,  \g_*\ , \  k=0,-1, \ldots 
\ee

We can now use this information, together with the inverse Mellin transform \eqref{heatfromzeta} to obtain the small-$t$ expansion of $K_-(t,x,x)$ as follows. Choosing e.g. $c>1$  and assuming  that\footnote
{
This should follow from the fact that   $\G(s)$ vanishes exponentially for $s=c\pm i \s$ and $\s\to\infty$, and, more generally on the dotted part of the integration contour. To estimate the $\G$-function for large arguments we use Stirling's formula
\be\label{Stirling}
\G(z+1)\sim_{|z|\to\infty} z^z e^{-z} =e^{z(\log z-1)} \ ,
\ee
which remains valid even if the argument is complex as long as $\arg z \ne \pm\pi$, since one must avoid the poles on the negative real axis. If we let $z=a\pm i\s$ with $\s\to\infty$ it is straightforward to see that
\be
\G(a+1\pm i \s)\sim_{\s\to\infty} e^{\pm i \big( \frac{\pi a}{2} +\s\log\s -\s\big) }  e^{-\frac{\pi}{2} \s + a \log\s +{\cal O}(\frac{1}{\s})}
\quad \Rightarrow\quad
\big\vert \G(a+1\pm i \s)\big\vert \sim_{\s\to\infty} \s^a e^{-\frac{\pi}{2} \s} \ .
\ee
One can show a similar result if also $a\to -\infty$, so that the  $\G$-function vanishes exponentially on the dotted part of the contour in the above figure as this contour is taken to infinity. Note that  the left vertical part of the contour must, of course, avoid the poles at the negative integers, but it remains true, even on the negative real axis between any two negative integers $-k-1$ and $-k$, that $|\G|$ vanishes exponentially as $k\to\infty$.
}
\be
\lim_{\s\to\infty} |\G(c'\pm i\s)\zeta_-(c'\pm i\s,x,x)|  = 0 \ , \forall c'\le c \ ,
\ee
we can close the integration contour by an infinite rectangle to the left, cf the figure, so that the integral is given by the sum of the residues of all the poles enclosed.
\vskip2.mm
\begin{center}
\includegraphics[width=6.cm]{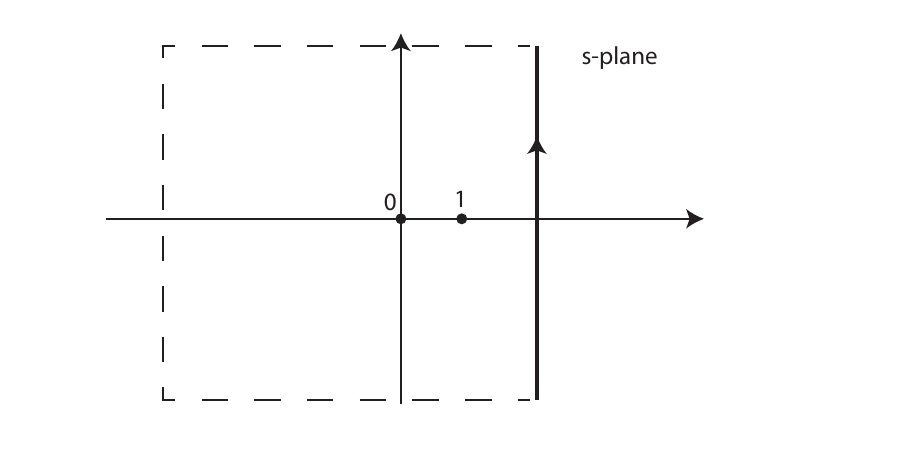} \\
The integration contour in the complex $s$-plane
\end{center}
There is the pole at $s=\frac{1}{2}$ from $\zeta_-(s,x,x)$ and there are the poles from $\G(s)$ at $s=-n,\ n=0,1,2,\ldots$ with residues $\frac{(-)^n}{n!}$. We get
\ba\label{Kmsmall-t-from-zeta}
K_-(t,x,x)&\sim_{t\to 0}& \frac{1}{2\pi i} \int_{\rm inf. rect.} \d s\, t^{-s}\, \G(s) \zeta_-(s,x,x)
=t^{-1/2} \G(\frac{1}{2}) \frac{im}{4\pi}\, \g_*\  + \sum_{n=0}^\infty t^n \frac{(-)^n}{n!} \zeta_-(-n,x,x) \nonumber\\
&=& im\, \g_* \Big( \frac{1}{4\sqrt{\pi\, t}} \, \g_* + \sum_{n=0}^\infty t^n \frac{(-)^n}{n!} \zeta_+(-n+\frac{1}{2},x,x) \Big)\ .
\ea
Just as for $\zeta_-(s,x,x)$, the explicit factor of $m$ on the right-hand side does not mean that $K_-(t,x,x)$ vanishes for $m=0$ since, in general,  the $\zeta_+(-n+\frac{1}{2},x,x)$ are expected to contain contributions $\sim m^{2n-1}$. However, we see that for $m=0$ the only term involving a half-integer power of $t$, namely $\frac{1}{\sqrt{t}}$ is absent (as are probably also the terms with $n\ge 1$).

Below, we will estimate the small-$t$ behaviour of $K_-$ on the flat torus where the eigenvalues and eigenfunctions are explicitly known, and we will find that
\be\label{Kmexplicittorus}
K_-(t,x,x)\sim_{t\to 0} \Big( \frac{m}{4\sqrt{\pi t}} + \frac{1}{4\pi^2}\Big) i\g_* + \ldots \ ,
\ee
in  agreement with the general formula \eqref{Kmsmall-t-from-zeta}~!

\subsubsection{Green's functions}\label{Green}

The Green's function $S(x,y)$ of the Dirac operator $D$ is a $2\times 2$-matrix solution of
\be\label{Sdef}
D_x S(x,y) = \frac{\dd^{(2)}(x-y)}{\sqrt{g}}\ {\bf 1}_{2\times 2} \ ,
\ee
while we denote $G$ the (also $2\times 2$-matrix) Green's function of $D^2$:
\be\label{Gdef}
D^2_x G(x,y) =\frac{\dd^{(2)}(x-y)}{\sqrt{g}}\ {\bf 1}_{2\times 2} \ .
\ee
In terms of the eigenfunctions and eigenvalues we have
\ba\label{SGeigen}
S(x,y)&=-i&\sum_n \frac{1}{\l_n} \big(\chi_n(x)\f_n^\dag(y)-\f_n(x)\chi_n^\dag(y)\big)
=\sum_n \frac{1}{\l_n} \big( \p_n(x)\p_n^\dag(x)-\p_n^*(x){\p_n^*}^\dag(y) \big)
 \ , \nonumber\\
G(x,y)&=&\sum_n \frac{1}{\l_n^2} \big(\chi_n(x)\chi_n^\dag(y)+\f_n(x)\f_n^\dag(y)\big)
=\sum_n \frac{1}{\l_n^2} \big( \p_n(x)\p_n^\dag(x)+\p_n^*(x){\p_n^*}^\dag(y) \big) \ .
\ea
They are indeed solutions of \eqref{Sdef}, resp. \eqref{Gdef} as one sees by applying $D$, resp. $D^2$, and using the completeness relation \eqref{completeness}.
It trivially follows from either \eqref{Sdef} and \eqref{Gdef}, or from \eqref{SGeigen},  that
\be\label{SGrel}
S(x,y)=D_x G(x,y) \ .
\ee
Comparing with \eqref{Zetadefs}, one sees that
\be\label{SGzeta}
S(x,y)=-i\,\zeta_-(\frac{1}{2},x,y) \quad , \quad G(x,y)=\zeta_+(1,x,y) \ .
\ee
Note that this is consistent with \eqref{zetapmrel} and \eqref{SGrel}.
Recall that $\zeta_-(s,x,x)$ has a pole at $s=\frac{1}{2}$ and $\zeta_+(s,x,x)$ at $s=1$, consistent with the fact that the Green's functions are singular as $x\to y$.

It is also obvious from the definitions that $G(x,y)$ is real and $S(x,y)$ purely imaginary. Furthermore,
\be\label{GShermconj}
(G(x,y))^\dag=G(y,x) \quad , \quad (S(x,y))^\dag=S(y,x)\ .
\ee
It follows from the orthonormality of the $\chi_n$ and $\f_n$ that
\be\label{SSGrel}
\int \d^2 z\sqrt{g(z)} \,S(x,z) S(z,y) = G(x,y) \ .
\ee
Note that this is also consistent with the differential equations obeyed by $S$ and $G$. Indeed, we have
\ba
D_x^2 G(x,y)&=& D_x \int \d^2 z\sqrt{g(z)} \ (D_x S(x,z)) S(z,y) = D_x \int \d^2 z\sqrt{g(z)}\    \frac{\dd^{(2)}(x-z)}{\sqrt{g(z)}}S(z,y) \nonumber\\
&=&D_x \,S(x,y)=  \frac{\dd^{(2)}(x-y)}{\sqrt{g}} {\bf 1}_{2\times 2} \ .
\ea

What can be said about the matrix structure of these Green's functions ? 
As before,   the matrix structure of $D^2$ is ${\bf 1} (\ldots) + \g_* (\ldots)$ (cf \eqref{Dsquared} and \eqref{spinorLapl2}), implying that $G$ must have the same structure:

\be\label{matrixstrG}
G(x,y)= G_0(x,y) \, {\bf 1}+ G_*(x,y) \, \g_*\ ,
\ee
which also follows from \eqref{SGzeta} and \eqref{zetaxymatrixstr}.
Combining this with the reality of $G$ and the property \eqref{GShermconj} we see that
$G_0(x,y)$ is real and symmetric and $G_*(x,y)$ is imaginary and antisymmetric. 
The matrix structure of $S$ is somewhat less trivial, but follows from $S(x,y)=(i\Nsl_x +m\g_*)G(x,y)$ along the same lines as discussed above for the zeta-functions.
We find
\be\label{StraceG}
\tr S(x,y)=m\tr \g_* G(x,y)\quad , \quad
\tr \g_* S(x,y) =  m\tr G(x,y) \ ,
\ee
which translates the corresponding relation \eqref{zeta-g*trace} between the traces of $\zeta_+$ and $\zeta_-$.
For vanishing mass, one must exclude the zero-mode contributions from $S$ and from $G$ (we then call them $\wt S$ and $\wt G$), but one still has the corresponding relation $\wt S=i\Nsl\, \wt G$ and it then follows that, for vanishing mass, 
\be\label{Stildetraces}
\tr \wt S(x,y) = \tr \g_* \wt S(x,y) = 0 \quad \text{for $m=0$}\ .
\ee

\subsection{Variation of  $\zeta(s)$\label{pertsec}}

Now we have the tools and definitions to express the variation of the zeta-function $\zeta(s)$ that appeared in the variation of the effective gravitational action. In the previous section we did the perturbation theory of the eigenvalues and had obtained the formula \eqref{zetavar}. It can now be nicely rewritten in terms of the local zeta functions $\zeta_\pm(s,x,x)$ as (recall that $\int f(x)$ is meant to be $\int \d^2 x\, \sqrt{g(x)} \, f(x)$)
\be\label{zetavar2}
\dd\zeta(s)=2s \int\dd\s  \Big( \tr \zeta_+(s,x,x) + im \tr \g_* \zeta_-(s+\frac{1}{2},x,x) \Big) \ .
\ee
We may  use \eqref{zetapmrel} to trade $\zeta_-(s+\frac{1}{2},x,x)$ for $im\g_* \zeta_+(s+1,x,x)$ so that
\be\label{zetavar3}
\dd\zeta(s)=2s \int\dd\s  \Big( \tr \zeta_+(s,x,x) - m^2 \tr \zeta_+(s+1,x,x) \Big) \ .
\ee
As we have  discussed above,  if zero-modes of $i\Nsl$ are present, we have $\l_0=m$ and $\dd\l_0=0$ so that there is no contribution from these zero-modes to the above formula. We can then equivalently rewrite it in terms of the $\wt\zeta_+(s,x,x)$  which are the local zeta functions with the zero-modes of $i\Nsl$ excluded from the sum~:
\be\label{zetavar3tilde}
\dd\zeta(s)=2s \int\dd\s  \Big( \tr \wt\zeta_+(s,x,x) - m^2 \tr \wt\zeta_+(s+1,x,x) \Big) \ .
\ee
Note that the zero-modes alone cannot contribute to the singularities and thus $\wt\zeta_+(s,x,x)$ has the same pole and residue at $s=1$ as $\zeta_+(s,x,x)$, only the regular parts differ by the zero-mode contribution.
For the derivative of \eqref{zetavar3} we obviously get
\ba\label{zetaprimevar}
\dd\zeta'(s)&=&2 \int\dd\s  \Big( \tr \zeta_+(s,x,x) - m^2 \tr \zeta_+(s+1,x,x) \Big)  \nonumber\\
&&+2 s\int\dd\s  \Big( \tr \zeta'_+(s,x,x) - m^2 \tr \zeta'_+(s+1,x,x) \Big) \ .
\ea
We want to evaluate both $\dd \zeta$ and $\dd \zeta'$ at $s=0$. Now $\zeta_+(0,x,x)$ is regular (cf e.g. \eqref{zetapluszero},  and\footnote{
If a meromorphic function is regular at a given point, then its derivative necessarily is also regular at this point.} so is $\zeta'_+(0,x,x)$. 
It follows that
\ba\label{deltazetaatzero}
&&\hskip-1.cm\dd\zeta(0)=-2 m^2 \int \dd\s\, \lim_{s\to 0} s\, \tr\zeta_+(s+1,x,x) \ ,\nonumber\\
&&\hskip-1.cm\dd\zeta'(0)=2 \int\dd\s  \tr \zeta_+(0,x,x) -2m^2  \int\dd\s\, \lim_{s\to 0} \tr   \Big(\zeta_+(s+1,x,x) + s \, \zeta'_+(s+1,x,x) \Big) \ .
\ea
Now $\zeta_+(\wt s,x,x)$ has a pole at 
$\wt s=1$~: $\zeta_+(\wt s,x,x)\sim \big(\frac{1}{4\pi} \frac{1}{\wt s-1} +C_+\big){\bf 1}_{2\times 2}+\zeta_+^{\rm reg}(\wt s,x,x)$, where the (finite) value of the constant $C_+$ depends on the exact definition of $\zeta_+^{\rm reg}$, cf eqs \eqref{zetaplussing3} and \eqref{zetaplusreg} below. It follows  that $\zeta_+(s+1,x,x)$ has a pole at $s=0$ and then 
\be\label{zetaplusresidue}
\lim_{s\to 0} s\, \zeta_+(s+1,x,x)=\lim_{s\to 0} s\, \wt\zeta_+(s+1,x,x)=\frac{1}{4\pi}\, {\bf 1}_{2\times 2} \ .
\ee
Note that we get the same expression whether we use $\zeta_+$ or $\wt\zeta_+$. In any case
\be\label{zetaatzero}
\dd\zeta(0)=-2m^2  \int\dd\s  \tr  \frac{1}{4\pi}\, {\bf 1}_{2\times 2} = -\frac{m^2}{\pi} \int \dd\s =-\frac{m^2}{2\pi} \dd A\ ,
\ee
where we used that 
\be
\int\dd\s\equiv\int \d^2 x\, \sqrt{g}\,\dd\s =\int \d^2 x \sqrt{\wh g} \,e^{2\s}\,\dd\s =\frac{1}{2}\, \dd \int \d^2 x \,\sqrt{\wh g}\, e^{2\s}=\frac{1}{2} \,\dd A \ .
\ee

Similarly, for the terms appearing in $\dd\zeta'(0)$ we have as $s\to 0$ 
\ba\label{zetazerolimitpert}
&&\hskip-2.cm  \lim_{s\to 0} \big(\zeta_+(s+1,x,x) + s\, \zeta'_+(s+1,x,x)\big) \nonumber\\
&&=\lim_{s\to 0} \Big[
\big(\frac{1}{4\pi\, s}+C_+\big) {\bf 1}_{2\times 2} +\zeta_+^{\rm reg}(s+1,x,x) +  s \frac{(-1)}{4\pi\, s^2} {\bf 1}_{2\times 2}
+ s (\zeta_+^{\rm reg})'(s+1,x,x) \Big] \nonumber\\
&&= \zeta_+^{\rm reg}(1,x,x) +C_+ \, {\bf 1}_{2\times 2}\ .
\ea
(We could have argued similarly about $\zeta_-(s+\frac{1}{2},x,x) + s\, \zeta'_-(s+\frac{1}{2},x,x)$ as $s\to 0$.)
Thus
\ba\label{zetprimeatzero}
\dd\zeta'(0)
&=& 2 \int\dd\s  \tr \zeta_+(0,x,x)  -2m^2 \int\dd\s  \tr \zeta_+^{\rm reg}(1,x,x)  - 2 m^2 C_+ \dd A \ ,\nonumber\\
&=& 2 \int\dd\s  \tr \wt\zeta_+(0,x,x)  -2m^2 \int\dd\s  \tr \wt\zeta_+^{\rm reg}(1,x,x)  - 2 m^2 \wt C_+ \dd A \ ,
\ea
In the second line we have written $\wt C_+$, but our definitions of $\zeta_+^{\rm reg}$ and $\wt \zeta_+^{\rm reg}$ below will be such that $C_+=\wt C_+=\frac{\g-\log\m^2}{4\pi}$.
Thus we can express $\dd\zeta'(0)$ either in terms of $\zeta_+(0,x,x)$ and $\zeta_+^{\rm reg}(1,x,x)$ or in terms of $\wt\zeta_+(0,x,x)$ and $\wt\zeta_+^{\rm reg}(1,x,x)$.


\subsection{Small-$t$ asymptotics of the  heat kernels and singularity structure of the local zeta-functions for the flat torus}\label{singstr}

We have seen that the singularities (poles) of the local zeta-functions are determined by the small-$t$ asymptotics of the  heat kernels.
We have also seen that we could relate $\zeta_+$ and $\zeta_-$ and, in particular, at coinciding points we could deduce from this relation the small-$t$ expansion of $K_-(t,x,x)$, see \eqref{Kmsmall-t-from-zeta}, from the one of $K_+(t,x,x)$, which is established in appendix A. However, we find it useful (and pedagogical) to try to explicitly compute these quantities for the example of the flat torus and see that we have indeed a perfect agreement with the results from the general statements. In this subsection we will thus establish the results for the flat ``square" torus with further detailed computations for $K_-$ presented in appendix~B.

From the explicit form of the normalised eigenfunctions $\p_{\vec{n}}$ of $D$ on the flat square torus (with periods $2\pi$ in both directions) given in \eqref{eigentorussol} we find 
\ba\label{calKtorus}
{\cal K}(t,x,y)&=&\sum_{n_1,n_2} e^{-\l^2_{\vec{n}}t} \p_{\vec{n}}(x)\p_{\vec{n}}^\dag(y) \nonumber\\
&=&\frac{1}{8\pi^2} \sum_{n_1,n_2} e^{-(n_1^2+n_2^2+m^2)t} e^{in_1 z^1+in_2 z^2}
\frac{1}{\l_{\vec{n}}} \begin{pmatrix} \l_{\vec{n}}-n_2 & -n_1-im \\  -n_1+im & \l_{\vec{n}}+n_2 \end{pmatrix} \ ,
\ea
where here, and throughout this subsection
$z^1=x^1-y^1, \ z^2=x^2-y^2$.
Note that  the terms odd under $n_1\to - n_1$ or $n_2\to - n_2$ drop out of the sum, so that (2 times) the real part is
\be\label{Kplustorus}
K_+(t,x,y)={\cal K}(t,x,y)+{\cal K}^*(t,x,y) = \frac{1}{4\pi^2} \sum_{n_1,n_2} e^{-(n_1^2+n_2^2+m^2)t} e^{in_1 z^1+in_2 z^2}
 \ {\bf 1}_{2\times 2} \ ,
\ee
while (-2 times) the imaginary part of ${\cal K}$ is
\ba\label{K--torus}
K_-(t,x,y)&=&i({\cal K}(t,x,y)-{\cal K}^*(t,x,y))\nonumber\\
&&\hskip-2.5cm =-\frac{1}{4\pi^2} \sum_{n_1,n_2} e^{-(n_1^2+n_2^2+m^2)t} 
\Big( \cos\big(n_1 z^1+n_2 z^2\big)
\frac{1}{\l_{\vec{n}}} \begin{pmatrix}0 &-m \\ m &0\end{pmatrix} 
+ \sin\big(n_1 z^1+n_2 z^2\big)
\frac{1}{\l_{\vec{n}}} \begin{pmatrix} -n_2 & -n_1 \\  -n_1 & n_2 \end{pmatrix} \Big)
\nonumber\\
&&\hskip-2.5cm =\frac{i}{4\pi^2} \sum_{n_1,n_2} e^{-(n_1^2+n_2^2+m^2)t} e^{in_1z^1+i n_2 z^2}
\frac{1}{\l_{\vec{n}}}\, \big( -n_1\s_x-n_2\s_z+m\g_*\big)
 \ .
\ea
Note that for $t=0$ one correctly finds that $K_+(0,x,y)= \dd^{(2)}(x-y){\bf 1}_{2\times 2}$, while nothing like this is true for $K_-$.
Also note that the a priori possible $\g_*$-terms are absent in \eqref{Kplustorus} and, correspondingly, there are no terms $\sim{\bf 1}_{2\times 2}$ in \eqref{K--torus}.

Obviously, at coinciding points $x=y$, i.e. $z=0$, these heat kernels simplify:
\ba\label{Kpmx=y}
K_+(t,x,x)&=&\frac{1}{4\pi^2} \sum_{n_1,n_2} e^{-(n_1^2+n_2^2+m^2)t} 
 \ {\bf 1}_{2\times 2} \ , \nonumber\\
K_-(t,x,x)&=&\frac{i\,m}{4\pi^2} \sum_{n_1,n_2} \frac{1}{\l_{\vec{n}}}\, e^{-(n_1^2+n_2^2+m^2)t} 
\ \g_*\ \ .
\ea
We see that $K_+(t,x,x)\sim  \ {\bf 1}_{2\times 2}$ and $K_-(t,x,x)\sim \g_*$, in agreement with the general statements above, cf.~\eqref{Kxxmatrix}.
Note that for $K_+(t,x,y)$ the double sum factorises into a sum over $n_1$ and another one over $n_2$, while for $K_-(t,x,y)$ or $K_-(t,x,x)$ there is no such factorisation since $\l_{\vec{n}}=\sqrt{n_1^2+n_2^2+m^2}$.
In fact, $K_+(t,x,y)$ can be expressed in terms of the theta function $\t_3$, (cf \cite{Erdelyi}) as
\be\label{theta3K+}
K_+(t,x,y)=\frac{1}{4\pi^2} \, e^{-m^2 t}\  \t_3\big(\frac{z^1}{2\pi}\big\vert i\frac{t}{\pi}\big) \ 
 \t_3\big(\frac{z^2}{2\pi}\big\vert i\frac{t}{\pi}\big)\  {\bf 1}_{2\times 2} 
 \quad , \quad \t_3(\n\vert\tau) = \sum_{n=-\infty}^\infty e^{i\pi\tau n^2+2\pi i \n n} \quad , \quad \Im\tau>0\ .
\ee
On the other hand,  $K_-(t,x,y)$ can be expressed in terms of the auxiliary sum  $L$ already defined in \eqref{Ldef},  or equivalently another related sum ${\cal L}$, as follows~:
 \be\label{KmLcalL1}
K_-(t,x,y)= D_x L(t,x,y) \equiv (i \s_x \del_1 + i \s_z \del_2 + m\g_*) L(t,x,y)\ ,
\ee
where
\be\label{KmLcalL2}
L(t,x,y)=\frac{i}{4\pi^2} \,  {\cal L}(t,z)   {\bf 1}_{2\times 2}
\quad , \quad
 {\cal L}(t,z)=
\sum_{n_1,n_2} \frac{e^{-\l_{\vec{n}}^2\,t} }{\l_{\vec{n}}}\ e^{i\vec{n}\cdot\vec{z}}
\equiv \sum_{n_1,n_2} \frac{e^{-(n_1^2+n_2^2+m^2)t} e^{in_1 z^1+in_2 z^2}}{\sqrt{n_1^2+n_2^2+m^2}} \ .
\ee
Note that $K_-$ at coinciding points is directly given in terms of this ${\cal L}$ as (cf \eqref{Kpmx=y})
\be\label{LKmxxagain}
K_-(t,x,x)=\frac{i}{4\pi^2}\,  m\g_*\  {\cal L}(t,0)\ .
\ee
For the present case of the torus, we find  even for $x\ne y$ that $L(t,x,y)\sim {\bf 1}_{2\times 2}$, a fact that is not necessarily true in general.

The explicit expression of $K_+$ in terms of  the theta function $\t_3$ makes it easy to study its properties. The periodicity property $\t_3(\n+1\vert \tau)=\t_3(\n\vert\tau)$ reflects the periodicity of the eigenfunctions on the torus.
$\t_3$ has a well-known modular transformation property \cite{Erdelyi} which can  be written as 
\be\label{modular2}
\t_3(\frac{z}{2\pi}\vert i\frac{t}{\pi}) = \sqrt{\frac{\pi}{t}}e^{-z^2/(4t)}\  \t_3(-i\frac{z}{2t},i\frac{\pi}{t}) \ ,
\ee
giving immediately  the small-$t$ asymptotics as follows~:
\be\label{t3small-t}
\t_3(\frac{z}{2\pi}\vert i\frac{t}{\pi})=\sqrt{\frac{\pi}{t}}\, e^{-z^2/(4t)}\ 
\sum_n e^{-\pi^2 n^2 / t + \pi z n/t}
= \sqrt{\frac{\pi}{t}}\, \sum_n e^{-\frac{1}{4t}(z-2\pi n)^2}\ .
\ee
This clearly exhibits again the $2\pi$-periodicity in $z$. Let us then assume that $z\in (-\pi,\pi]$ and consider $t\to 0$. For $z=\pi$, both $n=0$ and $n=1$ contribute equally to the sum so that one gets $\sqrt{\frac{\pi}{t}}\, 2\, e^{-\pi^2/(4t)} (1+{\cal O}(e^{-2\pi^2/t}))$. For all other values of $z\in (-\pi,\pi)$ there is a single dominant term that contributes $\sqrt{\frac{\pi}{t}}\, e^{-z^2/(4t)}$ and all other terms are exponentially smaller by some factor $e^{-a/t}$ for some $a>0$.
Hence the small-$t$ asymptotics is
\be\label{t3small-t-bis}
\t_3(\frac{z}{2\pi}\vert i\frac{t}{\pi}) \sim_{t\to 0} \sqrt{\frac{\pi}{t}}\, e^{-z^2/(4t)} \, \Big( 1 + {\cal O}\big( e^{-a/t}\big) \Big) \quad , \quad a>0\ , \quad \text{for} |z|<\pi \ .
\ee
One
immediately deduces  the small-$t$ behaviour of $K_+(t,x,y)$ (for $z^i\ne \pm\pi$) as
\be\label{Kplusastorus}
K_+(t,x,y) = \frac{1}{4\pi t} \ \exp\Big( - \frac{(z^1)^2+(z^2)^2}{4t}\Big)\   e^{-m^2 t}\  {\bf 1}_{2\times 2} \ \Big( 1 + {\cal O}\big( e^{-a/t}\big) \Big) \ .
\ee
The leading piece coincides, of course, with the well-known answer on ${\bf R}^2$.
For $x=y$ one simply has
\be\label{Kplusastorusxx}
K_+(t,x,x) = \frac{1}{4\pi t} \,  e^{-m^2 t}\  {\bf 1}_{2\times 2} \ \Big( 1 + {\cal O}\big( e^{-a/t}\big) \Big) \ .
\ee
\vskip2.mm
\noindent

On the other hand, the small $t$ asymptotics of $K_-(t,x,y)$, or equivalently the small-$t$ asymptotics of $L(t,x,y)$ or ${\cal L}(t,z)$ is much more difficult to obtain directly. We have devoted some effort to try to estimate their asymptotic behaviour. These results are presented in appendix B.
At coinciding points we have been able to estimate rather easily (cf \eqref{calLde0-5}) that
\be\label{calLasymptoticsatz=0}
{\cal L}(t,0) 
\simeq \sqrt{\frac{\pi^3}{t}}  + \frac{1}{m} +{\cal O}(t^0,m^0)\ .
\ee
This implies in turn that $K_-(t,x,x)$ is given by (cf \eqref{LKmxxagain})
\be\label{Kmoinstxxleading}
K_-(t,x,x)\sim_{t\to 0} 
\Big(\frac{m}{4\sqrt{\pi t}} + \frac{1}{4\pi^2}\Big)\ i\g_*\ +\ldots \ .
\ee
where the unwritten terms $+\ldots$ vanish as $t\to 0$, a result already cited above.
To estimate ${\cal L}(t,z)$ for $z\ne 0$ turned out to be much more difficult. The reader can find more details in the appendix B. But we will not need them here to proceed further.

As discussed above, the small-$t$ behaviours of $K_\pm(t,x,x)$  translate into possible poles of the corresponding local zeta-functions $\zeta_\pm(s,x,x)$. Using the above torus results for coinciding points one immeadiately finds that
\ba\label{toruszeta}
&&\zeta_+(s,x,x) \sim_{s\to 1} \ \frac{{\bf 1} }{4\pi (s-1)} + {\rm finite}\ 
\ , \quad
\zeta_+(0,x,x) = {\rm finite}\times {\bf 1} \ , \nonumber\\
&&\zeta_-(s,x,x) \sim_{s\to 1/2}\  \frac{m i \g_* }{4\pi (s-\frac{1}{2} )}  + {\rm finite} 
\ , \quad \zeta_-(0,x,x) = {\rm finite}\times  i\g_*\ ,
\ea
again in agreement with our general results.


\subsection{General statements about the singularity structure for  $x\ne y$}

In subsection \ref{defandbasicheat} we already discussed many general properties of the local heat kernels and local zeta functions, in particular  at coinciding point. Now, we will make some further general statements for\break $x\ne y$. These statements will be based on the small-$t$ behaviour of $K_+(t,x,y)$ as follows from the differential (heat) equation, and worked out in appendix $A$. For $K_-(t,x,y)$ we can deduce the corresponding
statements by using the fundamental relation \eqref{zetapmrel} between the corresponding $\zeta_+(s,x,y)$ and $\zeta_-(s,x,y)$.

As already discussed above, since $D^2$ only contains the matrix structures ${\bf 1}_{2\times 2}$ and $\g_*$, the same is true for  $G(x,y)$ and $K_+(t,x,y)$. However, the leading small-$t$ singularity of $K_+(t,x,y)$ is only $\sim {\bf 1}_{2\times 2}$ and is universal, cf \eqref{Kseries} of the appendix\footnote{
Actually, in $n=2$ dimensions, equation \eqref{Kseries}  gives 
$K(t,x,y)= \frac{1}{4\pi t} \exp\left( -{\ell^2(x,y)\over 4 t}\right) \big( F_0(x,y) + {\cal O}(t)\big)$. Now, $F_0(x,y) = 1+{\cal O}(\ell^2)$ and the exponential forces $\ell^2$  to be of order $t$ so that ${\cal O}(\ell^2)\simeq {\cal O}(t)$ and we indeed have $K(t,x,y)= \frac{1}{4\pi t} \exp\left( -{\ell^2(x,y)\over 4 t}\right) \big( 1 + {\cal O}(t)\big)$.
}~:
\be\label{K+txygen}
K_+(t,x,y) \sim_{t\to 0} \frac{1}{4\pi t}\, e^{-\ell^2(x,y)/(4t)} \, {\bf 1}_{2\times 2} \ ,
\ee
where,   $\ell(x,y)$ is the geodesic distance between $x$ and $y$. This  generalizes the corresponding formula  \eqref{Kplusastorus} for the torus. We can then deduce the singularity structure of $\zeta_+(s,x,y)$ solely form this universal leading behaviour of $K_+$.
This universality is due to the (leading) 2-derivative part of $D^2$ being always $-{\bf 1}_{2\times 2}\, g^{\m\n}\del_\m\del_\n$ which is the same as for the scalar Laplacian $-{\bf 1}_{2\times 2} \D_{\rm scalar}$. Indeed, as discussed above, we may obtain $K_+(t,x,y)$ solely from the differential equation \eqref{heateq}. 

For $K_-(t,x,y) = D_x L(t,x,y)$ we expect that the {\it leading} small-$t$ singularity of $L$ is again generic and hence given by the obvious generalization of the torus result, cf \eqref{calLadhoc}.
However, instead of relying on this ``expected" formula, we will instead use the proven relation \eqref{zetapmrel} between $\zeta_+$ and $\zeta_-$, together with \eqref{K+txygen}, to deduce the singularity structure of $\zeta_-(s,x,y)$. 

Let us then establish the singularity structures of both, $\zeta_\pm(s,x,y)$. To begin with,  \eqref{K+txygen} defines what we call the singular part of the heat-kernel~:
\be\label{K+sing}
K^{\rm sing}_+(t,x,y)=\frac{1}{4\pi t}\, e^{-\ell^2(x,y)/(4t)} \, {\bf 1}_{2\times 2} \ .
\ee
From this we define the singular part of the zeta-function as
\be\label{Zetaplussing}
\zeta^{\rm sing}_+(s,x,y) =\frac{1}{\G(s)}\int_0^{1/\m^2} \d t\, t^{s-1} K^{\rm sing}_\pm(t,x,y)\ .
\ee
Indeed, any singularity of the zeta-function must come from the small-$t$ part of the integral, and as we have seen before, only the $\frac{1}{t}$-part of the heat kernel $K_+$ can lead to a pole of the $\zeta_+(s,\cdot,\cdot)$, while the ``would-be" poles due to the sub-leading terms in $t$ are cancelled by the zeros of $\frac{1}{\G(s)}$.
For general $x,y$ we explicitly have (setting $\frac{1}{t}=\m^2 u$)
\be\label{zetaplussing2}
\zeta^{\rm sing}_+(s,x,y) 
=\frac{1}{\G(s)}\int_0^{1/\m^2} \d t\, t^{s-1} \frac{e^{-\ell^2(x,y)/(4t)}}{4\pi t} \, {\bf 1}_{2\times 2}
=\frac{(\m^2)^{1-s}}{4\pi \G(s)}\, E_s\big(\frac{\m^2 \ell^2(x,y) }{ 4}\big)\,  {\bf 1}_{2\times 2}\ ,
\ee
where $E_s$ is the exponential integral function defined as 
\be\label{expint}
E_s(z)=\int_1^\infty \d u\, u^{-s}e^{-z u}\ . 
\ee
Its asymptotic expansions are well known and, in particular, if we first set $s=1$ and then let $x\to y$, i.e. $\ell(x,y)\to 0$, we have
\be\label{Es1}
E_1\Big(\frac{\m^2\ell^2}{4} \Big) \sim_{\ell^2\to 0} \ \ -\g -\log \frac{\m^2\ell^2 }{4} +{\cal O}\big( \m^2\ell^2 \big) \ ,
\ee
(where $\g\simeq 0.57$ is Euler's constant, not to be confused with the matrix $\g_*$)
so that
\be\label{zetaplussing2-bis}
\zeta^{\rm sing}_+(1,x,y) 
\sim_{\ell\to 0}\  \frac{1}{4\pi} \Big(  -\log \frac{\m^2\ell^2(x,y) }{4}  -\g +{\cal O}\big( \m^2\ell^2 \big) \Big) 
\,  {\bf 1}_{2\times 2}\ ,
\ee
Of course, $\zeta_+(1,x,y)$ is the Green's function $G(x,y)$ and the term $-\frac{1}{4\pi}\log \frac{\m^2\ell^2 }{4}\, {\bf 1}_{2\times 2}$  is just the (leading)  short-distance singularity of $G(x,y)$ which is identical to the well-known short-distance singulartity of the Green's function of the scalar Laplace operator in two dimensions.
On the other hand, if we first set $x=y$ so that $\ell(x,y)=0$ we have, assuming $\Re s>1$, $E_s(0)=\int_0^1 \d v\, v^{s-2}=\frac{1}{s-1}$~:
\be\label{Es}
E_s(0)=\frac{1}{s-1} \ .
\ee
Expanding $\G(s)$ and $(\m^2)^{1-s}$ we have
\be\label{zetaplussing3}
\zeta^{\rm sing}_+(s,x,x)=\frac{1}{4\pi} \Big( \frac{1}{s-1} -\log\m^2 +\g +{\cal O}(s-1) \Big) \, {\bf 1}_{2\times 2}\ .
\ee
It follows that if we define
\be\label{zetaplusreg}
\zeta_+^{\rm reg}(s,x,y)=\zeta_+(s,x,y)- \zeta^{\rm sing}_+(s,x,y) 
\ee
we have subtracted all the potential singularities of this zeta-function, namely  the pole at $s=1$ that is present for $x=y$ and the short-distance singularity of the Green's function which is present for $s=1$. 
In particular, we see that the following limit exists~:
\ba\label{zetapluslimit}
\lim_{s\to 0}\Big( \zeta_+(s+1,x,x) + s\,  \zeta_+'(s+1,x,x)\Big) 
&=& \zeta_+^{\rm reg}(1,x,x)+\frac{1}{4\pi} \big( \g-\log\m^2\big)\, {\bf 1}_{2\times 2} \nonumber\\
&\equiv&\zeta_+^{\rm reg}(1,x,x)+C_+\, {\bf 1}_{2\times 2} \ .
\ea
The existence of this limit was already anticipated above in \eqref{zetazerolimitpert}, but now we have also determined the value of finite constant $C_+$.

It is straightforward to repeat this reasoning for the  $\wt \zeta_\pm(s,x,y)$ where the zero-modes are excluded. Since the finitely many zero-modes cannot contribute to the singular parts the latter are unchanged, $\wt \zeta^{\rm sing}_\pm(s,x,y)=\zeta^{\rm sing}_\pm(s,x,y $. Then whole discussion can be repeated with the result
\ba\label{wtzetapluslimit}
\lim_{s\to 0}\Big( \wt\zeta_+(s+1,x,x) + s\,  \wt\zeta_+'(s+1,x,x)\Big) 
&=& \wt\zeta_+^{\rm reg}(1,x,x)+\frac{1}{4\pi} \big( \g-\log\m^2\big)\, {\bf 1}_{2\times 2} \nonumber\\
&\equiv&\wt\zeta_+^{\rm reg}(1,x,x)+C_+\, {\bf 1}_{2\times 2} \ ,
\ea
where, in particular, the constant $C_+$ is the same as before.


We can also translate all these statements to $\zeta_-(s,x,y)$ by using \eqref{zetapmrel}~:
\be\label{zetamoinssing}
\zeta_-^{\rm sing}(s,x,y)= i D_x \zeta_+^{\rm sing}(s+\frac{1}{2},x,y)
= i\frac{(\m^2)^{\frac{1}{2}-s}}{4\pi \G(s+\frac{1}{2})} \, (i\Nsl_x + m\g_*)\, E_{s+\frac{1}{2}}\big(\frac{\m^2 \ell^2(x,y) }{ 4}\big) \ . \ \ 
\ee
In particular, for $s=\frac{1}{2}$ we have again $E_1\big(\frac{\m^2 \ell^2(x,y) }{ 4}\big)$ with its small-$\ell$ behaviour given in \eqref{Es1} and, for $x=y$, obviously \eqref{Es} translates into  $E_{s+\frac{1}{2}}(0)=\frac{1}{s-\frac{1}{2}}$. However, to evaluate $\zeta_-^{\rm sing}(s,x,y)$ at $x=y$ we must first act with $\Nsl_x$ and then set $x=y$~: $\Nsl_x\, E_{s+\frac{1}{2}}\big(\frac{\m^2 \ell^2(x,y)}{4}\big)\big\vert_{x=y}= \frac{\m^2}{4} \big( \Nsl_x \, \ell^2(x,y)\big)\big\vert_{x=y} E'_{s+\frac{1}{2}}(0)$.
What does it mean to have a spinorial covariant derivative $\Nsl=\g^\m \nabla_\m^{\rm sp}=\g^\m(\del_\m-\frac{i}{4}\o_\m\g_*)$ acting on $\ell^2(x,y)$~? One has $\nabla_\m^{\rm sp} \ell^2(x,y)=\del_\m \ell^2(x,y)-\frac{i}{4}\o_\m\g_*\ell^2(x,y)$. This can be evaluated more explicitly, for example, in Riemann normal coordinates around $y$, see the appendix A.1. One then sees that $\del_\m\ell^2(x,y)=2(x-y)^\m$ and $\o_\m=-\frac{\cR}{2} \e_{\m\n}(x-y)^\n$ ($\e_{12}=-\e_{21}=1$). In particular, we see that for $x\to y$ this quantity vanishes. The ``invariant" statement we can make is that
\be\label{Nslonell2}
\Nsl_x \ell^2(x,y)\big\vert_{x=y}=0 \ .
\ee
It follows that
\be\label{zetamoinssingx=y}
\zeta_-^{\rm sing}(s,x,x)
= i m\g_*\ \frac{(\m^2)^{\frac{1}{2}-s}}{4\pi \G(s+\frac{1}{2})} \,  E_{s+\frac{1}{2}}(0) \ . 
\ee
Again, just as for $\zeta_-(s,x,x)$, only the $\g_*$-matrix part remains at coinciding points.
In particular, for $s\to\frac{1}{2}$ one finds, much as above, 
\be\label{zetaminussingx=y2}
\zeta^{\rm sing}_-(s,x,x)=\frac{im}{4\pi}\, \g_*\,  \Big( \frac{1}{s-\frac{1}{2}} -\log\m^2 +\g +{\cal O}(s-\frac{1}{2}) \Big) \ .
\ee
On the other hand, if we first set $s=\frac{1}{2}$ and then let $x\to y$, using \eqref{Es1}, we have
\ba\label{zetamoinssing3}
\zeta_-^{\rm sing}(\frac{1}{2},x,y)\hskip-3.mm&\hskip-3.mm=& \hskip-3.mm\frac{i}{4\pi} \, (i\Nsl_x + m\g_*)\, E_1\big(\frac{\m^2 \ell^2(x,y) }{ 4}\big) 
\sim_{x\to y}
 \frac{i}{4\pi} \, (i\Nsl_x + m\g_*)\, \Big( -\g -\log \frac{\m^2\ell^2 }{4} +{\cal O}\big( \m^2\ell^2 \big) \Big) \nonumber\\
&\sim_{x\to y}& -  \frac{i}{4\pi}  \, \Big( \frac{i\Nsl\, \ell^2}{\ell^2} + m \g_*\, \big( \log \frac{\m^2\ell^2 }{4}  + \g \big) \Big) \ .
\ea
Of course, this is the short-distance behaviour of the fermionic Green's function $i S(x,y)$ and
we recognise the usual leading singularity $\simeq  i\Nsl\, \ell^2/\ell^2 $ of the Dirac Green's function, as well as a sub-leading logarithmic singularity. Just as for $\zeta_+$, It follows that if we define
\be\label{zetaminusreg}
\zeta_-^{\rm reg}(s,x,y)=\zeta_-(s,x,y)- \zeta^{\rm sing}_-(s,x,y) 
\ee
we have subtracted all the potential singularities of this zeta-function, namely  the pole at $s=\frac{1}{2}$ that is present for $x=y$ and the short-distance singularity of the Green's function which is present for $s=\frac{1}{2}$. 
In particular, we see that the following limit exists~:
\be\label{sonehalf}
\lim_{s\to 0}\Big( \zeta_-(s+\frac{1}{2},x,x) + s\,  \zeta_-'(s+\frac{1}{2},x,x)\Big) = \zeta_-^{\rm reg}(\frac{1}{2},x,x)+\frac{m}{4\pi} \big( \g-\log\m^2\big)\, i\g_* \ .
\ee
Using $\zeta_-(s,x,x)=im\g_*\zeta_+(s+\frac{1}{2},x,x)$, cf \eqref{zetaxx-g*}, we conclude that also
\be\label{zetapmregrel}
\zeta_-^{\rm reg}(\frac{1}{2},x,x)=im\g_*\, \zeta_+^{\rm reg}(1,x,x)\ .
\ee

Note that, contrary to the singular parts, to evaluate the regular parts $\zeta_\pm^{\rm reg}$ the knowledge of the small-$t$ asymptotics of the heat kernels is {\it not} enough, but requires some knowledge about all eigenvalues and eigenfunctions. One way this information is coded is in the regularized and so-called renormalized Green's functions which we will discuss next.



\subsection{Renormalized Green's functions}

\subsubsection{$G_{\rm R}$, $G_\zeta$, $S_{\rm R}$ and $S_\zeta$}

For the Green's function $G(x,y)=\zeta_+(1,x,y)$ of $D^2$, we may define a regularized Green's function $G^{\rm reg}(x,y)$ by subtracting the short-distance singularity, cf \eqref{zetaplussing2-bis}
\be\label{Greg}
G^{\rm reg}(x,y)= G(x,y)-G^{\rm sing}(x,y)= G(x,y)+ \frac{1}{4\pi} \log \frac{\m^2\ell^2(x,y)}{4} \ {\bf 1}_{2\times 2}  \ .
\ee
Note that this is similar to, but different from $\zeta_+^{\rm reg}(1,x,y)=G(x,y)-\frac{1}{4\pi} E_1\big(\frac{\m^2\ell^2(x,y)}{4}\big)\, {\bf 1}_{2\times 2}$.
But just as $\zeta_+^{\rm reg}(1,x,y)$, the regulartized Green's function $G^{\rm reg}(x,y)$ has a well-defined limit as $x\to y$.
The so-called renormalized Green's function at coinciding points $G_{\rm R}$  then is simply defined as 
\be\label{GR}
G_{\rm R}(y)=\lim_{x\to y} G^{\rm reg}(x,y) \ .
\ee
In complete analogy, we define $S^{\rm reg}(x,y)$ and $S_{\rm R}(y)$~:
\be\label{Sreg}
S^{\rm reg}(x,y)=S(x,y)-S^{\rm sing}(x,y)= S(x,y)+ \frac{1}{4\pi}\, \big( i \Nsl_x + m\g_*\big) \log \frac{\m^2\ell^2(x,y)}{4} \ {\bf 1}_{2\times 2}  \ .
\ee
Again,  this is similar to, but different from $-i\zeta_-^{\rm reg}(\frac{1}{2},x,y)=S(x,y)-\frac{1}{4\pi} (i\Nsl_x + m\g_*)E_1\big(\frac{\m^2\ell^2(x,y)}{4}\big)$.
The renormalized Green's function at coinciding points $S_{\rm R}$ then is defined as
\be\label{SR}
S_{\rm R}(y)=\lim_{x\to y} S^{\rm reg}(x,y) \ .
\ee
Recall from \eqref{StraceG} that $\tr \g_* S(x,y)=m\tr G(x,y)$. The same relation obviously also holds between $S^{\rm sing}$ and $G^{\rm sing}$. Hence
\be\label{SregGregtracerel}
\tr \g_* S^{\rm reg}(x,y) = m\tr G^{\rm reg}(x,y) \ .
\ee
If we take the limit $x\to y$ of the last relation we get
$\tr \g_* S_{\rm R}(y)= m \tr G_{\rm R}(y)$, but we will prove the more general  relation without the traces below in \eqref{GRzetaSRzetarel}.

It is also easy to express $G_{\rm R}$ and $S_{\rm R}$ in terms of $\zeta_+^{\rm reg}(1,x,x)$ 
and $\zeta_-^{\rm reg}(\frac{1}{2},x,x)$. Indeed,
\ba\label{Gregzetareg}
G^{\rm reg}(x,y)&=& \zeta_+^{\rm reg}(1,x,y) +\frac{1}{4\pi} \Big( E_1\big( \frac{\m^2\ell^2(x,y)}{4}\big) + \log\frac{\m^2\ell^2(x,y)}{4} \Big) {\bf 1}_{2\times 2}\nonumber\\
S^{\rm reg}(x,y)&=&-i \zeta_-^{\rm reg}(\frac{1}{2},x,y) +\frac{1}{4\pi}(i\Nsl_x+m\g_*) \Big( E_1\big( \frac{\m^2\ell^2(x,y)}{4}\big) + \log\frac{\m^2\ell^2(x,y)}{4} \Big) \ .
\ea
Taking the limit $x\to y$ and using the asymptotics \eqref{Es1} of $E_1$ we get
\ba\label{Grenormalizedzetareg}
G_{\rm R}(x)&=& \zeta_+^{\rm reg}(1,x,x) -\frac{\g}{4\pi} {\bf 1}_{2\times 2} \nonumber\\
S_{\rm R}(x)&=&-i \zeta_-^{\rm reg}(\frac{1}{2},x,x) -\frac{m\g}{4\pi} \g_* \ .
\ea
For completeness, let us mention that one sometimes defines  different ``renormalized" Green's functions $G_\zeta(x)$ and $S_\zeta(x)$ from the $\zeta_\pm (s,x,x)$ by subtracting simply the poles and then letting $s\to 1$ or $s\to \frac{1}{2}$~:
\ba\label{Gzetadef}
G_\zeta(x)&=&\lim_{s\to 1} \Big( (\m^2)^{s-1} \zeta_+(s,x,x)  - \frac{{\bf 1}_{2\times 2}}{(4\pi)\, (s-1)} \Big) \ ,\nonumber\\
S_\zeta(x)&=&\lim_{s\to\frac{1}{2}}\Big(-i\, (\m^2)^{s-1/2} \zeta_-(s,x,x) -\frac{m \g_*}{4\pi (s-\frac{1}{2})} \Big) \ .
\ea
It follows, using \eqref{zetaplussing3} and \eqref{zetaplusreg}
that
\ba\label{Gzeta+regxx}
G_\zeta(x)&=&\zeta_+^{\rm reg}(1,x,x)+\frac{\g}{4\pi} \, {\bf 1}_{2\times 2} 
\quad \Rightarrow\quad G_\zeta(x)=G_{\rm R}(x) +\frac{\g}{2\pi} \, {\bf 1}_{2\times 2} \ ,\nonumber\\
S_\zeta(x)&=&-i\zeta_-^{\rm reg}(\frac{1}{2},x,x) +\frac{\g}{4\pi} \,m\,  \g_*
\quad \Rightarrow\quad S_\zeta(x)=S_{\rm R}(x) +\frac{\g}{2\pi}\, m\,\g_* \ .
\ea
We can now re-express the relation \eqref{zetapluslimit} in terms of $G_{\rm R}$ or $G_\zeta$ and equivalently the relation \eqref{sonehalf} in terms of $S_{\rm R}$ or $S_\zeta$ as follows
\ba\label{zeta-pluszeta-prime}
&&\lim_{s\to 0}\Big( \zeta_+(s+1,x,x) + s\,  \zeta_+'(s+1,x,x)\Big) 
= \zeta_+^{\rm reg}(1,x,x)+\frac{1}{4\pi} \big( \g-\log\m^2\big)\, {\bf 1}_{2\times 2}
\nonumber\\
&&\hskip1.cm = G_{\rm R}(x) +  \frac{1}{4 \pi} \,\big(2\g -\log\m^2 \big) \ {\bf 1}_{2\times 2}
= G_\zeta(x) -  \frac{1}{4 \pi} \,\log\m^2\  {\bf 1}_{2\times 2} \  , \nonumber\\
&&\lim_{s\to 0}\Big( \zeta_-(s+\frac{1}{2},x,x) + s\,  \zeta_-'(s+\frac{1}{2},x,x)\Big) 
= \zeta_-^{\rm reg}(\frac{1}{2},x,x)+\frac{m}{4\pi} \big( \g-\log\m^2\big)\, i\g_* \nonumber\\
&&\hskip1.cm = i S_{\rm R}(x) +  \frac{im}{4 \pi} \,\big(2\g -\log\m^2 \big) \ \g_* 
= i S_\zeta(x) -  \frac{im}{4 \pi} \,\log\m^2\  \g_* \ .
\ea

Moreover, it follows from \eqref{zetaxx-g*} and \eqref{Gzetadef} that $S_\zeta(x)=m\g_* G_\zeta(x)$ and then by \eqref{Gzeta+regxx} also that $S_{\rm R}(x)=m\g_* G_{\rm R}(x)$ (which also follows from \eqref{Grenormalizedzetareg} and \eqref{zetapmregrel})~:
\beb\label{GRzetaSRzetarel}
S_\zeta(x)=m\g_* G_\zeta(x) \quad , \quad S_{\rm R}(x)=m\g_* G_{\rm R}(x) \ .
\eeb

Finally note that if the metric has sufficiently many isometries these renormalized Green's function must be constants. This is the case for the round sphere and for the flat torus. Of course, they still depend on the geometric data of these surfaces, like the radius of the round sphere, or the periods of the flat torus, but not on the point $x$. Their explicit form can be quite non-trivial even for such simple geometries. Let us show this for our preferred example of the flat torus, now  with periods $2a$ and $2b$.

\subsubsection{Green's function $\wt G(x,y)$ and  $\wt G_{\rm R}(x)$ on the flat torus for $m=0$}

Let us first show how one can obtain the Green's function  of the squared Dirac operator in the massless case on a ``rectangular" flat torus with periods $2a$ and $2b$, i.e. with $\tau=i\frac{b}{a}$. Since we look at the massless case, we will determine the Green's function with the zero-mode part excluded, i.e.~$\wt G(x,y)$.  Being flat, there is no spin-connection and no curvature and we have, as above, that $D^2={\bf 1}_{2\times 2}(-\del_\n \del^\n)=-{\bf 1}_{2\times 2}\D_{\rm scalar}$. Hence, the problem is identical to finding the Green's function of the scalar Laplace operator on the flat torus. The non-trivial part  is to reconcile the appropriate short-distance singularity and the periodicity of the torus.  But this problem has a well-known solution in terms of theta functions and the corresponding computation can e.g.~be found in  \cite{BdL2}. Using complex coordinates $z=z^1+i z^2$, where $z^i=x^i-y^i$ one finds $\wt G(x,y)=g(z,\zb) {\bf 1}_{2\times 2}$ with
\ba\label{torusg}
g(z,\zb)= \frac{(\Im z)^2}{8ab}-\frac{1}{4\pi} \log\Big\vert\t_1\big(\frac{z}{2a}\big\vert i\frac{b}{a}\big)\Big\vert^2 
\equiv\frac{(\Im z)^2}{8ab}-\frac{1}{4\pi} \log\Big[\t_1\big(\frac{z}{2a}\big\vert i\frac{b}{a}\big) \t_1\big(\frac{\zb}{2a}\big\vert i\frac{b}{a}\big) \Big] \ ,
\ea
where the theta function $\t_1$ is defined by \cite{Erdelyi}
\ba\label{theta1def}
\t_1(\n\vert\tau)&=& 2 q^{1/4} \sum_{n=0}^\infty (-)^n q^{n(n+1)} \sin(2n+1)\pi \n
\ , \quad q=e^{i\pi\tau}\ .
\ea
Indeed, the scalar Laplace operator on flat space in the complex coordinates is simply $\D_{\rm scalar}=4\del_z \del_{\ov z}$ and then, for $z,\zb\ne 0$, we have $-4 \del_z \del_{\ov z}\, g(z,\zb)=-4 \del_z \del_{\ov z} \,\frac{(\Im z)^2}{8ab}=\frac{1}{4ab}=\frac{1}{A}$. On the other hand, $g(z,\zb)\sim_{|z|\to 0}\, -\frac{1}{4\pi} \log |z|^2$ which is the required short-distance singularity. Hence, one has correctly$-4 \del_z \del_{\ov z}\, g(z,\zb)=\dd^{(2)}(z) -\frac{1}{A}$.

As familiar by now, to get the renormalized Green's function one must subtract the short-distance singularity $-\frac{1}{4\pi} \log\frac{\m^2|z|^2}{4}\,{\bf 1}_{2\times 2}$ and take the limit $z\to 0$ with the result \be\label{GRflattorus}
\wt G^{m=0}_{\rm R}(x)=-\frac{1}{2\pi}  \log\Big( \frac{2\pi}{a\m} \,q^{\frac{1}{4}}\sum_{n=0}^\infty (-)^n (2n+1)q^{n(n+1)}\Big)\ {\bf 1}_{2\times 2}
\quad , \quad
q=e^{-\pi b/a}
\quad , \quad \text{(flat torus)} \ .
\ee
Obviously, this does not depend on the point $x$ on the torus, but it depends quite non-trivially on the shape of the torus (i.e. on $b/a$), as well as on its volume ($(2\pi)^2 a b$).

What about $\wt S_{\rm R}^{m=0}(x)$~? Equation \eqref{GRzetaSRzetarel} states that in the massive theory  $S_{\rm R}=m\g_* G_{\rm R}$. Now
from our explicit study of the flat torus we know that the zero-mode parts of $G$ and $S$ are related as $S^{(0)}=\frac{1}{m}\big(\p_{\vec{0}}\p_{\vec{0}}^\dag - \p^*_{\vec{0}}(\p^*_{\vec{0}})^\dag \big)
=\frac{1}{m}\g_*\big(\p_{\vec{0}}\p_{\vec{0}}^\dag + \p^*_{\vec{0}}(\p^*_{\vec{0}})^\dag \big)
=m\g^* G^{(0)}$, a relation we will also prove in more generality below. Subtracting this relation from 
 $S_{\rm R}=m\g_* G_{\rm R}$ we get  $\wt S_{\rm R}=m\g_* \wt G_{\rm R}$. Now both, $\wt S_{\rm R}$ and $ \wt G_{\rm R}$ have a smooth limit as $m\to 0$, and we can conclude that $\wt S_{\rm R}^{m=0}(x)=0$. Below, in subsection \ref{allgSvar}, we will study the zero-mode parts of the Green's functions more generally and show that $\wt S_{\rm R}^{m=0}(x)=0$ always holds.

\newpage

\setlength{\baselineskip}{.59cm}
\section{Determining the gravitational action}

We now want to determine the variation of the effective gravitational action under infinitesimal conformal variations and write this variation in such a form that it can be integrated to obtain the effective action $S_{\rm grav}[g,\wh g]$. We will do this in an expansion in powers of $m^2$. For this expansion to make sense,  all quantities entering this expansion must have a well-defined limit as $m\to 0$, i.e.~the expansion is done in terms of quantities defined in the massless theory. Thus, for genus one and larger, one must carefully subtract the zero-mode parts from all Green's functions and zeta-functions. 

We will  first rewrite the variation of the gravitational action in terms of the renormalized Green's function $G_{\rm R}$ and its variation $\dd G_{\rm R}$. The latter can be re-expressed in  an expansion to all orders in $m^2$ involving higher Green's functions   and zeta functions, still of the massive theory. Proceeding this way first, provides a more streamlined presentation while being  already a useful result for spherical topology where the $m\to 0$ limit exists for all Green's functions and zeta-functions without the need to subtract any zero-mode parts. Then, however,  we repeat the analogous computation on arbitrary Riemann surfaces, taking into account the presence of zero-modes that must first be subtracted from these quantities. This uses the properties of the zero-mode projectors worked out in subsection \ref{zeromodesec}. The corresponding computations is a bit more cumbersome, but the final result will be only slightly more complicated. Finally, we obtain the gravitational action as an infinite expansion in powers of $m^2$, with a finite radius of convergence, valid on a manifold of arbitrary genus. In particular, at order $m^2$,  we obtain a Mabuchi-like term, as well as terms that are multi-local in $\s$ and involve the Green's functions for the reference metric $\wh g$, as well as a finite number of area-like parameters.
 
\subsection{Expressing the variation of $S_{\rm grav}$ in terms of $G_{\rm R}$}

Recall our formula \eqref{gravac4} for the gravitational action  in terms of $\zeta(0)$ and  $\zeta'(0)$ which gives
\be\label{sgravvar}
\dd S_{\rm grav}=\frac{1}{4}\dd\zeta'(0)+\frac{1}{4}\log \m^2\ \dd\zeta(0) \ .
\ee
The variations of the zeta functions under infinitesimal conformal rescalings with $\dd\s(x)$ have been worked out  above and are given in \eqref{zetaatzero} and \eqref{zetprimeatzero} or \eqref{deltazetaatzero}. 
$\dd\zeta(0)$ was shown to be simply $-\frac{m^2}{\pi} \int\sqrt{g}\,\dd\s=-\frac{m^2}{2\pi} \dd A$. 
Combining \eqref{deltazetaatzero} with \eqref{zetapluszero} 
and \eqref{zeta-pluszeta-prime} we get
\be\label{deltazetazeroprime2}
\dd\zeta'(0)=-\int\sqrt{g}\dd\s \Big( \frac{\cR}{12\pi}+\frac{m^2}{\pi}\Big) 
- 2 m^2 \int\sqrt{g}\, \dd\s \Big( \tr G_{\rm R} +\frac{1}{2\pi}(2\g-\log\m^2) \Big)\ .
\ee
Thus
\be\label{sgravvar-bis}
\dd S_{\rm grav}= -\frac{2\g+1}{8\pi}\, m^2\dd A    -\frac{1}{48\pi} \int \sqrt{g}\, \dd\s\, \cR 
-\frac{m^2}{2} \int\sqrt{g}\,\dd\s  \tr  G_{\rm R}\ .
\ee
The term $\sim \dd A$ is usually referred to as the variation of a cosmological constant term, while the term $\sim \int \sqrt{g}\, \dd\s\, \cR$ is the variation of the Liouville action  \eqref{Liouvaction}, see  \eqref{Liouvillevar1}~:
 \be\label{Liouville}
\int\sqrt{g}\, \dd\s\, \cR = \dd S_{\rm Liouville} \ .
\ee
Note that the coefficient of $\dd S_{\rm Liouville}$ in $\dd S_{\rm grav}$  is exactly $\frac{1}{2}$ the one that occurs in the gravitational action of a bosonic scalar matter field. It is interesting to trace back how this coefficient occurs. 
The term $\sim \cR$ originates from $\zeta_+(0,x,x)$ which was given in terms of the heat kernel coefficient  $a_1\equiv F_1(x,x)$.
For the bosonic scalar field and scalar Laplacian $a_1=\frac{\cR}{6}$. At present, $a_1$ is $\frac{\cR}{6}-\frac{\cR}{4}=-\frac{\cR}{12}$. But because we are dealing with fermions, the gravitational action has an overall minus sign, so that in the end one gets a $+\frac{\cR}{12}$, i.e. exactly one half the bosonic result.\footnote{Of course, we also had the $\frac{1}{4}$ multiplying the $\dd\zeta'(0)$ and $\dd\zeta(0)$ instead of a $\frac{1}{2}$ in the bosonic case. But this extra $\frac{1}{2}$ is offset by a factor $2$ coming from the Dirac traces.}
Of course, this is consistent with the well-known central charges 1 and $\frac{1}{2}$ of the conformal algebra for a single massless scalar and massless Majorana fermion. We conclude that
\beb\label{sgravvar4}
\dd\Big[ S_{\rm grav} +\frac{1}{48\pi} S_{\rm Liouville} +\frac{m^2}{8\pi} \big(2\g+1\big) A \Big]
=-\frac{m^2}{2} \int\sqrt{g}\,\dd\s  \tr G_{\rm R} \ .
\eeb
It remains to characterise the right-hand side of this equations and express it as a total variation of some appropriate quantity. Obviously, we have (using $\dd\sqrt{g}=2\sqrt{g}\, \dd\s$)
\be\label{ddsigmaGR}
 \int\sqrt{g}\,\dd\s  \tr G_{\rm R} 
 = \frac{1}{2} \ \dd  \int\sqrt{g}\,  \tr G_{\rm R} -\frac{1}{2}  \int\sqrt{g}\,  \dd\tr G_{\rm R} \ ,
 \ee
 so that we can rewrite \eqref{sgravvar4} as
 \be\label{sgravvar4-2}
\dd\Big[ S_{\rm grav} +\frac{1}{48\pi} S_{\rm Liouville} +\frac{m^2}{8\pi} \big(2\g+1\big) A  +\frac{m^2}{4}  \int\sqrt{g}\,  \tr G_{\rm R}\Big]
= \frac{m^2}{4} \int\sqrt{g}\,\dd \tr G_{\rm R} \ .
\ee


\subsection{Conformal variations of the Green's functions}

We will need to study the variations of the Green's functions $S(x,y)$ and $G(x,y)$ under conformal transformations. Since $G(x,y)=\int d^2 z\sqrt{g(z)}\, S(x,z)S(z,y)$ it will be enough in the first place to obtain the variation of $S(x,y)$, from which the variation of $G(x,y)$ can then be deduced. In this subsection we assume again that $m\ne 0$ so that there are no zero-modes of the Dirac operator $D=i\Nsl +m\g_*$. Generically there are, of course, zero-modes of $i\Nsl$ as discussed  in subsection \ref{zeromodesec}. In the next subsection, when we want to do a small mass expansion,  we will explicitly separate  the contributions of these zero-modes of $i\Nsl$, that contribute terms of order $\frac{1}{m}$ to the Green's function $S(x,y)$ and terms of order $\frac{1}{m^2}$ to the Green's function $G(x,y)$.  But at present, we will not need to do this separation, as long as $m\ne 0$.

\subsubsection{Conformal variations of $S(x,y)$, $G(x,y)$ and $G_{\rm R}(y)$}

Recall that $S(x,y)$ is the solution of $D_x S(x,y)\equiv (i\Nsl_x+m\g_*)S(x,y)=\dd^{(2)}(x-y)/\sqrt{g}$, see eq. \eqref{Sdef}. We want to determine  the variation of $S$ under a conformal rescaling. Consider two metrics $g$ and $g'$ related by an infinitesimal conformal rescaling $g'=e^{2\dd\s}g$. Then, of course $\sqrt{g'}=e^{2\dd\s}\sqrt{g}$ and from \eqref{spinorderconfgeneraldim-4},  ${\Nsl}{\ '}=e^{-3\dd\s/2}\,\Nsl \ e^{\dd\s/2}$. Thus the corresponding Green's functions $S$ and $S'$ satisfy
\ba\label{SandShatsol}
(i\Nsl_x+m\g_*)S(x,y)&=&\frac{\dd^{(2)}(x-y)}{\sqrt{g}} \ , \nonumber\\
\big( i e^{-3\dd\s(x)/2}\,\Nsl \ e^{\dd\s(x)/2} + m \g_* \big) S'(x,y) &=&\frac{\dd^{(2)}(x-y)}{\sqrt{g}} \, e^{-3\dd\s(x)/2}e^{-\dd\s(y)/2}\ .
\ea
It looks as if for $m=0$ one would simply have $S'(x,y)=e^{-\dd\s(x)/2} S(x,y) e^{-\dd\s(y)/2}$. But for $m=0$ we must precisely subtract the zero-mode contribution and deal with $\wt S$ and $\wt S'$ instead and then the relation becomes again more complicated due to the conformal transformation of the zero-mode projector. This will be dealt with in the next two subsections. But this remark motivates the following definition of $\wh\dd S$~:
\ba\label{hatddS}
S'(x,y)\equiv S(x,y)+\dd S(x,y)&=&e^{-\dd\s(x)/2} S(x,y) e^{-\dd\s(y)/2} +\wh\dd S(x,y)\nonumber\\
\Leftrightarrow\quad
\dd S(x,y)&=&- \frac{1}{2} \big( \dd\s(x)+\dd\s(y)\big) S(x,y)+\wh\dd S(x,y)\ .
\ea
We then insert this $S'(x,y)$ into the second equation \eqref{SandShatsol} and develop to first order in the variation. This yields
\be\label{hatdeltaSeq}
D_x\,  \wh\dd S(x,y) \equiv (i\Nsl_x+m\g_*) \wh\dd S(x,y)=-m \dd\s(x) \, \g_* S(x,y) \ ,
\ee
which is solved by multiplying with the Green's function of $D$ and integrating~:
\be\label{deltaSt1-2}
\int \d^2 z\, \sqrt{g(z)}\, S(x,z)\, D_z\,  \wh\dd S(z,y)=
-m\int \d^2 z\, \sqrt{g(z)} \, \dd\s(z)\, S(x,z)\, \g_*\,  S(z,y) \ .
\ee
Using the hermiticity of $D_z$  the terms on the left-hand side is $\int \d^2 z \sqrt{g(z)}\, (D_z S(z,x))^\dag\,  \wh\dd S(z,y)$ $=\wh\dd S(x,y)$, so that
\be\label{hatddS-2}
\wh\dd S(x,y)=-m\int\d^2 z \sqrt{g}\, \dd\s(z)\, S(x,z)\g_*  S(z,y) \ ,
\ee
and
\be\label{deltaST3}
\dd S(x,y) = - \frac{1}{2} \big( \dd\s(x)+\dd\s(y)\big) S(x,y)
-m\int\d^2 z \sqrt{g}\, \dd\s(z) \, S(x,z)\g_* S(z,y) \ .
\ee

One can now continue from this to obtain  the variation of $S_{\rm R}(x)$ by subtracting the variation of the singular part of $S(x,y)$, and then deduce the variation of $\tr \g_* S_{\rm R}=\tr G_R$.  We have tried this avenue, but it turns out not to be the most straightforward one. Instead, we will determine the variation of $G(x,y)$ from the variation of $S(x,y)$ and then deduce the variation of $G_{\rm R}(x)$  by subtracting the variation of the singular part of $G(x,y)$. Recall that
$G(x,y)=\int\d z\, S(x,z) S(z,y)$,
where here and in the following, to simplify the notations, we abbreviate the integration measure $\d^2 z \sqrt{g(z)}$ by writing simply $\d z$. Of course, one should not forget then that $\d z$ changes under a conformal rescaling~:
\be\label{condensedint}
\int\d z \simeq \int \d^2 z \sqrt{g(z)} 
\quad , \quad 
\dd\, \d z = \d z\  2 \dd\s(z) \ .
\ee
The variation of $G(x,y)$ is then given by
\ba\label{Gxyvar}
&&\hskip-1.cm\dd G(x,y)= \dd \int d z\, S(x,z)S(z,y)
=\int\d z\, \Big[ 2\dd\s(z) S(x,z)S(z,y) + \dd S(x,z) S(z,y) + S(x,z) \dd S(z,y) \Big] \nonumber\\
&&=\int\d z\, \Big[ \big(\dd\s(z)-\frac{\dd\s(x)}{2}-\frac{\dd\s(y)}{2}\big) S(x,z)S(z,y) + \wh\dd S(x,z) S(z,y) + S(x,z) \wh\dd S(z,y) \Big] \ .\nonumber\\
\ea
Now the short-distance singularity of $G(x,y)$ is $G^{\rm sing}(x,y)=-\frac{1}{4\pi} \log\frac{\m^2\ell^2(x,y)}{4} {\bf 1}_{2\times 2}$ and we need to compute its variation
\be\label{Gsingvar}
\dd G^{\rm sing}(x,y)=-\frac{1}{4\pi}  \frac{\dd\ell^2(x,y)}{\ell^2(x,y)}\ {\bf 1}_{2\times 2} \ .
\ee
Actually, we only need this in the limit $x\to y$, i.e. when $x$ and $y$ are only separated by an infinitesimal $\d x$. But then the (geodesic) distance between the two points is just given by $\ell^2(x, x+\d x)=g_{\m\n}\d x^\m \d x^\n= e^{2\s(x)} \wh g_{\m\n}\d x^\m \d x^\n$. Obviously then $\dd\ell^2(x,x+\d x)=2 \dd\s(x) \ell^2(x,x+\d x)$. Somewhat more generally, for $x$ close to $y$, one can show that (see e.g. the appendix A1 of \cite{BF})~:
\be\label{Gsingvar2}
\frac{\dd\ell^2(x,y)}{\ell^2(x,y)} =  \dd\s(x)+\dd\s(y)+ {\cal O}(\ell^2) \ .
\quad \Rightarrow\quad
\dd G^{\rm sing}(x,y)=-\frac{1}{4\pi} \big(\dd\s(x)+\dd\s(y)+ {\cal O}(\ell^2)\big)\ {\bf 1}_{2\times 2} \ .
\ee
We see that the variation of the singular part is non-singular. Hence it must be that $\dd G(x,y)$ as given in \eqref{Gxyvar} also is non-singular. Indeed, the terms involving $\wh\dd S$ are easily seen to be non-singular, so only the first term in \eqref{Gxyvar} could potentially be singular as $x\to y$. Indeed  $S(y,z)S(z,y)$ is divergent as $\sim \frac{1}{\ell^2(z,y)}$ when the integration variable $z$ gets close to $y$. But then the factor $\dd\s(z)-\dd\s(y)$ vanishes as $z-y\sim\ell(z,y)$ so that the integrand behaves as $\frac{1}{\ell(z,y)}$ which is an integrable singularity in 2 dimensions, leading to a finite integral over $\d^2 z$. We may thus take the limit $x\to y$~:
\be\label{Gxyvarcoinc}
\lim_{x\to y} \dd G(x,y)
=\int\d z\,\Big[ \big(\dd\s(z)-\dd\s(y)\big) S(y,z)S(z,y) + \wh\dd S(y,z) S(z,y) + S(y,z) \wh\dd S(z,y) \Big] \ ,
\ee
and subtracting \eqref{Gsingvar2} with $x=y$ we get
\be\label{GRvar}
\dd G_{\rm R}(y)
=\int\d z\, \Big[ \big(\dd\s(z)-\dd\s(y)\big) S(y,z)S(z,y) + \wh\dd S(y,z) S(z,y) + S(y,z) \wh\dd S(z,y) \Big] 
+\frac{\dd\s(y)}{2\pi} \ {\bf 1}_{2\times 2}\ .
\ee
If we now take the trace and integrate over $y$, the first term in the integrand is odd under the exchange of $y$ and $z$ and thus does not contribute. Using the cyclicity of the trace and exchanging again $y$ and $z$, the second and third terms are easily seen to be identical. We arrive at
\be\label{GRvarint}
\int\d y\,\, \dd \tr G_{\rm R}(y)
=2 \int\d y\, \d z\, \tr S(y,z) \,\wh\dd S(z,y) 
+\frac{\dd A}{2\pi}\ .
\ee
We now insert the explicit expression \eqref{hatddS-2} for $\wh\dd S$ to get
\ba\label{GRvarint2}
\int\d y \,\dd \tr G_{\rm R}(y)
&=& \frac{\dd A}{2\pi}  -2m \int\d y\, \int\d z\,\int\d u \,\dd\s(u)
\tr S(y,z)  S(z,u)\g_* S(u,y)  \nonumber\\
&=& \frac{\dd A}{2\pi}  -2 m \int\d y\, \d u \, \dd\s(u) \tr G(y,u) \g_* S(u,y)  \ . 
\ea
In the previous section, we have shown in \eqref{StraceG} that $\tr \g_* S(x,y) =  m\tr G(x,y)$. The argument was based on the relation  $S(x,y)=(i\Nsl_x +m\g_*)G(x,y)$  and the fact that $G$ only contains the matrices ${\bf 1}_{2\times 2}$ and $\g_*$ and that in the end only  ${\bf 1}_{2\times 2}$ has a non-vanishing trace. One can show in completely the same way that
\be\label{SGtraceGG}
\tr \g_* S(x,y) G(u_1,v_1)\ldots G(u_n,v_n)=  m\tr G(x,y) G(u_1,v_1)\ldots G(u_n,v_n)\ .
\ee
This allows us to rewrite \eqref{GRvarint2} as
\be\label{GRvarint2-2}
\int\d y \,\dd \tr G_{\rm R}(y)
= \frac{\dd A}{2\pi}  -2 m^2 \int\d y\, \d u \, \dd\s(u) \tr G(u,y)   G(y,u) 
= \frac{\dd A}{2\pi}  -2 m^2 \int \d u \, \dd\s(u) \tr G_2(u,u)\ ,
\ee
where we used the orthonormality of the eigenfunctions to rewrite the right-hand side in terms of a  higher Green's function $G_2$~:
\be\label{GGG2}
\int\d y\,  G(u,y)G(y,v) =\sum_n \frac{1}{\l_n^4} \big(\chi_n(u)\chi_n^\dag(v) + \f_n(u)\f_n^\dag(v)\big) \equiv G_2(u,v) \equiv \zeta_+(2,u,v) \ .
\ee
Let us note for later reference that one can equally easily show that
\be\label{SGSG2}
\int\d u\, \d v\, S(x,u)G(u,v) S(v,y)= 
G_2(x,y) \equiv \zeta_+(2,x,y) \ .
\ee
Finally, inserting \eqref{GRvarint2-2} into \eqref{sgravvar4-2},  we get
 \be\label{sgravvar4-3a}
\dd\Big[ S_{\rm grav} +\frac{1}{48\pi} S_{\rm Liouville} +\frac{\g}{4\pi}  m^2 A  +\frac{m^2}{4}  
\int\d y\,  \tr G_{\rm R}(y)\Big]
= -\frac{m^4}{2}  \int\d u\, \dd\s(u) \tr G_2(u,u)   \ ,
\ee 
If there are no zero-modes of $i\Nsl\,$, which is the case for spherical topology, then $S$ and $G$ have finite limits as $m\to 0$ and one can assert that the term on the right-hand side of \eqref{sgravvar4-3a} is of order $m^4$. In this case, if we satisfy ourselves with an order $m^2$ calculation, we may just drop the right-hand side, and moreover replace $G_{\rm R}$ by the corresponding quantity in the massless theory. However, for genus one or larger, we have zero-modes of $i\Nsl$ and one must separate the zero-mode and the non zero-mode parts to isolate the contributions that are truly of order $m^2$. This separation will be the subject of   subsection \ref{allgSvar}.

\subsubsection{Conformal variation of the higher Green's function $G_n,\ n\ge 2$}

But let us go  on and determine $ \int\d u\,\dd \tr G_n(u,u)$ for $n\ge 2$. We have
\be\label{Gsubn}
G_n(u_1,u_1) = \int \d u_2 \dots \d u_n\, G(u_1,u_2) G(u_2,u_3) \ldots G(u_n,u_1) \ ,
\ee
\vskip-3.mm
\noindent
so that
\vskip-7,mm
\ba\label{Gnvar1}
&&\hskip-0.5cm\int\d u_1\ \dd \tr G_n(u_1,u_1))
=\int\d u_1\ldots \d u_n \Big[ 2\sum_{i=2}^n \dd\s(u_i) \tr  G(u_1,u_2)  \ldots G(u_n,u_1)
 \nonumber\\
&&\hskip6.cm+ \sum_{i=1}^n \tr G(u_1,u_2) \ldots \dd G(u_i,u_{i+1})\ldots G(u_n,u_1) \Big] 
\nonumber\\
&&\hskip-0.5cm=
\int\d u_1 \Big[ 2(n-1) \dd\s(u_1) \tr  G_n(u_1,u_1) 
+ n \int \d u_1\, \d u_2\, \tr \dd G(u_1,u_2)  G_{n-1}(u_2,u_1) \Big] ,
\ea
where in the last step we used the cyclicity of the trace.
The variation  $\dd G(z,y)$ was determined in \eqref{Gxyvar} and it follows that
\ba\label{G2var2}
&&\int\d u_1\, \d u_2\,  \tr \big(\dd G(u_1,u_2) \big) \, G_{n-1}(u_2,u_1)\nonumber\\
&&=
\int \d v \,d u_1\, \d u_2   \big( \dd\s(v)-\frac{\dd\s(u_1)+\dd\s(u_2)}{2}\big) \tr S(u_1,v)S(v,u_2)   G_{n-1}(u_2,u_1)\nonumber\\
&&+\int \d v \,d u_1\, \d u_2  \tr \Big( \wh\dd S(u_1,v)S(v,u_2)+S(u_1,v)\wh\dd S(v,u_2)\Big) 
G_{n-1}(u_2,u_1)
\ .
\ea
The different pieces of the first term on the right-hand side can be rewritten, using  the cyclicity of the trace and an identity analogous to \eqref{SGSG2}, as 
$\int \d v \d u_1\d u_2\,   \dd\s(v) \tr S(u_1,v)S(v,u_2)   G_{n-1}(u_2,u_1)
= \int \d v\, \dd\s(v)\tr G_n(v,v)$, and similarly for  $\int \d v\d u_1\d u_2\,   \dd\s(u_1) \tr S(u_1,v)S(v,u_2)   G_{n-1}(u_2,u_1) =$\break $\int \,d u_1 \dd\s(u_1) \tr G_n(u_1,u_1)$, etc, and we see that the terms on the right-hand side of \eqref{G2var2} involving the explicit $\dd\s$  cancel. Hence only the terms involving $\wh\dd S$ remain.
Thus
\ba\label{G2var3}
&&\hskip-0.7cm\int\d u_1\, \d u_2 \tr \big(\dd G(u_1,u_2) \big) \, G_{n-1}(u_2,,u_1)\nonumber\\
&&\hskip-0.7cm=
-m\hskip-1,mm\int\hskip-1.mm \d v \, \d w\,d u_1 \d u_2  \dd\s(w) \tr \Big(  S(u_1,w)\g_* S(w,v)S(v,u_2)+S(u_1,v)S(v,w)\g_* S(w,u_2)\Big)
G_{n-1}(u_2,u_1)\nonumber\\
&&=
-m\int   \d w\,d u_1 \dd\s(w)\, \tr \Big( \g_* G_n(w,u_1) S(u_1,w) + \g_* S(w,u_1) G_n(u_1,w))\Big) \nonumber\\
&&= -2m^2 \int\d w\, \dd\s(w)  \tr G_{n+1}(w,w)
\ .
\ea
We see that we can rewrite \eqref{Gnvar1} as
\be\label{Gnvar4}
\int\d u\ \dd \tr G_n(u,u)
- 2(n-1) \int\d u\, \dd\s(u) \tr  G_n(u,u) 
=-2 n \,m^2 \int\d u\, \dd\s(u)  \tr G_{n+1}(u,u)\ .
\ee
Now, on the left-hand side we combine $\int\d u\ \dd \tr G_n+2 \int\d u\, \dd\s(u) \tr  G_n
=\dd \int\d u\, \tr  G_n$  so that, after dividing by $2n$~:
\be\label{Gnvar5}
 \int\d u\,\dd\s(u) \tr  G_n=\dd \int\d u\, \frac{1}{2n} \tr G_n
 + m^2 \int\d u\, \dd\s(u)  \tr G_{n+1} \ ,
\ee
where we dropped the arguments, all Green's functions being obviously evaluated at coinciding points $u$.
If we multiply this equation by $m^{2n}$ and iterate it $N-n+1$ times we get
\be\label{Gnvar6}
m^{2n} \int\d u\,\dd\s(u) \tr  G_n=\dd \int\d u\, \sum_{k=n}^N\frac{m^{2k}}{2k} \tr G_k
 + m^{2(N+1)} \int\d u\, \dd\s(u)  \tr G_{N+1} \ .
\ee
We now use this relation for $n=2$ to rewrite the right-hand side of \eqref{sgravvar4-3a} to get
\ba\label{sgravvar4-5}
\dd\Big[ S_{\rm grav} +\frac{1}{48\pi} S_{\rm Liouville} +\frac{\g}{4\pi}  m^2 A  +\frac{1}{4}  
\int\d^2 y \sqrt{g}\,  \Big(  m^2\tr G_{\rm R} (y)
+\sum_{k=2}^N \frac{m^{2k}}{k}  \tr G_k(y,y)\Big) \Big]&& \nonumber\\
 =-\frac{m^{2(N+1)}}{2}  \int\d^2 y\, \sqrt{g}\, \dd\s(y) \tr G_{N+1}(y,y)   \ ,&&
\ea
where on the left-hand side one can replace $\int\d^2 y \sqrt{g} \tr G_k(y,y)=\zeta(k)$.

\subsubsection{Variation of the gravitational action to all orders in $m^2$ for spherical topology}

Contrary to what this expansion \eqref{sgravvar4-5} might suggest, it is important to realise that the quantities $G_{\rm R}$ and $\zeta(k)$ are the quantities computed in the massive theory. However, in the absence of zero-modes of $i\Nsl\,$ (i.e.~for spherical topology), they all have a finite limit as $m$ goes to zero. Below, in subsubsection \ref{confvarhighersec}, we will give a precise bound on the remainder term on the right-hand side of \eqref{sgravvar4-5}. In particular, 
for $\frac{m^2 A}{a_*^2}<1$, where $a_*$ is the smallest eigenvalue of $D$ for zero mass and unit area $A=1$,  this term goes to zero as $N\to\infty$. Hence
 \be\label{sgravvar4-8-sphere}
\dd\Big[ S_{\rm grav}^{\rm sphere} +\frac{1}{48\pi} S_{\rm Liouville} +\frac{\g}{4\pi} m^2 A  +  \frac{m^2}{4}  \int\sqrt{g}\,  \tr G_{\rm R}(y) 
+\sum_{k=2}^\infty \frac{m^{2k}}{4k}   \zeta(k) \Big] =0
\ee
It is relatively straightforward to re-express $G_{\rm R}$ and $ \zeta(k)$ in terms of the same set of quantities in the massless theory. We will do this below for the case of arbitrary genus.


\subsection{Variation of the gravitational action for arbitrary genus\label{allgSvar}}


\subsubsection{Zero-mode and non-zero-mode parts of the Green's functions}

As discussed in subsection \ref{zeromodesec}, for genus one or larger, there are zero-modes $\p_{0,i}$ of $i\Nsl\,$. They may be chosen as having definite chirality, i.e.~be eigenstates of $\g_*$. There are as many positive as negative chirality zero-modes. For the arbitrary torus there is exactly one zero-mode of each chirality and we have obtained them explicitly. In general, since $i\Nsl\, \p_{0,i}=0$ we have $D\p_{0,i}=m\g_* \p_{0,i}=\pm m \p_{0,i}$, i.e~$\l_0=m$. Recall the spectral decomposition \eqref{SGeigen} of the Green's functions in terms of the eigenfunctions and eigenvalues. It follows that the zero-mode parts of the Green's functions $S$ and $G$ behave as $\frac{1}{m}$ and $\frac{1}{m^2}$ and do not have a limit as $m\to 0$. Since we want to do a small-$m$ expansion, we have to separate the Green's functions into their zero-mode and their non-zero-mode parts. In terms of the projectors  $P_0$ and $Q_0$ on the zero-modes defined in \eqref{zeromodeproj} and \eqref{zeromodeQ}, the zero-mode contribution to the Green's functions $G$ and $S$ simply are $\frac{P_0}{\l_0^2}=\frac{P_0}{m^2}$ and $\frac{Q_0}{\l_0}=\frac{Q_0}{m}$. Obviously then, one has the decomposition
\ba\label{StildeGtildegen}
G(x,y) &=& G_{(0)}(x,y)+\wt G(x,y) \quad , \quad  
G_{(0)}(x,y)=\frac{P_0(x,y)}{m^2} \ , \nonumber\\
S(x,y) &=& S_{(0)}(x,y)+\wt S(x,y) \quad , \quad 
S_{(0)}(x,y)=\frac{Q_0(x,y)}{m} \ ,
\ea
where $\wt G(x,y)$ and $\wt S(x,y)$ are the contributions of the non-zero-modes for which the $\l_n$ have finite limits as $m\to 0$. Hence, $\wt G(x,y)$ and $\wt S(x,y)$ have finite limits as $m\to 0$, which obviously is not the case for $G^{(0)}$ and $S^{(0)}$.

As already noted, the zero-modes are orthogonal to the non-zero modes so that
\ba\label{zero-nonzeroortho}
&&\int\d z\,S_{(0)}(x,z)\wt S(z,y) = 0 \quad  , \quad
\int\d z\, G_{(0)}(x,z)\wt G(z,y) = 0  \ , \nonumber\\
&&\int\d z\, S_{(0)}(x,z)\wt G(z,y) = 0 \quad  , \quad
\int\d z\, G_{(0)}(x,z)\wt S(z,y) = 0 \ .
\ea
Furthermore (cf \eqref{PQrel})
\be\label{S0S0isG0}
\int\d z\,S_{(0)}(x,z) S_{(0)}(z,y)
=\frac{1}{m^2} \int\d z \, Q_0(x,z) Q_0(z,y)
=\frac{1}{m^2} P_0(x,y)=G_{(0)}(z,y) \ .
\ee
We also have $G(x,y)=\int \d z\, S(x,z) S(z,y)$. If we substitute $S=\wt S+ S_{(0)}$ and use the orthogonality between the zero and non-zeromodes, eq.~\eqref{zero-nonzeroortho}, we get
\be\label{StildeStildeGtilde}
\wt G(x,y)=\int\d z\, \wt S(x,z) \wt S(z,y)\ .
\ee
We can similarly separate all higher Green's functions as $G_n(x,y)=G_n^{(0)}(x,y) + \wt G_n(x,y)$ and we have separate relations for the zero-mode and the non-zero-mode parts~:
\ba\label{Gntilderels}
\wt G_{n+1}(x,y)=\int\d z\, \wt G(x,z) \wt G_n(z,y) 
&=&\int\d z\, \d u\, \wt S(x,z) \wt G_n(z,u) \wt S(u,y)\ ,\nonumber\\
G_{n+1}^{(0)}(x,y)=\int\d z\,  G_{(0)}(x,z) \wt G_n^{(0)}(z,y) 
&=&\int\d z\, \d u\, S_{(0)}(x,z) G_n^{(0)}(z,u) S_{(0)}(u,y) .\ \ \
\ea
Finally note that the zero-mode parts satisfy $i\Nsl \,S_{(0)}=i\Nsl\, G_{(0)} =0$ and recall from \eqref{PQrel} that $Q_0=\g_* P_0$ so that 
$S_{(0)} =  m\g_* G_{(0)} = (i\Nsl + m\g_*) G_{(0)}$. Subtracting this  identity from 
$S=(i\Nsl + m\g_*) G$ results in
\be\label{StildeGtilderel}
\wt S(x,y)=(i\Nsl_x + m\g_*) \wt G(x,y)\equiv D_x \wt G(x,y) \ ,
\ee
so that the zero-mode parts and the non-zero-mode parts satisfy this same relation separately. Of course, all these relations also follow directly from the spectral representations of the Green's functions.

We now want to rewrite the various relations of the previous subsection in terms of these $\wt S(x,y)$ and $\wt G(x,y)$ for which the $m\to 0$ limits exist. In the end we will find that we obtain almost the same result as in \eqref{sgravvar4-5} or \eqref{sgravvar4-8-sphere}  except that all quantities are replaced by the the corresponding quantities with tildes referring to the non-zero-modes. This result can be understood as being due to the fact that when varying the gravitational action the eigenvalue $\l_0$ corresponding to the zero-modes does not change and that the zero-mode and non-zero-mode contributions are orthogonal.

But let us proceed.
First note that the zero-mode pieces do not contribute to the short-distance singularities. Hence  $G(x,y)$ and $\wt G(x,y)$ have the same short-distance singularity $-\frac{1}{4\pi} \log\frac{\m^2\ell^2(x,y)}{4}\, {\bf 1}_{2\times 2}$. Similarly $S(x,y)$ and $\wt S(x,y)$ have the same short-distance singularity  $-\frac{1}{4\pi} (i\Nsl_x+m\g_*)\log\frac{\m^2\ell^2(x,y)}{4}$. Thus
\ba\label{Sregtilde}
S^{\rm reg}(x,y) &=& S_{(0)}(x,y)
+\wt S(x,y) +\frac{1}{4\pi} (i\Nsl_x+m\g_*)\log\frac{\m^2\ell^2(x,y)}{4}
= \frac{Q_0(x,y)}{m} + \wt S^{\rm reg}(x,y) \ , \nonumber\\
G^{\rm reg}(x,y) &=& G_{(0)}(x,y) 
+\wt G(x,y) +\frac{1}{4\pi} \log\frac{\m^2\ell^2(x,y)}{4}\, {\bf 1}_{2\times 2}
=\frac{P_0(x,y)}{m^2}  + \wt G^{\rm reg}(x,y) \ .
\ea
By taking the coincidence limit $x\to y$ we then get
\be\label{SRSRtilde}
S_{\rm R}(y)= \frac{Q_0(y,y)}{m}  + \wt S_{\rm R}(y) \quad , \quad
G_{\rm R}(y)= \frac{P_0(y,y)}{m^2}  + \wt G_{\rm R}(y) \ .
\ee
Also, since $Q_0=\g_* P_0$, the relation \eqref{GRzetaSRzetarel}  translates into
\be\label{SRGRtilderel}
\wt S_{\rm R} =m \g_* \wt G_{\rm R} \ .
\ee
But both, $\wt S_{\rm R}$ and $\wt G_{\rm R}$ have a smooth limit as $m\to 0$, and we find that $\wt S_{\rm R}$ vanishes in the massless theory~:
\be\label{SRtildem=0}
\wt S^{m=0}_{\rm R}(x)=0 \ .
\ee

We will need the variations of the zero-mode parts of the Green's functions $G_{(0)}(x,y)$ and $S_{(0)}(x,y)$ under conformal transformations. As we have just seen, the latter are given in terms of the zero-mode projectors $P_0(x,y)$ and $Q_0(x,y)=\g_* P_0(x,y)$. But the conformal variations of $P_0(x,y)$ have been studied in some detail in subsection \ref{confvarzeroproj}, see in particular \eqref{deltaP0} and \eqref{P0confvar2}.

\subsubsection{Expressing the variation of the gravitational action in terms of $\wt G_{\rm R}$}

The variation of the gravitational action, eq.~\eqref{sgravvar4} contains the term 
$-\frac{m^2}{2} \int\sqrt{g}\,\dd\s  \tr G_{\rm R}$ on its right-hand side. We now use \eqref{SRSRtilde} and \eqref{P0confvar2} to rewrite this as
\ba
&&-\frac{m^2}{2} \int\d y\,\dd\s(y)  \tr G_{\rm R}(y)
=-\frac{m^2}{2} \int\d y\,\dd\s(y) 
\tr \Big( \frac{P_0(y,y)}{m^2}  + \wt G_{\rm R}(y)\Big) \nonumber\\
&&= -\dd\ \Big( \log\det {\cal P}_0
+\frac{m^2}{4} \int\d y\, \tr \wt G_{\rm R}(y)\Big) 
+\frac{m^2}{4} \int\d y\,\dd \tr \wt G_{\rm R}(y)
\ .
\ea
We can then rewrite our initial equation \eqref{sgravvar4-2} as
 \be\label{sgravvar4-tilde-1}
\dd\Big( S_{\rm grav} +\frac{1}{48\pi} S_{\rm Liouville} +  \log\det {\cal P}_0 +\frac{m^2}{8\pi} (2\g+1) A  +\frac{m^2}{4}  \int\sqrt{g}\,  \tr \wt G_{\rm R}\Big)
= \frac{m^2}{4} \int\sqrt{g}\,\dd \tr \wt G_{\rm R} \ .
\ee
This is as \eqref{sgravvar4-2} but with $G_{\rm R}$ replaced by $\wt G_{\rm R}$ and the extra term $\dd\,  \log\det {\cal P}_0$ on the left-hand-side.

At present, this term $\dd\,  \log\det {\cal P}_0$ has emerged from separating the zero-mode piece from $G_{\rm R}$. But we have seen in section 3 that, since $\dd \l_0=0$, we may write $\dd S_{\rm grav}$ from the beginning in terms of quantities that do not involve the zero-modes. 
Then in eq.~\eqref{sgravvar4}  it is directly $\wt G_{\rm R}$ that appears. So where does this term $\dd\,  \log\det {\cal P}_0$ then come from?
Recall that the variation of the gravitational action was $\dd S_{\rm grav}=\frac{1}{4}\dd\zeta'(0)+\frac{1}{4}\log \m^2\ \dd\zeta(0)$, cf \eqref{sgravvar}. First,  $\dd\zeta(0)$ only involved the residue of the pole of
$\zeta_+(s,x,x)$ which is the same as the residue of the pole of $\wt\zeta_+(s,x,x)$. Second,
when expressing $\dd\zeta'(0)$ directly in terms of the quantities without zero-modes
 as given in the second line of \eqref{zetprimeatzero}, namely
\be\label{deltazetaprimetilde}
\frac{1}{4}\dd\zeta'(0)=\frac{1}{2} \int\dd\s  \tr \wt\zeta_+(0,x,x)  -\frac{m^2}{2} \int\dd\s  \tr \wt\zeta_+^{\rm reg}(1,x,x)  - 2 m^2 C_+ \dd A \ ,
\ee
the last two terms on the right-hand side will be expressed  in terms of $\int\dd\s \wt G_{\rm R}$ with {\it no} zero-mode contribution subtracted. 
On the other hand, $\zeta_+(0,x,x)$ was given in terms of the heat kernel coefficient $\frac{1}{4\pi}F_1(x,x)$, cf \eqref{zetapluszero}, as $\zeta_+(0,x,x)=-\frac{1}{4\pi}  \big(\frac{\cR(x)}{12}+m^2\big)\, {\bf 1}_{2\times 2}$. Now the corresponding 
$\wt \zeta_+(0,x,x)$ is  obtained from the heat kernel $\wt K_+(t,x,x)$ with its zero-mode part subtracted. But the latter is just $e^{-m^2 t}P_0(x,x)= P_0(x,x) \big( 1+{\cal O}(t)\big)$. It follows that $\frac{1}{4\pi}\wt F_1(x,x)=\frac{1}{4\pi}F_1(x,x)- P_0(x,x)$ and, hence,
\be\label{tildezetapluszero}
\wt\zeta_+(0,x,x)=-\frac{1}{4\pi}  \big(\frac{\cR(x)}{12}+m^2\big)\, {\bf 1}_{2\times 2} -P_0(x,x)\ ,
\ee
so that
\be
\frac{1}{2}\int \d x\, \dd\s(x) \tr\wt\zeta_+(0,x,x)=-\frac{1}{48 \pi} \dd S_{\rm Liouville} -\frac{m^2}{8\pi} \dd A - \dd\, \log\det {\cal P}_0\ ,
\ee
where we used \eqref{P0confvar2}. Thus one gets again all the terms of  \eqref{sgravvar4-tilde-1}.
This latter computation also shows that this zero-mode related term $ \log\det {\cal P}_0$ is actually already present in the massless theory and should have appeared together with the Liouville action. But, as  mentioned earlier, in the massless theory one may redefine the matter partition function by an appropriate factor to precisely cancel this term.

Starting from \eqref{sgravvar4-tilde-1}, we can now proceed as in the previous subsection and  express $\int\sqrt{g}\,\dd \tr \wt G_{\rm R}$ in terms of a total variation and $\int\sqrt{g}\,\dd \tr \wt G_2(y,y)$, which then is again expressed as a total variation and a term   $\int\sqrt{g}\,\dd \tr \wt G_3(y,y)$, etc.  We have to be careful to subtract any zero-mode contributions where they might appear, but the orthogonality \eqref{zero-nonzeroortho} of the zero and non-zero mode parts of the Green's functions  essentially ensures that the variations of the non-zero-mode quantities only involve other non-zero-mode quantities.

To begin with, we have
\be\label{GRtildevar-2-a}
\dd \wt G_{\rm R}(y) =\lim_{x\to y} \ \dd \, \Big(\wt G(x,y) +\frac{1}{4\pi} \log\frac{\m^2\ell^2(x,y)}{4} \ {\bf 1}_{2\times 2} \Big) \ .
\ee
We already observed above that $\dd\big( \frac{1}{4\pi} \log\frac{\m^2\ell^2(x,y)}{4}\big) \sim_{x\to y} \frac{\dd\s(y)}{2\pi}$ is non-singular and, hence, the same must be true for $\dd \wt G(x,y)$. From \eqref{StildeStildeGtilde} we know that
$\wt G(x,y)=\int\d z\, \wt S(x,z) \wt S(z,y)$,  and we can set $x=y$ {\it after} taking the variation, so that
\be\label{GRtildevar-2}
\dd \wt G_{\rm R}(y) =\int\d z \Big( 2\dd\s(z) \wt S(y,z) \wt S(z,y)  + \dd\wt S(y,z) \wt S(z,y)  + \wt S(y,z) \dd\wt S(z,y) \Big)
+\ \frac{\dd\s(y)}{2\pi}  \, {\bf 1}_{2\times 2} \ .\ \ \ 
\ee

\subsubsection{Determining $\dd \wt S$}

We then need to determine $\dd \wt S(x,y)$. To do this, we just repeat the steps leading to \eqref{hatddS-2} and \eqref{deltaST3}, but now subtracting the zero-mode parts.
We start with $(i\Nsl_x + m\g_*) S(x,y)=\dd^{(2)}(x-y)/\sqrt{g}$ and separate the non-zero and zero-mode parts~: $S=\wt S+S_{(0)}$. The zero-mode piece $S_{(0)}=\frac{Q_0}{m}$  satisfies 
$i\Nsl_x \, S_{(0)}(x,y) =0$ so that (recall $\g_* Q_0=P_0$)
\be\label{Stildediffeq}
(i\Nsl_x + m \g_*)  \wt S(x,y) =
 \frac{\dd^{(2)}(x-y)}{\sqrt{g}} - m\g_* S_{(0)}(x,y)
= \frac{\dd^{(2)}(x-y)}{\sqrt{g}} - P_0(x,y) .
\ee
Of course, this just expresses that  $D_x \wt S(x,y)$ gives the completeness relation without the zero-modes. Consider now a
metric  $g'$ related to $g$ by an infinitesimal conformal rescaling $g'=e^{2\dd\s}g$. Then, as before, $\sqrt{g'}=e^{2\dd\s}\sqrt{g}$ and ${\Nsl}{\ '}=e^{-3\dd\s/2}\,\Nsl \ e^{\dd\s/2}$. We have, of course, $\wt S'(x,y)=S'(x,y)-S'_{(0)}(x,y)$ where $i\Nsl{\ '} S'_{(0)}(x,y)=0$. Then \eqref{Stildediffeq} becomes
\be\label{Stildediffeqprime}
\big( i e^{-3\dd\s(x)/2}\,\Nsl \ e^{\dd\s(x)/2} + m \g_* \big) \wt S'(x,y) = \frac{\dd^{(2)}(x-y)}{\sqrt{g}} \, e^{-3\dd\s(x)/2}e^{-\dd\s(y)/2} - P_0'(x,y) \ .
\ee
Much as in \eqref{hatddS}, we set
\ba\label{tildehatddS}
\wt S'(x,y)&=& \wt S(x,y)+\dd \wt S(x,y)=e^{-\dd\s(x)/2} \Big(\wt S(x,y) +\wh\dd \wt S(x,y)\Big) e^{-\dd\s(y)/2} \nonumber\\
\Leftrightarrow\quad
\dd \wt S(x,y)&=&- \frac{1}{2} \big( \dd\s(x)+\dd\s(y)\big) \wt S(x,y)+\wh\dd \wt S(x,y)\ ,
\ea
where the equivalence between the two equations holds up to terms of second order in the variations.
Developing \eqref{Stildediffeqprime} to first order in the variations and comparing with \eqref{Stildediffeq} we get
\be\label{tildehatdeltaSeq}
D_x\,  \wh\dd \wt S(x,y) \equiv (i\Nsl_x+m\g_*) \wh\dd \wt S(x,y)=-m \dd\s(x) \, \g_* \wt S(x,y) 
-\big( \frac{3}{2}\dd\s(x)+\frac{1}{2}\dd\s(y)\big) P_0(x,y) -\dd P_0(x,y) \ ,
\ee
where $\dd P_0=P_0'-P_0$, so that
\ba\label{tildehatdeltaSeq-2}
&&\hskip-3.cm\int\d z\,\wt S(x,z) D_z\,  \wh\dd \wt S(z,y)
=-m \int\d z\,\wt S(x,z) \dd\s(z) \, \g_* \wt S(z,y) 
\nonumber\\
&&\hskip 1.cm-\int\d z\,\wt S(x,z)
\Big[ \big( \frac{3}{2}\dd\s(z)+\frac{1}{2}\dd\s(y)\big) P_0(z,y) +\dd P_0(z,y)\Big] \ .
\ea
On the left-hand side, as before,  we use the hermiticity of $D_z$ and let it act on $\wt S(x,z)$, according to \eqref{Stildediffeq}~:
\be
\int\d z\,\wt S(x,z) D_z\,  \wh\dd \wt S(z,y)
= \wh\dd \wt S(x,y) - \int\d z\, P_0(x,z)  \wh\dd \wt S(z,y) \ .
\ee
Hence,
\ba\label{tildehatdeltaSeq-3}
&&\hskip-2.cm  \wh\dd \wt S(x,y) 
=-m \int\d z\,\wt S(x,z) \dd\s(z) \, \g_* \wt S(z,y) 
\nonumber\\
&&\hskip -0.cm-\int\d z\,\wt S(x,z)
\Big[ \big( \frac{3}{2}\dd\s(z)+\frac{1}{2}\dd\s(y)\big) P_0(z,y) +\dd P_0(z,y)\Big] 
+ \int\d z\, P_0(x,z) \, \wh\dd \wt S(z,y) \ .
\ea
Recall the orthogonality relation between $\wt S$ and $G_{(0)}\sim P_0$ which we write for metric $g$ and for the metric $g'=e^{2\dd\s}g$ (with $\d z$ still standing for $\d^2 z \sqrt{g(z)}$)~:
\be\label{orthogandg'}
\int\d z\, P_0(x,z) \wt S(z,y)=0
\quad , \quad 
\int\d z\, e^{3\dd\s(z)/2} e^{-\dd\s(y)/2} P'_0(x,z)\Big( \wt S(z,y) +\wh\dd\wt S(z,y)
\Big) =0 \ .
\ee
First, this shows that the term $\sim \dd\s(y)$ in \eqref{tildehatdeltaSeq-3} does not contribute.
Next, developing the second equation to first order and using the first one we get
\be
 \int\d z\, P_0(x,z) \, \wh\dd \wt S(z,y) =
 - \int\d z\, \Big( \frac{3}{2}\dd\s(z)  P_0(x,z) \wt S(z,y)  + \dd P_0(x,z)\wt S(z,y)\Big)  \ .
 \ee
 \vskip-2.mm
 \noindent
Hence,
\ba\label{tildehatdeltaSeq-4}
&&\hskip-1.cm  \wh\dd \wt S(x,y) 
=-m \int\d z\, \dd\s(z) \, \wt S(x,z) \g_* \wt S(z,y) 
\nonumber\\
&&\hskip -0.2cm-\int\d z\,\Big[ \frac{3}{2}\dd\s(z)\big( \wt S(x,z) P_0(z,y)+P_0(x,z)\wt S(z,y) \big)+ \wt S(x,z)\dd P_0(z,y)+\dd P_0(x,z)\wt S(z,y)\Big]  . \ \ \ 
\ea
Now, $\dd P_0(x,z)$ has been worked out in \eqref{deltaP0}. It contains a piece $-\frac{1}{2} \big(\dd\s(x)+\dd\s(z)\big) P_0(x,z)$ as well as a piece $\sum_{i,j} a_{ij}\, \p_{0,i}(x)\p_{0,j}^{\, \dag}(z)$ where the $a_{ij}$ are some constants. Thus, when multiplied by $\wt S(z,y)$ and integrated over $\d z$ the second piece does not contribute. Similarly for $\int\d z\,\wt S(x,z) \dd P_0(z,y)$. Thus we can rewrite \eqref{tildehatdeltaSeq-4} as
\be\label{tildehatdeltaSeq-5}
 \wh\dd \wt S(x,y) 
=- \int\d z\, \dd\s(z) \, \Big[ m \,\wt S(x,z) \g_* \wt S(z,y) 
+ \wt S(x,z) P_0(z,y)+P_0(x,z)\wt S(z,y) \Big] \ .
\ee
This, together with the second equation \eqref{tildehatddS} determines $\dd\wt S$ entirely in terms of $\wt S$, $\dd\s$ and $P_0$.  Note that we cannot (yet) use the ``orthogonality" of $\wt S$ and $P_0$ since the integral over $\d z$  also involves $\dd\s(z)$. However, we will see in the sequel that the $P_0$ terms drop out from the relevant quantities we will compute here.

\subsubsection{Determining $\dd \wt G_{\rm R}$}

We have already given $\dd \wt G_{\rm R}$ in terms of $\dd\wt S$ and $\dd\s$  in \eqref{GRtildevar-2}.  Using now the second equation \eqref{tildehatddS} to express $\dd\wt S$ in terms of $\dd\s\, \wt S$ and $\wh\dd\wt S$, this becomes
\be\label{GRtildevar-4}
\dd \wt G_{\rm R}(y) 
=\int\d z\, \Big(\big(\dd\s(z)-\dd\s(y)\big) \wt S(y,z)\wt S(z,y)
+\wh\dd\wt S(x,z) \wt S(z,y) + \wt S(x,y) \wh\dd\wt S(z,y)\big)
+\frac{\dd\s(y)}{2\pi} \ {\bf 1}_{2\times 2}  \ .
\quad
\ee
Let us remark for further reference that if we do not take the $x\to y$ limit and discard the last term we also get
\be\label{Gtildevar}
\dd \wt G(x,y) 
= \int\d z\, \Big(\big(\dd\s(z)-\frac{\dd\s(x)}{2}-\frac{\dd\s(y)}{2}\big) \wt S(x,z)\wt S(z,y)+\wh\dd\wt S(x,z) \wt S(z,y) + \wt S(x,z) \wh\dd\wt S(z,y)\Big)\ .
\ee
If we take the trace of $\dd\wt G_{\rm R}$ and integrate, as before, the term $\dd\s(z)-\dd\s(y)$ vanishes by antisymmetry, and we are left with
\be\label{intGRtildevar}
\int\d y\,\dd\, \tr\wt G_{\rm R}(y) 
= 2\int\d y\,  \d z\,   \tr \hat \dd\wt S(y,z) \wt S(z,y) +\frac{\dd A}{2\pi}  \ .
\ee
Next, we insert $\wh\dd\wt S$ from \eqref{tildehatdeltaSeq-5}~:
\ba\label{intGRtildevar-2}
\hskip-1.cm\int\d y\,\dd\, \tr\wt G_{\rm R}(y) &=& 
-2\int\d y\, \d z\, \d u\,  \dd\s(u) \tr \Big( m \,\wt S(y,u) \g_* \wt S(u,z) \nonumber\\
&&\hskip3.cm+ \wt S(y,u) P_0(u,z)+P_0(y,u)\wt S(u,z)  \Big) \wt S(z,y) +\frac{\dd A}{2\pi}  \ .
\ea
We see that the terms involving $P_0$ now vanish : $\int \d z\, P_0(u,z)\wt S(z,y) = 0$ and (using the cyclicity of the trace)
$\int\d y\, \wt S(z,y)P_0(y,u)=0$. Furthermore, we can replace $\int\d y\, \wt S(z,y)\wt S(y,u)$ by $\wt G(z,u)$. We are left with
\ba\label{intGRtildevar-3}
&&\hskip-1.cm\int\d y \,\dd\, \tr\wt G_{\rm R}(y) 
=-2m \int \d z\, \d u\,  \dd\s(u) \tr  \g_* \wt S(u,z) \wt G(z,u) +\frac{\dd A}{2\pi} \nonumber\\ 
&&\hskip-1.cm=-2m^2 \int \d z\, \d u\,  \dd\s(u) \tr \wt G(u,z) \wt G(z,u) +\frac{\dd A}{2\pi}
=-2m^2 \int  \d u\,  \dd\s(u) \tr \wt G_2(u,u) +\frac{\dd A}{2\pi} \ ,
\ea
where we used, much as for $S$ and $G$, that
$\wt S(u,z)=(i\Nsl_u +m\g_*)\wt G(u,z)$ and the fact that only the $m\g_*\wt G$ survives in the trace. Inserting this into \eqref{sgravvar4-tilde-1} we  get
 \ba\label{sgravvar4-tilde-2}
&&\hskip-2.cm\dd\Big( S_{\rm grav} +\frac{1}{48\pi} S_{\rm Liouville} +  \log\det {\cal P}_0 +\frac{\g}{4\pi} \, m^2 A  +\frac{m^2}{4}  \int\d u\,  \tr \wt G_{\rm R}(u)\Big)\nonumber\\
&&\hskip6.5cm
= -\frac{m^4}{2} \int \d u\,\dd \s(u)\,  \tr \wt G_2(u,u) \ .
\ea
Of course, this is very similar to \eqref{sgravvar4-3a}, except that now all Green's functions have their zero-modes excluded and, instead, we have the zero-mode related piece $ \log\det {\cal P}_0$.

Since $\wt G_2$ has a finite limit as $m\to 0$, the right-hand side is a genuine order $m^4$-term and one can read the complete variation of $S_{\rm grav}$ up to and including the order $m^2$-terms from the left-hand side. But we also want to obtain the higher-order contributions which requires to study similarly the variation of the higher $\wt G_n$.

\subsubsection{Conformal variation of the higher $\wt G_n(x,y), n\ge 2$\label{confvarhighersec}}

As before, we now want to obtain the conformal variations of the non-zero-mode parts of the higher Green's functions. It follows immediately from the orthonormality of the zero-mode and non-zero-mode parts of $G$ that
\be\label{Gntilde}
\wt G_n(x,y)=G_n(x,y)-G_n^{(0)}(x,y)\equiv G_n(x,y)-\frac{P_0(x,y)}{m^{2n}}
=\int \d u_2\ldots \d u_{n} \, \wt G(x,u_2) \wt G(u_2,u_3)\ldots \wt G(u_n,y) \ .
\ee
Then, as before for $G_n$ in \eqref{Gnvar1}, we now have for $\wt G_n$
\ba\label{Gntildeintvar}
\int\d u_1\, \dd\, \tr  \wt G_n(u_1,u_1)
&=&  (2(n-1) \int\d u\, \dd\s(u) \tr  \wt G_{(n)}(u,u)  \nonumber\\
&&+ n \int \d u_1 \d u_2 \tr \dd \wt G(u_1,u_2) \wt G_{(n-1)}(u_2,u_1) \ .
\ea
Let us evaluate the second term on the right-hand side of this equation. $\dd\wt G$ was given in \eqref{Gtildevar} so that
\ba\label{GNtildevarlast}
&&\hskip-2.cm\int \d x\, \d y\, \tr \dd \wt G(x,y) \wt G_{(n-1)}(y,x)\nonumber\\
&&= \int \d x\, \d y\, \d z\, \Big[ \big(\dd\s(z)-\frac{\dd\s(x)}{2}-\frac{\dd\s(y)}{2}\big)\tr  \wt S(x,z)\wt S(z,y)\wt G_{(n-1)}(y,x)\nonumber\\
&&\hskip3.cm+\tr \big(\wh\dd\wt S(x,z) \wt S(z,y) + \wt S(x,z) \wh\dd\wt S(z,y)\big)\wt G_{(n-1)}(y,x) \Big]  \ .
\ea
Using the cyclicity of the trace we can show again that the term involving explicitly the $\dd\s$ vanishes and only the terms involving $\wh\dd \wt S$ remain. Using \eqref{tildehatdeltaSeq-5} these terms are
\ba\label{deltahattermsGn}
&&\int \d x\, \d y\, \d z\, 
\tr \big(\wh\dd\wt S(x,z) \wt S(z,y) + \wt S(x,z) \wh\dd\wt S(z,y)\big)\wt G_{(n-1)}(y,x) 
\nonumber\\
&&=-\int \d x\, \d y\, \d u\, \d z\, \dd\s(z)
\tr \Big[ m \,\wt S(x,z) \g_* \wt S(z,u) 
+ \wt S(x,z) P_0(z,u)+P_0(x,z)\wt S(z,u) \Big]\times\nonumber\\
&&\hskip 5.cm \times \big[ \wt S(u,y) \wt G_{(n-1)}(y,x) + \wt G_{(n-1)}(u,y)  \wt S(y,x)  \big] \ .
\ea
Again, by the cyclicity of the trace, we see that the terms involving $P_0$ always appear as\break $\int \d u\, P_0(z,u) \wt S(u,y)=0$ or $\int\d x\, \wt G_{(n-1)}(y,x) P_0(x,z) =0$, etc. Hence they give vanishing contributions and \eqref{deltahattermsGn} reduces to
\ba\label{deltahattermsGn-2}
(\ref{deltahattermsGn})&&= - m \int\d x\, \d z \, \dd\s(z)\, \tr \big( \wt S(x,z)\g_* \wt G_n(z,x) +  \g_* \wt S(z,x) \wt G_n(x,z)\Big)
\nonumber\\
&&= - m^2 \int\d x\, \d z\, \dd\s(z)\, \tr \big( \wt G(x,z)\wt G_n(z,x) +   \wt G(z,x) \wt G_n(x,z)\Big)\nonumber\\
&&=-2 m^2 \int\d y\,  \dd\s(y)\, \tr \wt G_{n+1}(y,y) \ ,
\ea
where in the next to last step we used the by now familiar argument to replace $\wt S\g_*$ and $\g_* \wt S$ by $m\wt G$ inside the trace.
Putting the pieces together, using \eqref{Gntildeintvar}, \eqref{GNtildevarlast} and \eqref{deltahattermsGn-2}, we get
\be\label{Gntildeintvar-2}
\int\d y\, \dd \tr  \wt G_n(y,y)
= 2(n-1) \int \d y\, \dd\s(y)\, \tr  \wt G_n(y,y)
- 2n \, m^2 \int \d y\, \dd\s(y)\, \tr  \wt G_{n+1}(y,y) \ ,
\ee
exactly as in \eqref{Gnvar4} but now for the $\wt G_n$ instead of the $G_n$. One can then do the same manipulations, obtaining the analogue relations of \eqref{Gnvar5} (again we do drop the arguments $y$)
\be\label{Gntildevaridentity-3}
\int\d y\, \dd\s(y) \tr  \wt G_n
= \frac{1}{2n}\, \dd\, \int \d y\, \tr  \wt G_n
+ m^2 \int \d y\, \dd\s(y)\, \tr  \wt G_{n+1} \ .
\ee
This can be iterated to obtain a relation analogous to \eqref{Gnvar6}~:
\be\label{Gntildevaridentity-4}
m^{2n}\int\d y\, \dd\s(y) \tr  \wt G_n
= \dd\, \int \d y\, \sum_{k=n}^N \frac{m^{2k}}{2k}\,  \tr  \wt G_k
+ m^{2(N+1)} \int \d y\, \dd\s(y)\, \tr  \wt G_{N+1}\ .
\ee
This looks like an expansion in powers of $m^2$, but one should be aware that the Green's functions $\wt G_k$ are the Green's function of the massive theory. However, and this was the whole point of having subtracted the zero-modes, each $\wt G_k$ now has a finite limit as $m\to 0$ and can be expanded itself in powers of $m^2$, as we will do shortly.

Let us give an upper bound on the term on the right-hand side.
We have  $\int\d y\,\tr\wt G_{N+1}=\sum_{n\ne 0} \frac{2}{\l_n^{2(N+1)}} = \wt \zeta(N+1)$. Furthermore, $\l_n^2=\l_n^2\vert_{m=0}+m^2$ and the eigenvalues of the massless theory have a simple scaling with the area, namely\footnote{
This follows rather trivially from the fact that a conformal rescaling with {\it constant} $\s$ changes $A\to e^{2\s} A$ and $\Nsl \to e^{-\s}\Nsl\ $ so that ${\Nsl\ }^2\to e^{-2\s} {\Nsl\ }^2$ and, hence, $\l_n^2\to e^{-2\s} \l_n^2$.
} 
\be\label{lambdaareascaling}
\l_n^2\vert_{m=0}^A=\frac{1}{A} \l_n^2\vert_{m=0}^{A=1}\ .
\ee 
If we denote $M_{\dd\s}=\max |\dd\s |\ge 0$ we have the following bound 
\ba
\Big\vert  \int \d y\, \dd\s(y)\, \tr  \wt G_{N+1} \Big\vert
&\le& M_{\dd\s}\  \Big\vert  \int \d y\,  \tr  \wt G_{N+1} \Big\vert 
=  M_{\dd\s}\  \wt \zeta(N+1) \le M_{\dd\s}\  \wt \zeta_{m=0}(N+1) \nonumber\\
&=&A^{N+1}M_{\dd\s}\  \wt \zeta_{m=0}^{A=1}(N+1) \ .
\ea
Now, just as Riemann's zeta-function $\zeta_{\rm R}(s)$ has a finite limit as $s\to\infty$, our $ \wt \zeta_{m=0}^{A=1}(N+1)$ has a well-defined ``limit behaviour" as $N\to\infty$ and this  is given in terms of the smallest eigenvalue that contributes. 
If we denote  $a_0\equiv \l_1\vert_{m=0}^{A=1}$ is the smallest (positive) non-zero eigenvalue for zero mass and unit area $A=1$, and $d_{a_0}$ its degeneracy, then $ \zeta_{m=0}^{A=1}(N+1) \sim_{N\to\infty} 2 \frac{d_{a_0}}{a_0^{2(N+1)}}$. Hence
\be
\Big\vert m^{2(N+1)} \int \d y\, \dd\s(y)\, \tr  \wt G_{N+1} \Big\vert 
\le m^{2(N+1)} A^{N+1}M_{\dd\s}\  \wt \zeta_{m=0}^{A=1}(N+1) \sim_{N\to\infty} 2 d_{a_0}  M_{\dd\s}\, \Big(\frac{m^2 A}{a_0^2}\Big)^{N+1}
\ .
\ee
Thus for small enough mass, $m^2 A< a_0^2$ the right-hand side goes to zero as $N\to\infty$
and we may drop the corresponding term in \eqref{Gntildevaridentity-4}. We may similarly estimate the radius of convergence of the series
\be\label{Gtildeseries}
\int \d y\ \sum_{k=n}^\infty \frac{m^{2k}}{2k}\,  \tr  \wt G_k
=  \sum_{k=n}^\infty \frac{m^{2k}}{2k} \wt\zeta(k)
= \sum_{k=n}^\infty \frac{(m^2 A)^k}{2k} \wt\zeta^{A=1}(k) \ .
\ee
If we now denote by $a$ the smallest positive eigenvalue (of the massive theory) appearing in $\wt\zeta^{A=1}$ (such that its zero-mass limit is $a_0>0$), then the series converges for $\big\vert \frac{m^2 A}{a^2}\big\vert < 1$. We conclude that for small enough mass we can take the $N\to
\infty$ limit in \eqref{Gntildevaridentity-4} and obtain
\be\label{Gntildevaridentity-5}
m^{2n}\int\d^2 y\, \sqrt{g}\, \dd\s(y) \tr  \wt G_n
= \dd\, \int \d^2 y\, \sqrt{g}\ \sum_{k=n}^\infty \frac{m^{2k}}{2k}\,  \tr  \wt G_k \ .
\ee

\subsubsection{Variation of the gravitational action to all orders in $m^2$ for arbitrary genus}

We insert this result  \eqref{Gntildevaridentity-5} for $n=2$ into eq.~\eqref{sgravvar4-tilde-2} and get
\ba\label{sgravvar4-tilde-4}
&&\hskip-2.cm\dd\Bigg[ S_{\rm grav} +\frac{1}{48\pi} S_{\rm Liouville} + \log\det {\cal P}_0
+\frac{\g}{4\pi} \, m^2 A  \nonumber\\
&&\hskip1.cm+\frac{1}{4}  \int\d^2 u\sqrt{g}\,  \Big(m^2  \tr \wt G_{\rm R}(u)
+ \sum_{k=2}^\infty \frac{m^{2k}}{k}  \tr \wt G_k(u,u) \Big) \Bigg]  = 0 \ .
\ea
As before, the integrals of the traces of the higher Green's functions $\wt G_k$ at coinciding points are just the $\wt\zeta(k)$ so that equivalently
\be\label{sgravvar4-tilde-5}
\dd\Bigg[ S_{\rm grav} +\frac{1}{48\pi} S_{\rm Liouville} +  \log\det {\cal P}_0  +\frac{\g}{4\pi} \, m^2 A  
 +  \frac{m^2}{4} \int\d^2 u\sqrt{g}\,   \tr \wt G_{\rm R}(u)
+ \sum_{k=2}^\infty \frac{m^{2k}}{4k} \wt\zeta(k)  \Bigg] = 0 \ .
\ee
This looks very similar to the result \eqref{sgravvar4-8-sphere} where we did not separate the zero-mode contributions. The only difference (apart from replacing the $G$'s by the $\wt G$'s, or replacing the $\zeta(k)$ by the $\wt\zeta(k)$) is the presence of the zero-mode term $\log\det{\cal P}_0$ which only contributes at order $m^0$.
Indeed, we have seen in sect.~\ref{pertsec} that $\dd\l_0=0$ so that $\dd\zeta(s)=\dd\wt\zeta(s)$.

As already remarked above,  $\wt G_{\rm R}$ and the $\wt\zeta(k)$ are the quantities defined with non-vanishing mass. Since we subtracted the zero-mode parts, these quantities have a well-defined series expansion in $m^2$ involving quantities defined in the massless theory. We will now work out these expansions. First, we  have $\l_n^2=\l_{n,(m=0)}^2+m^2$, cf \eqref{lambdam=0andm}, so that
\be\label{lambdankexp}
\frac{1}{\l_n^{2k}}=\frac{1}{(\l_{n,(m=0)}^2+m^2)^k}=\frac{1}{\l_{n,(m=0)}^{2k}}\Big(1+\frac{m^2}{\l_{n,(m=0)}^2}\Big)^{-k}
=\frac{1}{\l_{n,(m=0)}^{2k}} \sum_{r=0}^\infty (-)^r \frac{(k+r-1)!}{(k-1)!\ r!} 
\Big(\frac{m^2}{\l_{n,(m=0)}^2}\Big)^r \ .
\ee
Then
\ba\label{sumidentity}
&&\hskip-1.cm\sum_{k=2}^\infty \frac{m^{2k}}{k} \frac{1}{\l_n^{2k}}
=\sum_{k=2}^\infty 
\sum_{r=0}^\infty (-)^r \frac{(k+r)!}{k!\ r!} \frac{1}{k+r}\Big(\frac{m^2}{\l_{n,(m=0)}^2}\Big)^{r+k}
=\sum_{l=2}^\infty \Big( \sum_{r=0}^{l-2} (-)^r\frac{l!}{(l-r)!r!} \Big) 
\frac{1}{l} \Big(\frac{m^2}{\l_{n,(m=0)}^2}\Big)^l \nonumber\\
&&\hskip-1.cm=\sum_{l=2}^\infty \Big[ \sum_{r=0}^{l} (-)^r\frac{l!}{(l-r)!r!} -(-)^{l-1} l -(-)^l\Big]
\frac{1}{l} \Big(\frac{m^2}{\l_{n,(m=0)}^2}\Big)^l
=\sum_{l=2}^\infty \Big[ 0 -(-)^{l-1} l -(-)^l\Big]
\frac{1}{l} \Big(\frac{m^2}{\l_{n,(m=0)}^2}\Big)^l \nonumber\\
&&\hskip-1.cm=\sum_{l=2}^\infty (-)^l \frac{l-1}{l} \Big(\frac{m^2}{\l_{n,(m=0)}^2}\Big)^l  \ .
\ea
If we now sum over $n\ne 0$ we get the corresponding relation for the sum of zeta-functions
\be\label{zetasumidentity}
\sum_{k=2}^\infty \frac{m^{2k}}{k}  \wt \zeta(k) = \sum_{k=2}^\infty (-)^k \frac{k-1}{k} m^{2k} \wt\zeta_{m=0}(k)\ .
\ee
Note that to lowest order in $m$ this is just $\frac{m^4}{2} \wt\zeta(2)=\frac{2-1}{2} \,m^4 \wt\zeta_{m=0}(2) +{\cal O}(m^6)$, which is indeed obviously correct, and it is similarly easy to directly check also the next order.

We may similarly proceed with $\wt G_{\rm R}$. First note that  the short distance singularity of $\wt G(x,y)$  is the same for vanishing and for non-vanishing mass. Indeed, this singularity depends on the 2-derivative part of $D^2$ and is not changed by adding the mass or not. On the other hand, as already used repeatedly, the $\wt G_r(x,y)$ have no short-distance singularities for $r\ge 2$ and the $\wt G_r(y,y)$ are finite. Recall the obvious fact that if $\p_n$ is an eigenfunction of $D^2$ for $m\ne 0$, i.e. $D^2\p_n=\l_n^2 \p_n$ where $D^2=-\Nsl^{\ 2} + m^2$, then this same $\p_n$ is also an eigenfunction of $D^2$ for $m=0$, i.e. of $-\Nsl^{\ 2}$ with eigenvalue $\l_n^2-m^2=\l^2_{n,(m=0)}$. 
(Note that this argument is correct for the eigenfunctions of $D^2$, but not for the eigenfunctions of $D$ and the following reasoning for $\wt G$ can not be transposed to $\wt S$.)
Hence, using \eqref{lambdankexp} for $k=1$
\ba\label{GandGm=0}
\wt G(x,y)&=&\sum_{n\ne 0} \frac{1}{\l_n^2}\big( \p_n(x)\p_n^\dag(y) +\p_n^*(x)(\p_n^*)^\dag(y)\big)
\nonumber\\
&=& \sum_{r=0}^\infty (-)^r m^{2r} \sum_{n\ne 0} \frac{1}{\l_{n,(m=0)}^{2r+2}}
\big( \p_n(x)\p_n^\dag(y) +\p_n^*(x)(\p_n^*)^\dag(y)\big)\nonumber\\
&=& \sum_{r=0}^\infty (-)^r m^{2r} \wt G_{r+1}^{m=0}(x,y)  = \wt G^{m=0}(x,y) + \sum_{r=1}^\infty (-)^r m^{2r} \wt G_{r+1}^{m=0}(x,y) \ .
\ea
Subtracting the short-distance singularity and letting then $x\to y$ gives
\be\label{GRandGRm=0}
\wt G_{\rm R}(y) = \wt G_{\rm R}^{m=0}(y) + \sum_{r=1}^\infty (-)^r m^{2r} \wt G_{r+1}^{m=0}(y,y)
 \ ,
\ee
so that
\be\label{GRandGRm=0zeta}
m^2 \int \d^2 y \sqrt{g}\, \wt G_{\rm R}(y)= m^2\int \d^2 y \sqrt{g}\,  \wt G_{\rm R}^{m=0}(y) - \sum_{k=2}^\infty (-)^k m^{2k}\wt\zeta_{m=0}(k) \ .
\ee
We can then rewrite \eqref{sgravvar4-tilde-5} as
\ba\label{sgravvar4-tilde-6}
&&\hskip-2.cm\dd\Bigg[ S_{\rm grav} +\frac{1}{48\pi} S_{\rm Liouville} + \log\det {\cal P}_0 
+\frac{\g}{4\pi} \, m^2 A 
\nonumber\\
&&\hskip2.cm
 +  \frac{m^2}{4} \int\d^2 u\sqrt{g}\,   \tr \wt G_{\rm R}^{m=0}(u)
+ \sum_{k=2}^\infty \frac{(-)^{k+1} }{4k} m^{2k}\, \wt\zeta_{m=0}(k) \Bigg] = 0 \ .
\ea
This looks exactly like \eqref{sgravvar4-tilde-5} - which was written in terms of the massive Green's function and massive zeta-function - except that now the terms in the sum over $k$ have an alternating sign~!

The last sum in this equation \eqref{sgravvar4-tilde-6} ressembles, of course, the Taylor expansion of $\log(1+x)$ and, indeed, formally we have
\ba\label{logeigenexp}
\log \Big(\frac{\Det' D^2}{\Det' D^2_{m=0}} \Big)
&=&  2\log \prod_{n\ne 0} \frac{\l_n^2}{\l^2_{n,m=0}}=2\sum_{n\ne 0} \log \Big( 1 + \frac{m^2}{\l^2_{n,m=0}}\Big)\
= 2 \sum_{n\ne 0} \sum_{k=1}^\infty \frac{(-)^{k+1}}{k}  \frac{m^{2k}}{\l^{2k}_{n,m=0}}\nonumber\\
&=&
\sum_{k=1}^\infty \frac{(-)^{k+1}}{k} m^{2k}\, \wt\zeta_{m=0}(k) \ .
\ea
It looks a bit as if we have been going in circles and we could have written this expansion right away at the beginning.
However, this formal manipulation results in a sum over $k$ including also the singular $k=1$ term $\wt\zeta_{\rm m=0}(1)$. Our more careful treatment has produced instead, the integral of the Green's function $\wt G^{m=0}_{\rm R}$, which can be rewritten in terms of $\wt G^{m=0}_\zeta$ which in turn is related to the regular part of $\wt\zeta_{m=0}$, i.e. $\wt\zeta^{\rm reg}_{m=0}(1)$, as well as the piece $ \log\det {\cal P}_0$ that originated from the zero-modes.


\subsection{Integrating the variation of  the  gravitational action}

We have obtained the variation of the gravitational action - to all orders in an expansion in powers of $m^2$ - under an infinitesimal variation of the conformal factor, corresponding to an infinitesimal variation of the metric. This variation was expressed in terms of variations of integrals of the renormalized Green's function $\wt G_{\rm R}^{m=0}$ of the massless theory and of the higher Green's functions $\wt G_n^{m=0}(x,x)$ ($n\ge 2$), or equivalently of the finite values of zeta function $\wt \zeta_{m=0}(n)$, also of the massless theory. We can then immediately ``integrate" these infinitesimal variations to obtain $S_{\rm grav}[g,\wh g]$.
Note that ${\cal P}_0$  as defined in \eqref{P0ijdef} depends  on $\wh g$ (through the zero-modes $\wh \p_{0,i}$) and $\s$. We write ${\cal P}_0[g]\equiv {\cal P}_0[\wh g,\s]\equiv {\cal P}_0[g,\wh g]$. Since the zero-modes $\wh \p_{0,i}$ are orthonormalized for the metric $\wh g$ we obviously have ${\cal P}_0[\wh g]\equiv {\cal P}_0[\wh g,\wh g]={\bf 1}$. It follows that
$ \dd\,\log\det {\cal P}_0$ integrates to $\log \frac{\det{\cal P}_0[g,\wh g]}{\det{\cal P}_0[\wh g,\wh g]}=\log\det{\cal P}_0[g,\wh g]$. 
We get
\ba\label{sgravfinite}
S_{\rm grav}[g,\wh g]&=&
 -\frac{1}{48\pi} S_{\rm Liouville}[g,\wh g] - \det{\cal P}_0[g, \wh g] -\frac{\g}{4\pi} \, m^2 (A -\wh A) \nonumber\\
 &&
 -  \frac{m^2}{4}\big( {\cal G}[g]-{\cal G}[\wh g]\big)
+ \sum_{k=2}^\infty \frac{(-)^{k} }{4k} m^{2k}\, 
\big(\wt\zeta_{m=0}^{\, g}(k) - \wt\zeta_{m=0}^{\, \wh g}(k)\big) \ ,
\ea
where we defined
\be\label{Gfunctional}
{\cal G}[g]= \int\d^2 u\sqrt{g(u)}\,   \tr \wt G_{\rm R}^{m=0}[g](u) \ ,
\ee
with the obvious notation that $\wt G_{\rm R}^{m=0}[g]$ is the renormalized Green's function of the massless theory, with the zero-mode subtracted, for the metric $g$.
Let us insist that the right-hand side of \eqref{sgravfinite} is  entirely expressed in terms of quantities defined in the massless theory. We  now want to determine, in particular, the dependence of the functional ${\cal G}[g]$ on the conformal factor $\s$.

Just as the Liouville action only involves the local quantities $\s,\  \wh g$ and $\wh R$, ideally one would like to express the higher-order terms similarly in terms of such local quantities. For the simpler case of the effective gravitational action obtained by integrating out a massive scalar field  \cite{FKZ} it turned out that the order $m^2$-term could be expressed  in terms of the Mabuchi action which is local in $\s$ and the K\"ahler potential $\F$, and some relatively simple term that involved $\wt G_{\rm R}^{m=0}[\wh g]$. (In this case $\wt G_{\rm R}^{m=0}$ is the renormalized  Green's function of the {\it scalar} Laplacian). For such a massive scalar field the higher-order (in $m^2$) terms were computed in \cite{BL2} and were expressed in terms of  higher Green'a functions which encode much non-local information about the manifold.

At present, $\wt G^{m=0}$ is the Green's function of $D^2=-\D_{\rm sp}+\frac{\cR}{4}$ which, in particular, involves the spinoral Laplacian which is different from the scalar Laplacian.
In order to express  ${\cal G}[g]$ in terms of local quantities and possibly $\wt G_{\rm R}^{m=0}[\wh g]$ we have tried to determine the variation of $\wt G_{\rm R}^{m=0}$ and tried to write it in such a form that one can integrate the infinitesimal variations to obtain its variation for finite $\s$. We tried to mimic the computation for the scalar field where it was useful to express the variation of the conformal factor as the scalar Laplacian of the variation of the K\"ahler potential and then integrate by parts this scalar Laplacian. However, at present, this turns out to be more complicated since $\wt G$ is {\it not} the Greens's function of the scalar Laplacian, and we did not manage to obtain a useful form of the variation of $\wt G_{\rm R}^{m=0}$ along this line. 

Here, we will instead proceed along a different avenue which, nevertheless,  will allow us to express ${\cal G}[g]-{\cal G}[\wh g]$ in terms of the (local) conformal factor $\s$, as well as several quantities pertaining to the metric $\wh g$ only, such as the Green's function $\wt S_{\wh g}$ or the zero-mode projector $\wh P_0$.

To simplify the notations in this subsection we will introduce further abbreviations for the different integration measures including various powers of $e^\s$~:
\be\label{intmeasures}
\d \wh z=\d^2 z \sqrt{\wh g(z)}
\quad , \quad
\d \wh z_\s=\d^2 z \sqrt{\wh g(z)}\, e^{\s(z)}
\quad , \quad
\d z =\d^2 z \sqrt{g(z)}\equiv \d^2 z \sqrt{\wh g(z)} \,e^{2\s(z)} \ .
\ee

\subsubsection{The variations of $\wt S$  under finite conformal rescalings for $m=0$}

We now want to relate $\wt S\equiv \wt S_g$, which is the Green's function for the metric $g$, to $\wt{\wh S}\equiv \wt S_{\wh g}$, which is the Green's function for the metric $\wh g$, where as usual $g=e^{2\s}\wh g$.

We will again exploit $\Nsl_x= e^{-3\s(x)/2}\wh\Nsl_x e^{\s(x)/2}$ but now for finite $\s(x)$. 
This relation will be particularly fruitful when relating the Green's function of the massless theory. 
Throughout the remainder of this subsection all Green's functions $\wt S$,  $\wt{\wh S}$ and $\wt G$ refer to the massless Green's functions, although we will  not indicate it explicitly any more.
For $\s=0$, i.e. for the metric $\wh g$, we have 
\be\label{Shateq}
i\wh\Nsl_x \wt{\wh S}(x,y)=\frac{\dd^{(2)}(x-y)}{\sqrt{\wh g}}-\wh P_0(x,y) \ ,
\ee
where $\wh P_0$ is the projector on the zero-modes of $i\wh\Nsl\,$, as given in \eqref{P0hatdef}. Recall also  from \eqref{P0sigma} that the projector on the zero-modes of $i\Nsl$ is
$P_0(x,y)=e^{-\s(x)/2} \Pt_0(x,y) e^{-\s(y)/2}$, where $\Pt_0$ was defined in \eqref{Ptdef}.
In particular it involves the ``area-like" ortho-normalisation factors ${\cal P}_{0,ij}=\int\d \wh z\, e^{\s(z)}\wh\p^\dag_{0,i}(z)\wh\p_{0,j}(z)$, cf \eqref{P0ijdef}. The  differential equation for $\wt S$ then is
\ba\label{Steq}
&&\hskip-1.cmi\Nsl_x \wt S(x,y)=\frac{\dd^{(2)}(x-y)}{\sqrt{g}}-P_0(x,y) \nonumber\\
&&\hskip-2.5cm\quad \Leftrightarrow\quad
i \wh\Nsl_x \big(e^{\s(x)/2}  \wt S(x,y) e^{\s(y)/2}\big)
=\frac{\dd^{(2)}(x-y)}{\sqrt{\wh g}}  - e^{\s(x)} \Pt_0(x,y) \ .\quad
\ea
We now set 
\be\label{SbarStilde}
\St(x,y)=e^{\s(x)/2}  \wt S(x,y) e^{\s(y)/2} \quad \Leftrightarrow\quad 
\wt S(x,y)=e^{-\s(x)/2}  \St(x,y) e^{-\s(y)/2} \ ,
\ee
so that \eqref{Steq} becomes
\be
i \wh\Nsl_x\St(x,y) =\frac{\dd^{(2)}(x-y)}{\sqrt{\wh g}}   - e^{\s(x)} \Pt_0(x,y) \ .
\ee
This is very similar to \eqref{Shateq} except for the term involving the zero-mode ``projector" $\Pt_0$. We then set
\be\label{Tdef}
\St(x,y)=\wt{\wh S}(x,y)+T(x,y) \ ,
\ee
and obtain the following equation for $T$~:
\be\label{Teq}
i \wh\Nsl_x T(x,y) =\wh P_0(x,y)   - e^{\s(x)} \Pt_0(x,y) \ .
\ee
Changing $x\to z$ and multiplying with $\wt{\wh S}(x,z)$ and integrating we get
\be\label{Teq2}
\int\d \wh z\, \wt{\wh S}(x,z) i \wh\Nsl_z T(z,y) =-\int\d\wh z\, \wt{\wh S}(x,z) e^{\s(z)}\Pt_0(z,y)
\equiv-\int\d\wh z_\s\, \wt{\wh S}(x,z) \Pt_0(z,y)\ ,
\ee
where the piece $\sim \wh P_0$ dropped out since $\int\d\wh z\, \wt{\wh S}(x,z) \wh P_0(z,y)=0$.
On the left-hand side we integrate by parts  the $i \wh\Nsl_z$ and use \eqref{Shateq} to get
\be\label{Teq3}
T(x,y)-\int\d\wh z\, \wh P_0(x,z) T(z,y) =-\int\d\wh z\, \wt{\wh S}(x,z) e^{\s(z)}\Pt_0(z,y) \ .
\ee
This determines $T$ up to a certain zero-mode part which is subtracted on the left-hand side. Just like $\St$ and $\wt{\wh S}$, $T$ must also satisfy $\big(T(x,y)\big)^\dag=T(y,x)$, and we can  also expand $T(x,y)$ on products of eigenfunctions of $i\wh\Nsl$. We separate the zero-mode and non-zero-mode parts, and we have schematically
\be
T(x,y)\ ``="\  \sum_{i,j} \wh \P_{0,i}(x) t_{ij} \wh \P_{0,j}^\dag(y) 
+ \sum_{i,n} \wh \P_{0,i}(x) c_{in} \wh \P_{n}^\dag(y)
+ \sum_{i,n} \wh \P_{n}(x) c_{in}^* \wh \P_{0,i}^\dag(y)
+ \sum_{n,k} \wh \P_{n}(x) d_{nm} \wh \P_{m}^\dag(y) \ ,
\ee
where $\wh\P_{0,i}$ stands generically for the zero-modes $\wh\p_{0,i}$ and $\wh\p^*_{0,i}$,
and $\wh\P_{n}$ stands generically for the non zero-modes $\wh\p_{n}$ and $\wh\p^*_{n}$.
Then, the left-hand side of equation \eqref{Teq3} only contains the terms with the $c_{in}^*$ and $d_{nm}$ while the right-hand side  contains terms of the form $\wh\P_n(x) \wh\P^\dag_{0,i}(y)$ only. This shows that  $d_{nm}=0$, and that  all  $c_{in}^*$ are determined, and hence also all $c_{in}$. Thus, \eqref{Teq3} determines $T$ uniquely, up to the pure zero-mode piece involving the $t_{ij}$. The latter corresponds to the arbitrary solution to the homogeneous equation $T(x,y)-\int\d\wh z\, \wh P_0(x,z) T(z,y)=0$ one may always add. Now, obviously, as for $\wt{\wh S}$, we do not want to include such a pure zero-mode piece in $T$ and we make the choice to set it to zero. Then $T$ only contains the pieces with $c_{in}$ and $c_{in}^*$ and, hence
\be\label{Tsol}
T(x,y)=-\int\d\wh z_\s\, 
\big( \wt{\wh S}(x,z) \Pt_0(z,y) + \Pt_0(x,z )\wt{\wh S}(z,y) \big) \ . 
\ee
Obviously, this satisfies $T(x,y)^\dag=T(y,x)$ and one may, of course, check directly that it is a solution\footnote{
Indeed, since $i\wh\Nsl_x \Pt_0(x,y)=0$ one has $i\wh\Nsl_x T(x,y)=-e^{\s(x)} \Pt(x,y) + \int\d\wh z\, e^{\s(z)}  \wh P_0(x,z))\Pt_0(z,y)$, and it is not difficult to show that $\int\d\wh z\, e^{\s(z)}  \wh P_0(x,z))\Pt_0(z,y)= \wh P_0(x,y) $.
} 
of \eqref{Teq}. 
Inserting the solution \eqref{Tsol} into \eqref{Tdef}, and using \eqref{SbarStilde},   we get
\be\label{Stildesol}
\wt S(x,y)=e^{-\frac{\s(x)}{2}} \Big(\wt{\wh S}(x,y) -\int\d\wh z_\s\, 
\big( \wt{\wh S}(x,z) \Pt_0(z,y) + \Pt_0(x,z )\wt{\wh S}(z,y) \big)\Big) e^{-\frac{\s(y)}{2}}  \ .
\ee
This can be equivalently rewritten in terms of
\be\label{Sprimedef}
{\cal S}(x,y)\equiv e^{-\frac{\s(x)}{2}} \wt{\wh S}(x,y) e^{-\frac{\s(y)}{2}}
\ee
as
\be\label{Stildesol-2}
\wt S(x,y)={\cal S}(x,y) -\int\d z\, 
\big({\cal S}(x,z) P_0(z,y) + P_0(x,z ) {\cal S}(z,y) \big) \ ,
\ee
which shows that in the absence of zero-modes (spherical topology) one simply has $\wt S={\cal S}$.

Both \eqref{Stildesol} and \eqref{Stildesol-2} express $\wt S$ solely in terms of the conformal factor $\s$ and quantities computed with the reference metric $\wh g$, i.e. $\wt{\wh S}$ and the zero-modes $\wh\p_{0,i}$. To make the dependencies slightly more explicit we introduce the notations
$\wt S\equiv \wt S_g$, $\wt{\wh S}\equiv \wt S_{\wh g}$ and
$\Pt_0\equiv \Pt_{0,\wh g,\s}$, so that \eqref{Stildesol} is rewritten once more as
\be\label{Stildesol-3}
\wt S_g(x,y)=e^{-\frac{\s(x)}{2}} \Big(\wt S_{\wh g}(x,y) -\int\d\wh z_\s\, 
\big( \wt S_{\wh g}(x,z) \Pt_{0,\wh g,\s}(z,y) + \Pt_{0,\wh g,\s}(x,z )\wt S_{\wh g}(z,y) \big)\Big) e^{-\frac{\s(y)}{2}}  \ .
\ee
Let us insist, that without the presence of the factor $e^{\s(z)}$ (hidden in the integration measure $\d\wh z_\s$) the integral of $\wt S_{\wh g}\Pt$ and of $\Pt_0 \wt S_{\wh g}$ would of course vanish due to the orthogonality of the zero and non zero-modes.

\subsubsection{The variations of  $\wt G$ and $\wt G_{\rm R}$ under finite conformal rescalings for $m=0$}

The dependence of the Green's function $\wt G_g(x,y)$ on the conformal factor can then be obtained straightforwardly by inserting the previous result for $\wt S_g$ into $\wt G_g(x,y)=\int \d^2 z \sqrt{\wh g}\, e^{2\s(z)}\, \wt S_g(x,z) \wt S_g(z,y)$~:
\ba\label{Gsigmadep}
&&\hskip-2.0cm\wt G_g(x,y)=e^{-\frac{\s(x)}{2}} 
\int\d\wh z_\s\,
\Big(\wt S_{\wh g}(x,z) -\int\d \wh u_\s\,
\big( \wt S_{\wh g}(x,u) \Pt_{0,\wh g,\s}(u,z) + \Pt_{0,\wh g,\s}(x,u)\wt S_{\wh g}(u,z) \big)\Big)
\nonumber\\
&&\hskip1.5cm \times \Big(\wt S_{\wh g}(z,y) -\int\d \wh v_\s
\big( \wt S_{\wh g}(z,v) \Pt_{0,\wh g,\s}(v,y) + \Pt_{0,\wh g,\s}(z,v)\wt S_{\wh g}(v,y) \big)\Big)
e^{-\frac{\s(y)}{2}} \ .\ea
Developing the product of the two brackets we get 9 terms. To further simplify the notation, we will  suppress the arguments in the multiple integrals. Then we find
\ba\label{Gsigmadep2}
\hskip-1.5cm  \wt G_g(x,y) &=&e^{\frac{-\s(x)}{2}}\Bigg[
\int\d \wh z_\s\, \wt S_{\wh g} \wt S_{\wh g} 
- \int\d  \wh z_\s\,d \wh u_\s\, \Big( \wt S_{\wh g} \wt S_{\wh g} \Pt_0
+ \wt S_{\wh g} \Pt_0\wt S_{\wh g}  + \wt S_{\wh g} \Pt_0 \wt S_{\wh g} 
+\Pt_0 \wt S_{\wh g} \wt S_{\wh g} \Big)\nonumber\\
&&\hskip3.mm+ \int\d  \wh z_\s\, \d \wh u_\s\, \d \wh v_\s\, \Big( \wt S_{\wh g} \Pt_0 \wt S_{\wh g} \Pt_0 + \wt S_{\wh g} \Pt_0 \Pt_0 \wt S_{\wh g} +\Pt_0 \wt S_{\wh g}  \wt S_{\wh g} \Pt_0 + \Pt_0\wt S_{\wh g} \Pt_0 \wt S_{\wh g}  \Big) \Bigg] e^{-\frac{\s(y)}{2}}\ .
\ea
Again, due to the presence of the $e^\s$ factors (hidden in $\d u_\s$ etc) for each integration, no simplification occurs\footnote{The only slight simplification comes from $\int\d \wh u_\s \, \Pt_0(z,u)\Pt_0(u,v)= \Pt_0(z,v)$ which one can show by rewriting $\Pt_0$ in terms of $P_0$ and using that $\int\d^2 u \sqrt{g}\, P_0(z,u) P_0(u,v)=P_0(z,v)$.
}
 in the integrals of products of $\wt S_{\wh g}$ and $\Pt_0$. 
 
The short-distance singularity of \eqref{Gsigmadep2} comes from the first term in the bracket and we know that it is $-\frac{1}{4\pi} \log \frac{\m^2 \ell^2_g(x,y)}{4}{\bf 1}_{2\times 2}$. As already discussed, we have for finite conformal transformations
\be\label{ellrescaling}
\ell_g^2(x,y) \sim_{x\to y} e^{\s(x)+\s(y)} \ell^2_{\wh g}(x,y)
\quad \Rightarrow\quad 
- \frac{1}{4\pi}\log \frac{\m^2 \ell^2_g(x,y)}{4} \sim_{x\to y}  - \frac{1}{4\pi}\log \frac{\m^2 \ell^2_{\wh g}(x,y)}{4} - \frac{\s(x)+\s(y)}{4\pi}\ .
\ee
We see again that the singular parts of $\wt G_g$ and $\wt G_{\wh g}$ are the same and it follows that $\wt G_g(x,y)-\wt G_{\wh g}(x,y)$ has no short-distance singularity.
We also see that
\be\label{GgGwhg}
\wt G_{\rm R}[g](y)-\wt G_{\rm R}[\wh g](y)=\lim_{x\to y} \Big( \wt G_g(x,y)-\wt G_{\wh g}(x,y) + \frac{\s(y)}{2\pi}\, {\bf 1}_{2\times 2}\Big)\ .
\ee
Combining this with \eqref{Gsigmadep2} we get
\ba\label{GgGwhg-2}
\hskip-10.mm\wt G_{\rm R}[g](y)&=&\wt G_{\rm R}[\wh g](y) + \frac{\s(y)}{2\pi}\, {\bf 1}_{2\times 2}\ + \int \d \wh z \Big( e^{\s(z)-\s(y)}-1\Big) \wt S_{\wh g}(y,z)\wt S_{\wh g}(z,y)\nonumber\\
&&\hskip-5.mm-e^{-\s(y)} \int\d \wh z_\s\,d \wh u_\s\, \Big( \wt S_{\wh g} \wt S_{\wh g} \Pt_0
+ \wt S_{\wh g} \Pt_0\wt S_{\wh g}  + \wt S_{\wh g} \Pt_0 \wt S_{\wh g} 
+\Pt_0 \wt S_{\wh g} \wt S_{\wh g} \Big)\Big\vert_{x=y}\nonumber\\
&&\hskip-5.mm+e^{-\s(y)}  \int\d \wh z_\s\, \d \wh u_\s\, \d \wh v_\s\, \Big( \wt S_{\wh g} \Pt_0 \wt S_{\wh g} \Pt_0 + \wt S_{\wh g} \Pt_0 \Pt_0 \wt S_{\wh g} +\Pt_0 \wt S_{\wh g}  \wt S_{\wh g} \Pt_0 + \Pt_0\wt S_{\wh g} \Pt_0 \wt S_{\wh g}  \Big) \Big\vert_{x=y}\ .
\ea
The functional ${\cal G}[g]=\int\d^2 y\sqrt{g} \tr \wt G_{\rm R}[g](y)= \int\d \wh y \,e^{2\s(y)} \tr \wt G_{\rm R}[g](y)\equiv \int\d \wh y_\s\, e^{\s(y)}  \tr \wt G_{\rm R}[g](y)$ then is easily obtained from the previous relation. In particular, one can now use the cyclicity of the trace to simplify the many terms involving $\Pt_0$. Writing out  explicitly all factors of $e^\s$ we get
\ba\label{calGfunctionalgwhg}
&&\hskip-1.5cm{\cal G}[g]
={\cal G}[\wh g] +\int \d \wh y\,\big(e^{2\s}-1\big)\tr \wt G_{\rm R}[\wh g](y) 
+ \frac{1}{\pi}\int\d \wh y\, e^{2\s}\s \nonumber\\
&&\hskip-5.mm
+ \int \d\wh y\, \d\wh z\,  \Big( e^{\s(z)+\s(y)}-e^{2\s(y)}\Big) \tr \wt S_{\wh g}(y,z)\wt S_{\wh g}(z,y)\nonumber\\
&&\hskip-5.mm- 4 \int\d \wh y\, \d\wh z\, \d\wh u\, e^{\s(y)+\s(z)+\s(u)}\tr \wt S_{\wh g}(y,z) \wt S_{\wh g}(z,u) \Pt_0(u,y)
\nonumber\\
&&\hskip-5.mm+ 2 \int \d \wh y\, \d\wh z\, \d\wh u\, \d\wh v\, e^{\s(y)+\s(z)+\s(u)+\s(v)}\, \tr \Big( \wt S_{\wh g} \Pt_0 \wt S_{\wh g} \Pt_0 + \wt S_{\wh g} \wt S_{\wh g} \Pt_0 \Pt_0 \Big)\ .
\ea
Again, in the absence of zero-modes, i.e.~for spherical topology, the last two integrals are absent.
Let us recall, once more,  that all quantities in this equation are those of the massless theory.

\subsubsection{The gravitational action to order $m^2$ as an explicit functional of $\s$}

We can then insert this result into \eqref{sgravfinite} to get the contributions to the effective gravitational action at order $m^2$~:
\vskip-9.mm
\ba\label{sgravfiniteorderm2}
&&\hskip-10.mm S_{\rm grav}[g,\wh g]=
 -\frac{1}{48\pi} S_{\rm Liouville}[g,\wh g] - \log\det {\cal P}_0[g,\wh g] -\frac{\g}{4\pi} \, m^2 (A -\wh A) \nonumber\\
&&\hskip5.mm-\frac{m^2}{4} \Bigg[\int \d^2 y \sqrt{\wh g}\big(e^{2\s}-1\big)\tr \wt G_{\rm R}[\wh g](y) 
+ \frac{1}{\pi}\int\d^2 z\, \sqrt{\wh g} e^{2\s}\s \nonumber\\
&&\hskip1.5cm
+ \int \d^2 y \sqrt{\wh g}\, \d^2 z \sqrt{\wh g} \Big( e^{\s(z)+\s(y)}-e^{2\s(y)}\Big) \tr \wt S_{\wh g}(y,z)\wt S_{\wh g}(z,y)\nonumber\\
&&\hskip1.5cm - 4 \int\d^2 y \sqrt{\wh g}\,\d^2 z \sqrt{\wh g}\,d^2 u\sqrt{\wh g}\, e^{\s(y)+\s(z)+\s(u)}\tr \wt S_{\wh g}(y,z) \wt S_{\wh g}(z,u) \Pt_0(u,y)
\nonumber\\
&&\hskip1.5cm+ 2 \int\d^2 y \sqrt{\wh g}\,\d^2 z \sqrt{\wh g}\,d^2 u\sqrt{\wh g}\,\d^2 v \sqrt{\wh g}\, e^{\s(y)+\s(z)+\s(u)+\s(v)}\, 
 \tr \Big( \wt S_{\wh g} \Pt_0 \wt S_{\wh g} \Pt_0 + \wt S_{\wh g} \wt S_{\wh g} \Pt_0 \Pt_0 \Big)\ \Bigg] \nonumber\\
&&\hskip5.mm+\ {\cal O}(m^4)\ .
\ea
\vskip-3.mm
\noindent 
We see that at order $m^0$ we get the Liouville action and a specific zero-mode contribution involving the area-like constants ${\cal P}_{0,ij}$. Of course, if present, the zero-modes are zero-modes of the massless Dirac operator $i\Nsl\, $, while the massive Dirac operator never has zero-modes. In the massless theory, one does not do the functional integral over these zero-modes and then there is a certain freedom how exactly we define the matter partition function.  In particular, one can then redefine the matter partition by dividing $Z_{\rm mat}[g]$ by $\det{\cal P}_0[g]$ and then the corresponding term would have been absent in the effective gravitational action. However, we are really computing in the massive theory and then the massive Dirac operator  has no zero-modes. Hence, there is no reason to change the definition of the matter partition function and the term $\log\det {\cal P}_0[g,\wh g] $ should be genuinely present.

At order $m^2$ we find the cosmological constant term $\sim m^2 (A-\wh A)$, but also a term $\int \sqrt{\wh g} e^{2\s}\s$ characteristic of the Mabuchi action
\vskip-4.mm
 \be\label{Mabuchi}
 S_{\rm M}[g,\wh g]=\frac{4}{A}\int \sqrt{\wh g} \, e^{2\s}\, \s + \ldots ,
 \ee
 Obviously, this term is local in $\s$.  
 If we trace things back, this term originated from the conformal transformation of the singular part of the Green's function $\wt G$ one had to subtract to get the renormalized Green's function $\wt G_{\rm R}$ (see e.g.~\eqref{GgGwhg}). It is interesting to note, that the Mabuchi action appears at present with a coefficient $-\frac{m^2 A}{16\pi}$ while in the case of the massive scalar field one exactly obtained the opposite coefficient $+\frac{m^2 A}{16\pi}$.
 
 The other  terms  present in \eqref{sgravfiniteorderm2} are multi-local in $\s$ and also involve the various Green's function for the reference metric $\wh g$, as well as the area-like parameters ${\cal P}_{0,ij}$. Such ``non-local" terms involving the Green's functions on the manifold also are present in the effective gravitational action for massive scalars \cite{BL2} at higher orders in $m^2$, starting at $m^4$. 
However, it is quite remarkable that, at least at order $m^2$, the gravitational action
$S_{\rm grav}[g,,\wh g]\equiv S_{\rm grav}[\s,\wh g]$ can be expressed as a functional that is local or multi-local in $\s$.
 
One might wonder why our gravitational action \eqref{sgravfiniteorderm2} at order $m^2$ involves the zero-mode projectors when we have shown that corresponding variation \eqref{sgravvar4-tilde-6} could be written solely in terms of ${\cal G}[g]=\int \sqrt{g} \,\wt G_{\rm R}[g]$ ? The reason is that upon integrating the variation from $\s=0$ to $\s$ (from $\wh g$ to $g$) the zero-mode projector changes in a non-trivial way, and subtracting the zero-modes from $G[g]$ is not the same as subtracting them from $G[\wh g]$.
 
 \setlength{\baselineskip}{.57cm}
 \section{Discussion and outlook\label{disandout}}
 
In these notes, we have studied the effective gravitational action for massive fermions in two dimensions, continuing and completing what had been initiated in \cite{BDE}.
The appropriate mass term is a Majorana type mass  term $\int \p^\dag m \g_* \p$, and the spectral analysis we performed was based on the Dirac operator $D=i\Nsl +m\g_*$ whose eigenfunctions necessarily are complex. What might have looked as a simple generalisation of the massive scalar case, actually turned out to be technically quite involved. We performed a detailed study of the corresponding Green's functions, local zeta-functions and local heat kernels of this Dirac operator $D$ and of  its square $D^2$. One of the crucial insights of the present paper was to obtain a simple relation, not only between the Greens's functions $S$ of $D$ and $G$ of $D^2$, but also between the corresponding local zeta-functions $\zeta_-(s,x,y)$ and $\zeta_+(s+\frac{1}{2},x,y)$. By the (inverse) Mellin transform this allowed us to obtain much information about the local heat kernel $K_-(t,x,y)$ from the  small-$t$ expansion of the heat kernel $K_+(t,x,y)$ which we worked out in the appendix. (This confirms some of the conjectures made in \cite{BDE} but also infirms some others.) In particular, we could obtain all the necessary information about the short-distance singularities and the poles of the local zeta-functions $\zeta_\pm(s,x,x)$ at coinciding points.  Most importantly we studied the variations of these quantities under infinitesimal conformal rescalings of the metric, and then ``integrated" these infinitesimal variations to get the finite effective gravitational action $S[g,\wh g]$. 

The gravitational action  was obtained by performing an expansion in powers of the (small) mass and, accordingly, we need to define all our Green's and local zeta functions also for vanishing mass. In this case there are generically (for genus larger or equal to one) zero-modes of the massless Dirac operator that must be excluded from the spectral decomposition of these Green's or local zeta functions and, hence, a detailed understanding of the contributions of these fermionic zero-modes was necessary. 
Of course the number and properties of the zero-modes crucially depend on the topology of the manifold. Nevertheless, we were able to sufficiently characterise these zero-mode parts of the Green's functions, allowing us to make general statements for arbitrary genus. In particular, the dependence of the zero-modes (and only of the zero-modes) on the conformal factor turned out to be rather simple, which allowed us to completely characterise the different projectors on these zero-modes we needed to introduce. This was one of the important new insights in these notes as compared to what was done in \cite{BDE}.  

We could thus obtain an expansion of the gravitational action, for arbitrary genus, to all powers in $m^2$, involving higher and higher Green's functions at coinciding points. At order $m^0$ the  effective gravitational action contains  the well-known Liouville action with the appropriate coefficient for a Majorana fermion, namely $\frac{1}{2}$ times the one for a single scalar field. There is also a certain contribution involving the normalisation factors of the zero-modes which depend on a set of ``area-like" parameters. While in the massless theory this contribution could be removed by a correspondingly different definition of the matter partition function from the outset, there is no reason to do this in the massive theory and then these terms constitute a genuinely new contribution at order $m^0$.

At order $m^2$ we found a local contribution  $\int \sqrt{\wh g}\, \s\, e^{2\s}$, characteristic of the Mabuchi action. This term appeared with the same coefficient as for a massive scalar field, but with the opposite sign, as one might perhaps have expected. 
But we also found, at this
order $m^2$, contributions that can be expressed in terms of integrals and multiple integrals involving  the conformal factor $\s$ at the different integration points, and the Green's function of $D$ for the background metric $\wh g$ and the renormalized Green's function of $D^2$ at coinciding points, also in the background metric $\wh g$, as well as  on the area-like parameters ${\cal P}_{0,ij}$. To appreciate the degree of non-locality of these additional terms, one must remember that the knowledge of a Green's function encodes much non-local information about the manifold. But the Green's function which appear are only those for metric $\wh g$, so in this sense these contributions are ``non-local in $\wh g\, $" only, and local or multi-local in the conformal factor $\s$. Such multi-local or non-local terms are to be expected and were also present  in the scalar case starting at order $m^4$ \cite{BL2}. Finally we argued  that   the area-like constants ${\cal P}_{0,ij}$ that appear in the zero-mode projectors  are not more or less non-local than a term $\frac{1}{A}$.

The effective gravitational action is of course to be used in the next step, if one wants to compute the functional integral over the geometries. This will involve in particular a functional integration over the conformal factor ${\cal D}\s$. The form of the effective gravitational action we have obtained explicitly displays this $\s$-dependence, having it separated from the quantities that only depend on the fixed ``background" metric $\wh g$. Our explicit separation should   simplify this subsequent integration over the geometries~! We hope to return to this issue elsewhere.

\vskip3.mm
\noindent{\bf Acknowledgements}

\vskip3.mm
\noindent
M.N. is supported by a ``Laboratoire d'Excellence" grant of the physics department of the ENS. 
 
\newpage

\vskip1.cm
\setlength{\baselineskip}{.50cm}

\begin{appendix}

\section{Appendix : Small-$t$ expansion for heat kernels of type $K_+(t,x,y)$}

In this appendix we will show how the small-$t$ expansion of heat kernels for general (positive) second-order differential operators with a standard initial condition can be obtained by a recursive scheme. We then specialise these results to the squared Dirac operator in two dimensions. This will yield the expansion of $K_+(t,x,y)$. 
There is an abundant literature about small-$t$ expansions of heat kernels, see \cite{Gilkey, Avramidi, Vassil}, which  focus much on the ``diagonal" elements, i.e.~$x=y$. Here we will follow the more pedestrian approach of references \cite{BF, Bdiff} from which many of the formulae are borrowed.
Our computation will be facilitated by using Riemann normal  coordinates which we discuss first.

\subsection{Riemann normal coordinates}\label{riemappsec}

It is often useful to go to a special coordinate system for which the metric in the vicinity of a given point is as close to the Euclidean metric as possible. In general relativity, these are the coordinates of a freely falling observer. More generally, they are called Riemann normal coordinates. We will discuss here their definition and properties in an arbitrary dimension $n$.

\subsubsection*{Definition and basic relations}

We fix any point as the origin (which would have coordinates $y$ in a general coordinate system). Riemann normal coordinates $x$ around this point are defined by taking $n$ independent geodesics through this point and defining the geodesic distance from the origin along the $k^{\rm th}$ geodesic as the new coordinate $x^k$. Taking independent geodesics means that their tangent vectors at the origin are $n$ linearly independent vectors.  We may furthermore chose these tangent vectors as orthogonal. Then by rescaling the $x^k$ appropriately one can achieve that in these new coordinates the geodesic distance of the point $x$ from the origin is\footnote{
In this appendix we use latin indices $i,\ j,\ k,\ l,\ldots$ with the present normal coordinates. Of course, they are ``coordinate indices", not to be confused with the ``flat" indices $a,\ b,\ c$ of the local orthonormal frames used in section~2.
}
\be\label{geodistnormal}
\ell^2(x,0)=x^i x^i\equiv x^2 \ .
\ee
This looks like the standard Euclidean distance, but it does not mean that the space is flat.  It only means that the metric at the origin is $\dd_{ij}$ and that its first derivatives, and hence the Christoffel symbols, vanish at the origin. However, this change of coordinates cannot remove the curvature and, indeed, one can show that the Taylor expansion of the metric around $x=0$ is given by (see e.g. \cite{Schubert})
\begin{align}\label{metricnormalexp}
g_{ij}(x)&= \delta_{ij}-{1\over 3} R_{ikjl}x^k x^l -{1\over 6}R_{ikjl;m}x^k x^l x^m 
\nonumber\\
&\hskip3.mm+\ \Big[{2\over 45} R_{ikrl}R^r_{\ mjn} 
-{1\over 20}R_{ikjl;mn}\Big] x^k x^l x^m x^n +O(x^5) \ ,
\end{align}
where $R^i_{\ jkl}$  is the Riemann curvature tensor, defined as
\be\label{Riemann tensor}
R^i_{\ jkl}=\del_k \G^i_{lj}-\del_l \G^i_{kj}+\G^i_{kr}\G^r_{lj}-\G^i_{lr}\G^r_{kj} \ ,
\ee
satisfying $R_{ijkl}=-R_{jikl}=-R_{ijlk}=R_{klij}$. Here $(\ldots)_{;mn}=\nabla_n\nabla_m(\ldots)$ denotes covariant derivatives, and 
\be\label{Ricci}
\cR_{kl}=R^j_{\ kjl} \equiv R_{kjl}^{\ \ \ j}\quad , \quad  \cR=\cR^k_{\ k} \ ,
\ee 
denote the Ricci tensor and Ricci curvature scalar. All curvature tensors in \eqref{metricnormalexp} are evaluated at $x=0$. In particular, since $g_{ij}(0)=\delta_{ij}$ we do not have to distinguish upper and lower indices on these tensors in this formula. The inverse metric is 
\begin{align}\label{inversemetricexp}
g^{ij}(x)=&\ \delta_{ij}+{1\over 3} R_{ikjl}x^k x^l +{1\over 6}R_{ikjl;m}x^k x^l x^m 
\nonumber\\
&\hskip-3.mm+\ \Big[{1\over 15} R_{ikrl}R^r_{\ mjn} 
+{1\over 20}R_{ikjl;mn}\Big] x^k x^l x^m x^n +O(x^5) \ .
\end{align}
Note that in these coordinates one has
$g_{ij}(x) x^j=x^i $ and $g^{ij}(x) x^j = x^i$
since all other terms involve symmetric products of the coordinates $x$ contracted with antisymmetric curvature tensors. The square root of the determinant of the metric is
\begin{align}\label{sqrtg}
\sqrt{g(x)}\ =\ &1-{1\over 6} \cR_{kl}x^k x^l -{1\over 12} \cR_{kl;m} x^k x^l x^m \nonumber\\
&+ \left[
{1\over 72}  \cR_{kl} \cR_{mn} -{1\over 40}\cR_{kl;mn} -{1\over 180} R^r_{\ kls}R^s_{\ mnr}
\right] x^k x^l x^m x^n +O(x^5) \ ,
\end{align}
Note that the terms $O(x^5)$ are terms $O(R^3, \nabla R^2)$, i.e. involving at least 3 curvature tensors or 2 curvature tensor and a covariant derivative.

\subsubsection*{Scalar Laplace operator}

The scalar Laplace operator is given in general by 
\be\label{Laplace}
\Delta={1\over \sqrt{g}}\del_i\left(\sqrt{g} g^{ij}\del_j \right) 
=g^{ij}(x)\del_i\del_j +f^j(x) \del_j 
\quad , \quad
f^j=-g^{kl}\G_{kl}^j =\del_i g^{ij} + g^{ij} \del_i \log  \sqrt{g} \ .
\ee
Note that the scalar Laplacian is a negative operator, i.e.~its eigenvalues are $\le 0$ with a single zero-mode given by the constant function. In order to have a positive (or rather non-negative) operator the Laplacian is sometimes defined with the opposite sign. We will keep our definition \eqref{Laplace} so that the non-negative operator is $(-\D)$. It is indeed $-\D$ that appears in our squared Dirac operator.
Using \eqref{inversemetricexp} and \eqref{sqrtg}, a straightforward computation yields for the function $f^j(x)$ in the present normal coordinates
\ba\label{ffct2}
\hskip-0.mm f^j(x)&=&-{2\over 3} \cR_{jk}x^k +\Big[ {1\over 12} \cR_{kl;j}-{1\over 2} \cR_{kj;l}\Big] x^k x^l 
-
\Big[{1\over 5} \cR_{jl;mn}+{1\over 40} \cR_{mn;lj}-{3\over 40} \cR_{mn;jl}\nonumber\\
&&\hskip0.5cm 
- \ {23\over 180} \cR_{ls} R^s_{\ mnj}+{4\over 45} R_{rjls}R_{smnr}\Big] x^l x^m x^n  
\ +\ O(x^4) \ .
\ea
This, together with \eqref{inversemetricexp}, yields the expansion of the scalar Laplace operator in the vicinity of $x=0$, up to terms $O(x^4)$ or $O(R^3,\nabla R^2)$. In particular, we have 
$\Delta\, \ell^2(x,0)=\hskip-2.mm \left[g^{ij}(x)\del_i\del_j + f^j(x)\del_j\right] (x^k x^k)
= 2\left[g^{jj}(x)+f^j(x) x^j\right]$, so that (in $n$ dimensions)
\be\label{Laplacegeodist2}
\Delta\, \ell^2(x,0)-2n =-{2\over 3} \cR_{kl} x^k x^l -{1\over 2} \cR_{kl;m}x^k x^l x^m
 -\left[{1\over 5} \cR_{kl;mn} +{2\over 45} R^r_{\ kls}R^s_{\ mnr}\right] x^k x^l x^m x^n  
\ +\ O(x^5) .
\ee

\subsubsection*{Spinorial Laplace operator in 2 dimensions}

We also need to work out similarly the spinorial Laplace operator \eqref{spinorLapl2} in normal coordinates.  We will do this only for the case of interest in this paper, namely in 2 dimensions and only up to  the order we need. Recall that we had 
\be\label{spinorLapl2-app}
 \D_{\rm sp} = {\bf 1}_{2\times 2}\Big( \D_{\rm scalar} -\frac{1}{16}\o^\m\o_\m \Big)
 -\frac{i}{4}\, \g_*\, \Big( (\nabla_\m \o^\m) + 2 \o^\m \del_\m \Big) \ .
 \ee
So the first thing we need to figure out is the expansion of the $\o_\m=2\o^{12}_\m$ around the origin in Riemann normal coordinates. Rather then using some general formula, we will determine the zwei-beins $e^a$ and then solve the zero-torsion condition to the order we are interested in. It will turn out that we only need to determine $\o_\m$ up to first order in $x$. This means that we will need the expansions of the $e^a$ up to and including terms of order $(x)^2$. From \eqref{metricnormalexp} we get
\ba\label{metricnormal2form}
e^a \otimes e^a\hskip-2.mm&=&\hskip-2.mm g_{ij}\ \d x^i \otimes \d x^j= \d x^i \otimes \d x^i - \frac{1}{3} R_{ikjl}x^k x^l \, \d x^i\otimes \d x^j +{\cal O}(x^3) \nonumber\\
&=&\hskip-2.mm \d x^i \otimes \d x^i - \frac{1}{3}\Big( R_{1212}x^2x^2 \d x^1\otimes \d x^1
+ R_{1221}x^2x^1 \d x^1\otimes \d x^2 \nonumber\\
&&\hskip2.5cm +  R_{2112}x^1x^2 \d x^2\otimes \d x^1
+ R_{2121}x^1x^1 \d x^2\otimes \d x^2\Big) +{\cal O}(x^3) \nonumber\\
&=&\hskip-2.mm \d x^i \otimes \d x^i - \frac{\cR}{6} \Big( x^2 x^2  \d x^1\otimes \d x^1
-  x^1 x^2  (\d x^1\otimes \d x^2+\d x^2\otimes \d x^1)
+x^1 x^1 \d x^2\otimes \d x^2 \Big) +\, {\cal O}(x^3)\nonumber\\
\ea
where we used $R_{1212}=-R_{2112}=-R_{1221}=R_{2121}=\frac{1}{2}\cR$. From this we read
\ba\label{eas}
e^1&=&\d x^1 - \frac{\cR}{12} \big(x^2 x^2 \d x^1-x^1 x^2 \d x^2\big) +\, {\cal O}(x^3) 
\ , \nonumber\\
e^2&=&\d x^2 - \frac{\cR}{12} \big(x^1 x^1 \d x^2-x^1 x^2 \d x^1\big) +\, {\cal O}(x^3) 
\ .
\ea
Next,
\be\label{dea}
\d e^1=\frac{\cR}{4}  \, x^2\, \d x^1 \wedge \d x^2 +\, {\cal O}(x^2) \quad , \quad
\d e^2=-\frac{\cR}{4}  \, x^1\, \d x^1 \wedge \d x^2  +\, {\cal O}(x^2) \ .
\ee
The zero-torsion conditions $\d e^1+\o^{12}\wedge e^2=0=\d e^2 -\o^{12}\wedge e^1$ yield \be\label{omeganormalexp}
\o^{12}=-\frac{\cR}{4}  \, x^2 \d x^1 + \frac{\cR}{4}  \, x^1 \d x^2 +\, {\cal O}((x)^2)
\quad \Leftrightarrow\quad 
\o^{12}_1=-\frac{\cR}{4}  \, x^2 +\, {\cal O}((x)^2) \ \ , \ \ \o^{12}_2=  \frac{\cR}{4}  \, x^1 +\, {\cal O}((x)^2)
\ .
\ee
Recall that we defined $\o_\m=2\o^{12}_\m$ and, hence, to this order
\be\label{omeganormalexp2}
\o_1=\o^1=-\frac{\cR}{2}  \, x^2 +\, {\cal O}((x)^2) \ \ , \ \ \o_2=\o^2=  \frac{\cR}{2}  \, x^1 +\, {\cal O}((x)^2)
\ .
\ee
Since the Christoffel symbols are ${\cal O}(x)$ it follows that
\be\label{omegaexpressions}
 \o^\m\o_\m=\, {\cal O}((x)^2) \quad , \quad
\nabla_\m \o^\m=\del_\m\o^\m +\, {\cal O}((x)^2) =\, {\cal O}(x) 
\quad , \quad
\o^\m \del_\m=\frac{\cR}{2} ( x^1\del_2-x^2\del_1) + {\cal O}((x)^2\del)
 \ .
\ee

\subsection{Small-$t$ expansion of the heat kernel\label{heatapp}}

We now turn to the small-$t$ expansion of the  heat kernels $K(t,x,y)$
for a general class of second-order differential operators $\cD$ on an $n$-dimensional manifold, satisfying the standard initial condition. This yields heat kernels generalising our $K_+(t,x,y)$.
We then specialise these results to the squared Dirac operator $\cD=D^2$ in 2 dimensions.

We will suppose that the second-order differential operator is (the negative) of a  generalised Laplace type, i.e. of the form
\be\label{gendiffD}
-\cD={\bf 1} \,g^{ij}(x)\del_i\del_j + A^i(x) \del_i + B(x) \quad , \quad i,j=1,\ldots n \ ,
\ee
with ${\bf 1}, A^i(x)$ and $B(x)$ being $p\times p$-matrices and $g^{ij}$ a positive (inverse) metric. Note the minus-sign on the left-hand side needed so that $\cD$ can be a positive differential operator for appropriate $A^i$ and $B$. 

We want to find the solution $K(t,x,y)$ of the corresponding heat equation
\be\label{heateqdiffgen}
 {\d\over \d t} K(t,x,y) =-\cD K(t,x,y) \ ,
\ee
with initial condition
\be\label{initial}
K(t,x,y) \sim \ \delta^{(d)}(x-y) \, [g(x) g(y)]^{-1/4}\,  {\bf 1} \quad {\rm as}\ t\to 0 \ .
\ee
If $\cD$ is hermitian and we denote by $\L_n$ and $\vf_n$ its eigenvalues and ($p$-component) eigenfunctions, then
\be\label{Keigensum}
K(t,x,y)=\sum_n e^{-\L_n t} \vf_n(x) \vf_n^\dag(y) 
\ee
is the solution to the heat equation with the correct initial condition. Indeed, ${\d\over \d t} K(t,x,y)$\break $=- \sum_n \L_n e^{-\L_n t} \vf_n(x) \vf_n^\dag(y) = - \cD \sum_n e^{-\L_n t} \vf_n(x) \vf_n^\dag(y) =-\cD K(t,x,y)$ and for $t\to 0$ one gets $K(t,x,y)\to \sum_n \vf_n(x) \vf_n^\dag(y)$ which by  the completeness relation equals the right-hand side of \eqref{initial}. As is clear from \eqref{initial} and \eqref{Keigensum}
this heat kernel generalises  indeed what we called $K_+$ in the main text. The other heat kernel $K_-$ will be studied to a certain extent in appendix B.

For $A^i=B=0,\ g^{ij}=\dd^{ij}$ the differential operator is just $p$ copies of the flat space Laplace operator and the well-known solution is $K(t,x,y)= \ (4\pi t)^{-n/2} \exp\left( -{(x-y)^2\over 4 t}\right)\,  {\bf 1}$.
This must also be the leading small $t$, small distance behaviour on a curved manifold, if $(x-y)^2$ is replaced by the square of the geodesic distance $\ell^2(x,y)$ between $x$ and $y$. Corrections to this leading behaviour can then be obtained as a perturbative expansion in $t$ and $\ell(x,y)$. 

\subsubsection*{Asymptotic expansion and recursion relations}

We will now show that the heat kernel $K(t,x,y)$ admits an asymptotic small $t$ expansion around the ``flat-space" solution. We work out the recursion relations between the expansion coefficients in general. Then we use normal coordinates to explicitly solve these recursion relations and obtain the first few coefficients explicitly. This discussion is essentially taken from \cite{BF, Bdiff}, but generalises it slightly.

We search for a solution of (\ref{heateqdiffgen}), (\ref{initial}) of the form
\be\label{Kseries}
K(t,x,y)= (4\pi t)^{-n/2} \exp\left( -{\ell^2(x,y)\over 4 t}\right)
F(t,x,y) \ .
\ee
where $\ell(x,y)$ is the geodesic distance between $x$ and $y$ and $F$ is expanded in integer non-negative powers of $t$,
\be\label{Fexp}
F(t,x,y)=\sum_{r=0}^\infty F_r(x,y) t^r \ .
\ee
$K$ being a $p\times p$-matrix, the same is true for the coefficient functions $F_r(x,y)$

Let us begin by computing $g^{ij}\del_i \del_j K$~:
\be\label{secondderivonKF}
g^{ij}(x) \del_i\del_j\left( {e^{-\ell^2/(4t)} F\over t^{n/2}} \right)
={e^{-\ell^2/(4t)}\over t^{n/2}}\Big[\frac {g^{ij}(x)\del_i\ell^2\del_j\ell^2}{16 t^2} F -\frac{g^{ij}(x)\del_i\del_j \ell^2}{4t} F - \frac{g^{ij}(x)\del_i\ell^2}{2t}  \del_j F +g^{ij}(x)\del_i\del_j F \Big]\, ,
\ee
where $\del_i\equiv {\del\over \del x^i}$.
It is not too difficult to show that
\be\label{ellder}
g^{ij}(x)\del_i\ell(x,y)\del_j\ell(x,y)=1 \quad \Rightarrow \quad g^{ij}(x)\del_i\ell^2 \del_j \ell^2 = 4 \ell^2 \ .
\ee 
(On the other hand, there is no such simple result for $g^{ij}(x)\del_i\del_j \ell^2$.)
The remaining pieces in $\cD$ give
\be\label{firstderivonKF}
\big(A^i(x) \del_i+B(x)\big)\left( {e^{-\ell^2/(4t)} F\over t^{n/2}} \right)
={e^{-\ell^2/(4t)}\over t^{n/2}}
\Big[ -\frac{A^i(x)\del_i\ell^2}{4t} F + A^i \del_i F +B(x)F \Big]\, ,
\ee
Adding the two parts  we get
\ba\label{cDonKF}
&&-\cD_x\, \left( {e^{-\ell^2/(4t)} F\over t^{n/2}} \right)
={e^{-\ell^2/(4t)}\over t^{n/2}}
\Big[\frac {\ell^2}{4 t^2} F -\frac{g^{ij}(x)\del_i\del_j \ell^2}{4t} F - \frac{g^{ij}(x)\del_i\ell^2}{2t}  \del_j F -\frac{A^i(x)\del_i\ell^2}{4t} F \nonumber\\
&&\hskip5.5cm +\underbrace{g^{ij}(x)\del_i\del_j F + A^i \del_i F +B(x)F}_{\cD_x F}
\Big]\, ,
\ea
On the other hand,
\be\label{dtKF}
{\d\over \d t} \left({e^{-\ell^2/(4t)} F\over t^{n/2}}\right)
={e^{-\ell^2/(4t)}\over t^{n/2}}\Big[{\ell^2\over 4 t^2}F-{n\over 2 t}F +{\d F\over \d t}\Big] \ ,
\ee
Upon inserting \eqref{cDonKF} and \eqref{dtKF} into the heat equation \eqref{heateqdiffgen} we see that the leading singular terms $\sim \frac{1}{t^2}$ in the square bracket cancel, and the heat equation amounts to
\be\label{heateqexp}
 \frac{g^{ij}(x)\del_i\del_j \ell^2-2n}{4t} F + \frac{g^{ij}(x)\del_i\ell^2}{2t}  \del_j F +\frac{A^i(x)\del_i\ell^2}{4t} F  +{\d F\over \d t}= \cD_x F \ .
\ee
The terms on the left-hand side either have an explicit $\frac{1}{t}$ or a derivative $\frac{\d}{\d t}$ which also decreases the power of $t$ by one unit. Inserting the expansion \eqref{Fexp} of $F$ yields the desired recursion relations~:
\begin{align}\label{foeq}
\frac{1}{4}\Big(g^{ij}(x)\del_i\del_j \ell^2-2n +A^i(x)\del_i\ell^2\Big) F_0 + \frac{1}{2}g^{ij}(x)\del_i\ell^2 \del_j F_0  &= 0
 \ ,
\\
\label{freq}
 \frac{1}{4}\Big(g^{ij}(x)\del_i\del_j \ell^2-2n +A^i(x)\del_i\ell^2\Big)F_r + \frac{1}{2}g^{ij}(x)\del_i\ell^2 \del_j F_r   + r F_r &= \cD_x F_{r-1} \ , \ (r\ge 1) \ ,
\end{align}
(where again $\del_i\equiv {\del\over \del x^i}$.) The recursion relation \eqref{freq} can be somewhat simplified by writing $F_r=F_0 \wt F_r$. One then gets equivalently
\be\label{freqbis}
\frac{1}{2}g^{ij}(x)\del_i\ell^2 \del_j \wt F_r + r \wt F_r=F_0^{-1} \cD_x (F_0 \wt F_{r-1}) \ .
\ee

Before we turn to solving these recursion relations, let us note the obvious fact that that $K$ will be exponentially small unless $\ell^2/t$ is not too large. This means that as an order of magnitude estimate we must have $\ell \le \sqrt{t}$, and the small $t$ expansion must also be a small $\ell$ expansion. Thus if one is interested in expanding $F$ up to order $t^2$ it is consistent to expand $F_0(x,y)$ in powers of $( x-y)\sim\ell$ up to fourth order, expand $F_1$  in powers of $( x-y)\sim\ell$ up to second order, and for $F_2$ only keep $F_2(y,y)$. For the purpose of the main text it will be sufficient to expand to order $t$ only, which means obtaining $F_0$ up to order $\ell^2$ and $F_1$ only to order 0, i.e.~$F_1(y,y)$. Since this small $t$  / small $\ell$ expansion only explores the vicinity of the point $y$, it is much useful to introduce Riemann normal coordinates around $y$ which will simplify the calculations quite a bit.

\subsubsection*{Solution to the recursion relations of the heat equation in normal coordinates}

Let us then introduce  normal coordinates centered in $y$, i.e $F_r(x,y)\to F_r(x,0)$. Recall that $\ell^2=x^i x^i$ so that $\del_i\ell^2=2x^i$, and $g^{ij}\del_j \ell^2=2 x^i$ with all other terms cancelling due to the antisymmetry properties of the Riemann tensor, cf \eqref{inversemetricexp}. From this same equation one also gets
\be\label{gdeldelell}
g^{ij}\del_i\del_j\ell^2=g^{ij} 2 \dd_{ij}=2g^{ii}=2\dd^{ii}+\frac{2}{3}R_{ikil}x^k x^l +{\cal O}(x^3)
=2n + \frac{2}{3} \cR_{kl}x^k x^l +{\cal O}(x^3) \ .
\ee
Then relation \eqref{foeq} for $F_0$ gives
\be\label{F0normal}
x^j\del_j F_0=-\Big( \frac{1}{6}  \cR_{kl}x^k x^l +\frac{1}{2} A^l x^l + {\cal O}(x^3) \Big) F_0\ .
\ee
Let us assume that
\be\label{Aiexp}
A^i(x)=a_{ij}x^j+{\cal O}(x^2) \ ,
\ee
with  matrix-valued coefficients $a_{ij}$.
Then $A^l x^l=a_{kl}x^k x^l + {\cal O}(x^3)$ and the $F_0$-equation is
\be\label{F0normal-2}
x^j\del_j F_0=-\Big( \frac{1}{6}  \cR_{kl}x^k x^l +\frac{1}{2} a_{kl} x^k x^l + {\cal O}(x^3) \Big) F_0\ ,
\ee
with solution
\be\label{F0normal-3}
F_0=1 -\frac{1}{12}  \cR_{kl}x^k x^l -\frac{1}{4} a_{kl} x^k x^l + {\cal O}(x^3) \ .
\ee
Next, the recursion relation for $F_1$ reads
\be\label{F1eq}
\Big( \frac{1}{6}  \cR_{kl}x^k x^l +\frac{1}{2} a_{kl} x^k  x^l + {\cal O}(x^3) \Big) F_1 + x^j\del_j F_1 + F_1 =\cD_x F_0 \ .
\ee
If we only want to get $F_1$ to order $(x-y)^0$ this becomes
\be\label{F1eq-2}
 F_1 + {\cal O}(x)=\cD_x F_0 \ .
\ee
We have
\be\label{DxF0}
\cD_x F_0=g^{ij}\big(-\frac{1}{6}\cR_{ij}-\frac{1}{2}a_{ij}\big)+B  + {\cal O}(x^2) 
=  -\frac{1}{6}\cR -\frac{1}{2} g^{ij}a_{ij} + B  + {\cal O}(x^2) \ ,
\ee
so that finally
\be\label{F1eq-2-2}
F_1= -\frac{1}{6}\cR -\frac{1}{2} g^{ij}a_{ij} + B  + {\cal O}(x)  \ .
\ee
As a simple check of these formula, let us choose $-\cD=\D_{\rm scalar}$, i.e.~$p=1$ (no matrices) and $A^i=f^i=-\frac{2}{3}\cR_{ij}x^j+{\cal O}((x)^2)$ and $B=0$, according to \eqref{Laplace} and \eqref{ffct2}. Then $a_{ij}=-\frac{2}{3} \cR_{ij}$ and 
\be\label{Frlapl}
F_0^{\D}=1 +\frac{1}{12}  \cR_{kl}x^k x^l + {\cal O}(x^3)
\quad , \quad
F_1^{\D}=+\frac{1}{6}\cR+{\cal O}(x) \ ,
\ee
which are indeed the correct well-known values for the scalar Laplace operator.

Note that in the formula for the $F_r$, the $\cR_{kl}$, $\cR$ and $a_{kl}$ are to be taken at $y=0$ which was the origin of our normal coordinates. More generally, if we use normal coordinates around an arbitrary point $y$ (without shifting the coordinates) the $\cR_{kl}$, $\cR$ and $a_{kl}$ are to be evaluated at $y$ and the $x^k$ appearing should be $(x-y)^k$. Thus
\ba\label{F01normal}
F_0(x,y)&=&1 -\big(\frac{1}{12}  \cR_{kl}(y)+\frac{1}{4} a_{kl}(y)\big)  (x-y)^k (x-y)^l + {\cal O}((x-y)^3) 
\nonumber\\
F_1(x,y)&=& -\frac{1}{6}\cR(y) -\frac{1}{2} g^{ij}a_{ij}(y) + B(y)  + {\cal O}(x-y) 
\ .
\ea
In particular, at coinciding points we have
\be\label{F01coinc}
F_0(y,y)=1 \quad , \quad F_1(y,y) =-\frac{1}{6}\cR(y) -\frac{1}{2} g^{ij}a_{ij}(y) + B(y) \ .
\ee
This is independent of the normal coordinates used and true in an arbitrary coordinate system.

\subsubsection*{The heat kernel expansion for the 2-dimensional Dirac operator $D^2$}

Let us finally specify our formula to 2 dimensions and $\cD=D^2$, the squared Dirac operator, given in \eqref{Dsquared-2d} and \eqref{spinorLapl2} (we now use again $\m,\n,\ldots$ for the coordinate indices):
\ba\label{Dsquared-2d-app}
D^2&=& -\D_{\rm sp} +\frac{1}{4} \cR + m^2 
=-\D_{\rm scalar} +\frac{1}{16}\o^\m\o_\m 
 +\frac{i}{4}\, \g_*\, \Big( (\nabla_\m \o^\m) + 2 \o^\m \del_\m \Big)  +\frac{1}{4} \cR + m^2  \nonumber\\
&=& -g^{\m\n}\del_\m\del_\n -\big( f^\m -\frac{i}{2}\o^\m  \g_*\big) \del_\m
+\frac{1}{16}\o^\m\o_\m 
 +\frac{i}{4}\, \g_*\,  (\nabla_\m \o^\m)  +\frac{1}{4} \cR + m^2
\ ,
\ea
where $f^\m=\del_\n g^{\m\n}+g^{\m\n}\del_\n \log \sqrt{g}$ was given in \eqref{Laplace}.
From this we identify
\be\label{AandB2d}
A^\m= f^\m -\frac{i}{2}\o^\m  \g_*
\quad , \quad
B=-\frac{1}{16}\o^\m\o_\m 
 -\frac{i}{4}\, \g_*\,  (\nabla_\m \o^\m)  -\frac{1}{4} \cR - m^2 \ .
\ee
We know from \eqref{ffct2} that in normal coordinates around $y=0$ we have $f^\m=-\frac{2}{3}\cR_{\m\n} x^\n+{\cal O}((x)^2)$ and from \eqref{omeganormalexp2} that $\o^1=-\frac{\cR}{2} x^2+{\cal O}((x)^2)$ and $\o^2=\frac{\cR}{2} x^1+{\cal O}((x)^2)$. Recall that $a_{\m\n}$ was defined as $A^\m(x)=a_{\m\n}x^\n +{\cal O}((x)^2)$, so that
\be\label{aij2d}
a_{11}=-\frac{2}{3} \cR_{11}
\quad , \quad
a_{12}=-\frac{2}{3} \cR_{12}+\frac{i}{4}\cR \g_*
\quad , \quad
a_{21}=-\frac{2}{3} \cR_{12}-\frac{i}{4}\cR \g_*
\quad , \quad
a_{22}=-\frac{2}{3} \cR_{22} \ .
\ee
It follows that (still in normal coordinates)
\be\label{gijaij}
g^{\m\n}a_{\m\n}=-\frac{2}{3} \cR +{\cal O}((x)^2) \ .
\ee
Also (see \eqref{omegaexpressions})
\be\label{Bexpr2d}
B= -\frac{1}{4} \cR - m^2 +{\cal O}(x) \ .
\ee
It is maybe a bit disappointing, but we see that, at this order, no effect of the spin-connection $\o^\m$ and of the $\g_*$-matrix structure survives. We finally find for the first two heat kernel coefficients at coinciding points  (cf~\eqref{F01coinc})~:
\be\label{F01coinc2d}
F_0(y,y)=1 \quad , \quad F_1(y,y) =-\frac{1}{12}\cR(y)-m^2 \ .
\ee
Let us emphasise again that these results concern the heat kernel $K_+$.


\section{Appendix : The small-$t$ asymptotics of $K_-(t,x,y)$ for the flat torus}

In this appendix, we will similarly try to obtain the small-$t$ asymptotics of $K_-$. Although $K_-(t,x,y)$ satisfies the same differential equation as $K_+(t,x,y)$, it does not have any ``useful" initial condition and the method of the previous appendix then does not apply. Instead, we will focus on the case of the flat (square) torus where we explicitly know the eigenvalues and eigenfunctions.

As discussed in sect.~4, it will be enough to determine the small-$t$ asymptotics of ${\cal L}$. However, as we will see, this turns out to be still rather non-trivial. Before discussing the more complicated case of $x\ne y$, let us first study  the case of coinciding points $x=y$. Then $K_-(t,x,x)$ is given by \eqref{LKmxxagain}
in terms of ${\cal L}(t,0)$ which is
\be\label{calLde0}
{\cal L}(t,0) = \sum_{n_1,n_2} \frac{e^{-\l_{\vec{n}}^2\, t} }{\l_{\vec{n}}}
\equiv \sum_{n_1,n_2} \frac{e^{-(n_1^2+n_2^2+m^2)\,t} }{\sqrt{n_1^2+n_2^2+m^2}} \ .
\ee
The exponential provides an effective cut-off to the sum, restricting it to values of $n_1, n_2$ that are less than a few $\frac{1}{\sqrt{t}}$.  For small $t$ these are still very many contributions. One may then try to estimate this sum by replacing it by the corresponding integral. For $m$ not too small this turns out to be quite a good approximation. However, as $m\to 0$, the sum is dominated by the single point $n_1=n_2=0$ contributing $\frac{1}{m}$. Separating this term we write
\be\label{calLde0-2}
{\cal L}(t,0) = \frac{e^{-m^2t}}{m} + \sum_{(n_1,n_2)\ne (0,0)} \frac{e^{-(n_1^2+n_2^2+m^2)t} }{\sqrt{n_1^2+n_2^2+m^2}} 
\simeq  \frac{e^{-m^2t}}{m} + \int_{(p_1,p_2)\notin D_1} \d^2 p \frac{e^{-(p_1^2+p_2^2+m^2)t} }{\sqrt{p_1^2+p_2^2+m^2}} 
\ ,
\ee
where $D_1$ denotes a disc of unit area (radius $r_0=\pi^{-1/2}$) around the origin, corresponding to the ``area" of the missing value $n_1=n_2=0$. Upon going to polar coordinates ($p_1=p\cos\f, \ p_2=p\sin\f$), this becomes
\ba\label{calLde0-3}
{\cal L}(t,0) 
&\simeq&  \frac{e^{-m^2t}}{m} + 2\pi \int_{r_0}^\infty \d p\,  \frac{p\ e^{-t (p^2+m^2)} }{\sqrt{p^2+m^2}} 
= \frac{e^{-m^2 t}}{m} + \pi \int_{r_0^2}^\infty \d \xi\,  \frac{e^{-t(\xi+m^2)} }{\sqrt{\xi+m^2}}
= \frac{e^{-m^2 t}}{m} + \pi \int_{r_0^2+m^2}^\infty \d \xi\,  \frac{e^{-t\xi} }{\sqrt{\xi}} \nonumber\\
&=& \frac{e^{-m^2 t}}{m} + 2\pi \int_{\sqrt{r_0^2+m^2}}^\infty \d p\,  e^{-t p^2} 
=\frac{e^{-m^2 t}}{m} + 2\pi \Big( \frac{1}{2} \sqrt{\frac{\pi}{t}} - \int_0^{\sqrt{r_0^2+m^2}} \d p\,  e^{-t p^2} \Big)\nonumber\\
&=&\frac{e^{-m^2 t}}{m} + \frac{\pi^{3/2}}{\sqrt{t}}-2\pi \sqrt{r_0^2+m^2}\,  
\sum_{n=0}^\infty \frac{ (-)^n (r_0^2+m^2)^n }{n!(2n+1)}t^n
 \ .
\ea
We see that ${\cal L}(t,0) $ contains  integer powers of $t$, as well as a term $\sim t^{-1/2}$. The integer powers and in particular the order $t^0$ term (except the $\frac{1}{m}$) are affected by the ambiguity in choosing $r_0$ and this reflects the error made in replacing the sum by an integral. So the only thing we can safely conclude is
\be\label{calLde0-5}
{\cal L}(t,0) 
\simeq  \frac{1}{m} + \sqrt{\frac{\pi^3}{t}} +{\cal O}(t^0,m^0)\ .
\ee
By eq.~\eqref{LKmxxagain} this directly gives  the corresponding asymptotics of $K_-(t,x,x)$ as given in
\eqref{Kmoinstxxleading}.

Instead of approximating the sum by an integral, we could try to use the Poisson summation formula. This formula gives the sum as another infinite sum of Fourrier coefficients $\wh f_k$. These Fourrier coefficients are given by integrals and the  Fourrier coefficient $\wh f_0$ actually is just the  integral approximation to the original sum. Thus, adding the (infinite number) of higher Fourrier coefficients could be viewed as providing the correction terms for having replaced the sum by the integral $\wh f_0$. Unfortunately we will not be able to compute exactly the $\wh f_k$ for $k\ne 0$. Let us nevertheless outline this computation. The computation will be the same for ${\cal L}(t,0)$ and ${\cal L}(t,z)$, so we will do the general case allowing non-vanishing $z$. We will use the notation $\vec{z}=(z^1,z^2)$,  $\vec{n}=(n_1,n_2)$, $\vec{p}=(p^1,p^2)$ as well as $\vec{k}=(k_1,k_2)$. Thus we write
\be\label{calLtorus-2}
 {\cal L}(t,z)=
\sum_{n_1,n_2} f(\vec{n}; t,\vec{z}) 
\quad , \quad 
 f(\vec{p};t,\vec{z}) =\frac{e^{-t(\vec{p}^2+m^2)} e^{i\vec{p}\cdot\vec{z}}}{\sqrt{\vec{p}^2+m^2}} \ .
 \ee
 By the Poisson summation formula  one then has
\be\label{LPoisson}
{\cal L}(t,z)=\sum_{k_1,k_2} \wh f(\vec{k}; t,\vec{z}) \ ,
 \ee
 where
 \be\label{fhatdef}
\wh f(\vec{k}; t,\vec{z})=\int \d^2 p\, e^{2\pi i \vec{k}\cdot\vec{p}}  f(\vec{p}; t,\vec{z}) 
=\int \d^2 p\,  \frac{e^{-t(\vec{p}^2+m^2)} e^{i\vec{p}\cdot (\vec{z}+2\pi\vec{k})}}{\sqrt{\vec{p}^2+m^2}} 
= \wh f(0;t,\vec{z}+2\pi \vec{k})\ .
\ee
We see that the $\wh f$ only depend on $\vec{z}$ and $\vec{k}$ through the combination $\vec{u}\equiv\vec{z}+2\pi \vec{k}$. This is why evaluating them at $\vec{z}=0$ or $\vec{z}\ne 0$ essentially presents the same difficulty since we need to evaluate the $\wh f(0,t,\vec{u})$ at non-vanishing $\vec{u}$. 
Using again polar coordinates $p^1=p\cos\f,\ p^2=p\sin\f$, and denoting $p=|\vec{p}|$ and similarly $u=|\vec{u}|$, one has
\be\label{fhatint1}
\wh f(0; t,\vec{u})=\int_0^\infty  \d p\, \int_0^{2\pi} \d\f \,\frac{p\, e^{- t (p^2+m^2)}}{\sqrt{p^2+m^2}}  
e^{ i p u \cos\f}
\ee
For $\vec{u}=0$ this is indeed the integral we computed above. For $\vec{z}\ne 0$ or $\vec{k}\ne 0$, however, we have $\vec{u}\ne 0$ and the angular dependence makes the integral much more difficult to compute.
The angular $\f$-integral can be done exactly as $\int_0^{2\pi} \d\f\, e^{i p u \cos\f}=2\pi\, J_0(p u)$, where $J_0$ is a Bessel function of first kind, so that
\be\label{fhatint1bis}
\wh f(0; t,\vec{u})=2\pi \int_0^\infty  \d p\, \frac{p\, e^{- t (p^2+m^2)}}{\sqrt{p^2+m^2}}  \, J_0(p u) \ .
\ee
The Bessel function $J_0(\xi)$ is real, with $J_0(0)=1$, and as $\xi$ increases it decreases and has its first zero around $\xi\simeq 2.5$, and it then oscillates with an amplitude that decreases $\sim \xi^{-1/2}$. These oscillations will strongly diminish the contribution to the integral of those
 values of $p$ that are larger than a few $\frac{1}{u}$. On the other hand, the exponential $e^{-t p^2}$ also restricts the $p$-values to less than a few $\frac{1}{\sqrt{t}}$. Roughly speaking,  there is an upper cutoff of the order of $\min(\frac{1}{u},\frac{1}{\sqrt{t}})$. 
There does not seem to be any known expression for the integral \eqref{fhatint1bis} for $\vec{u}\ne 0$.  For $\vec{u}=0$ the exact result for this integral can be read from our above computations in \eqref{calLde0-2} and \eqref{calLde0-3} (with $r_0=0$ and without the $\frac{1}{m}$-term)~:
\be\label{fhatzero}
\wh f(\vec{0},t,\vec{0})=\frac{\pi^{3/2}}{\sqrt{t}} -2\pi \int_0^m \d p\, e^{-tp^2}
= \frac{\pi^{3/2}}{\sqrt{t}} 
-2\pi m \sum_{n=0}^\infty \frac{(-)^n m^{2n}}{n!(2n+1)}\, t^n \ .
\ee

Let us now assume that $\vec{u}\ne 0$. We want to study again the small-$t$ asymptotics, at fixed $\vec{u}\ne 0$. This means in particular that we can assume $\frac{1}{\sqrt{t}}\gg \frac{1}{u}$ and thus the effective cut-off scale for the integral \eqref{fhatint1bis} is set by $\frac{1}{u}$.
The difficulty in evaluating the integral comes (among others) from the factor $\frac{p}{\sqrt{p^2+m^2}}$, which equals 0 at $p=0$ and is close to 1 for $p\gg m$. Thus, for small mass $m$ one  could expect that replacing this factor by 1 will induce an error of order $m$ only.
But this means that the leading term actually computes the result for $m=0$ and we know that the original sum has a term $\frac{1}{m}$ which becomes singular in this limit. One could then compute the integral $2\pi\int_0^\infty \d p\, e^{-t p^2} J_0(p u)$ exactly
(ref.~\cite{Erdelyi}, sect. 7.7.3, eq 23) and obtain
$\frac{\pi^{3/2}}{\sqrt{t}}\, e^{-z^2/(8t)} I_0\big(\frac{z^2}{8t}\big) +{\cal O}(m)$, 
where $I_0$ is the modified Bessel function of the first kind, and then 
$\wh f(0;t,\vec{u}) \sim_{t\to 0} \frac{2\pi}{u} +{\cal O}(m)$.
By the Poisson formula \eqref{LPoisson} and \eqref{fhatdef}, we need so sum all $\wh f(0;t,\vec{z}+2\pi \vec{k})$ to obtain ${\cal L}(t,z)$ and the resulting sum obviously diverges.

Let us then proceed differently~: If we are only interested in the leading small-$t$ result, in the integral \eqref{fhatint1bis} we may simply replace $e^{-t p^2}$ by 1. Indeed, recall that this integral is effectively cut off at $\frac{1}{u}\ll \frac{1}{\sqrt{t}}$. Then we may express the result in terms of a known integral~:
\be\label{fhatint2}
\wh f(0; t,\vec{u})
=2\pi \int_0^\infty  \d p\, \frac{p}{\sqrt{p^2+m^2}}  \, J_0(p u)  
\ \big( 1+{\cal O}(t)\big)
=\frac{2\pi }{u}e^{-m u}\  \big( 1+{\cal O}(t)\big)
\ .
\ee
In the full Poisson sum for ${\cal L}(t,z)$ the exponential factor $e^{-m u}$ ensures the convergence~:
\be\label{calLPoisson}
{\cal L}(t,z) \sim_{t\to 0} 
 2\pi   \sum_{(k_1,k_2)} \frac{e^{-m|\vec{z}+2\pi\vec{k}|}}{|\vec{z}+2\pi\vec{k}|}
\quad , \quad (z\ne 0)\ .
\ee
Now, on the torus, $z^i=x^i-y^i\in [0,2\pi)$ and $|\vec{z}+2\pi\vec{k}|=\sqrt{(x^1-y^1+2\pi k_1)^2+(x^2-y^2+2\pi k_2)^2}$. The sum then is dominated by the value of $\vec{k}$ such that this expression takes the smallest possible value, which corresponds to the geodesic distance $\ell(x,y)$ on the torus.
Hence
\be\label{calLPoisson2}
{\cal L}(t,z) \sim_{t\to 0} 
 2\pi   \frac{e^{-m \ell(x,y)}}{\ell(x,y)} \ \big(1+ {\cal O}(e^{-m a})\big) \quad , \ a>0 \quad , \quad x\ne y\ .
 \ee
Again, for $m\to 0$, the sub-leading terms no longer are sub-leading and we get back the divergence of ${\cal L}$.

We will be mainly interested in the regime where both $t$  and $\ell(x,y)$ are small. Then we have seen that ${\cal L}(t,0)\sim_{t\to 0} \frac{1}{m} +\frac{\pi^{3/2}}{\sqrt{t}}$ and, for $\ell(x,y)\ne 0$ but small, ${\cal L}(0,z)\sim_{t\to 0} \frac{2\pi}{\ell(x,y)}$. A somewhat ad hoc formula for ${\cal L}(t,z)$ that incorporates both behaviours is
\be\label{calLadhoc}
{\cal L}(t,z)\sim_{{\rm small}\ t,\ {\rm small} \ \ell(x,y)} \ \ \big(\sqrt{\frac{\pi}{t}}+\frac{1}{\pi m}\big)
\int_{-\pi/2}^{\pi/2} \d\f\, e^{-\frac{\ell^2(x,y)}{4 t} \sin^2\f} \ .
\ee
Indeed, for $x=y$, i.e. $\ell^2(x,y)= 0$, the $\f$-integral trivially gives $\pi$ and one gets correctly $\frac{\pi^{3/2}}{\sqrt{t}}+\frac{1}{m}$. On the other hand for small but fixed $\ell(x,y)\ne 0$ and $t\to 0$, the integral can be evaluated by a  saddle-point approximation and yields $\int_{-\infty}^\infty \d\f \, e^{-\frac{\ell^2(x,y)}{4 t}\f^2}=\sqrt{\frac{4\pi t}{\ell^2(x,y)}}$, so that ${\cal L}(t,z)\sim  \big( \sqrt{\frac{\pi}{t}}+\frac{1}{\pi m}\big)  \sqrt{\frac{4\pi t}{\ell^2(x,y)}}=\frac{2\pi}{\ell(x,y)}+{\cal O}(\sqrt{t})$, as it should.

Recall that $K_-(t,x,y)$ was given by $D_x L(t,x,y)=\big(i\s_x\frac{\del}{\del x^1} + i \s_z \frac{\del}{\del x^2} +m\g_*\big) L(t,x,y)$ where 
\be\label{Lfinally}
L(t,x,y)= \frac{i}{4\pi^2} {\cal L}(t,x,y)  {\bf 1}_{2\times 2} \sim_{t\to 0} \frac{i}{2\pi} \frac{e^{-m\ell(x,y)}}{\ell(x,y)} \,  {\bf 1}_{2\times 2}\ \big(1+ {\cal O}(e^{-m a})\big) \quad , \quad x\ne y \ ,
\ee
It is straightforward to work out the corresponding asymptotics of $K_-(t,x,y)$ for $x\ne y$, but we will not do it here. Instead, in the main text we obtain the relevant information directly from the relation $\zeta_-(s,x,y)=- D_x \zeta_+(s+\frac{1}{2},x,y)$.


\end{appendix}

\setlength{\baselineskip}{.48cm}

\end{document}